\documentclass{siamltex}

\usepackage{subfigure}
\usepackage{graphicx}

\usepackage{color}
\usepackage{float}
\usepackage{bm}
\usepackage{hyperref}
\usepackage{listings}
\usepackage{mathrsfs}
\usepackage{amsmath}
\usepackage{amssymb}

\providecommand{\eps}{\varepsilon}

\numberwithin{equation}{section}

\DeclareMathAlphabet{\itbf}{OML}{cmm}{b}{it}

\newcommand{\om}{\omega}

\newcommand{\la}{\lambda}

\newcommand{\bP}{{\itbf{P}}}

\newcommand{\bx}{\mathbf{x}}
\newcommand{\by}{\mathbf{y}}

\newcommand{\vy}{\vec{\mathbf{y}}}

\newcommand{\vx}{\vec{\mathbf{x}}}

\newcommand{\dsp}{\displaystyle}
\newcommand{\IKM}{{\cal J}^{\tiny \mbox{KM}}}
\newcommand{\ICINT}{{\cal J}^{\tiny \mbox{CINT}}}

\newcommand{\nc}{\newcommand}
\nc{\bequ}{\begin{equation}}
\nc{\eequ}{\end{equation}}
\nc{\barr}{\begin{array}}
\nc{\earr}{\end{array}}
\nc{\RR}{\mathbb{R}}

\nc{\bka}{\boldsymbol{\kappa}}

\begin{document}
\title{Time and direction of arrival detection and filtering for imaging in strongly scattering random media}
  
 \author{L.  Borcea \footnotemark[1], G. Papanicolaou \footnotemark[2]  and C.
  Tsogka \footnotemark[3]}
  
  \maketitle

\renewcommand{\thefootnote}{\fnsymbol{footnote}}

\footnotetext[1]{Department of Mathematics, University of Michigan,
  Ann Arbor, MI 48109 (borcea@umich.edu)} 

\footnotetext[2]{Mathematics Department,
Stanford University,
Stanford, CA 94305.
(papanicolaou@stanford.edu)}

\footnotetext[3]{Mathematics and Applied Mathematics,
  University of Crete and IACM/FORTH, GR-71409 Heraklion,
  Greece. (tsogka@uoc.gr)}

\maketitle 
\begin{abstract}
We study detection and imaging of small reflectors in heavy clutter, using an array of transducers that 
emits and receives sound waves.  Heavy clutter means that multiple scattering of the waves in the heterogeneous host 
medium is strong and overwhelms the arrivals from the small reflectors. Building on the adaptive time-frequency filter of 
\cite{BPT-LCT}, we propose a robust method for detecting the direction of arrival of  the direct echoes from the small reflectors,  
 and suppressing the unwanted clutter backscatter. This  improves the resolution of  imaging.  We illustrate the performance of 
 the method  with realistic numerical simulations in a non-destructive testing setup. 
\end{abstract}

\section{Introduction}
\label{sect:intro}
We study detection and imaging of remote small reflectors in a 
strongly scattering medium, aka heavy clutter, using an array of $N$ transducers that emit and receive sound waves.  
This is a difficult inverse problem because 
the echoes arriving directly from the  reflectors are weak by the time they reach the array 
and are overwhelmed by the  waves multiply scattered in clutter. We call these waves clutter backscatter and note that  they  arrive at the array long before and after the direct echoes.

The array probes sequentially the medium 
with pulses emitted from one transducer at a time, and records the resulting  acoustic pressure waves at all the $N$ transducers. These recordings 
form the $N \times N$ array response matrix $\bP(t)$, which is a function of time $t$. 
The detection problem is to distinguish in  $\bP(t)$, which is dominated by clutter backscatter, the time and 
direction of arrival of the weak echoes from the small reflectors. For imaging we need to extract these echoes from  
$\bP(t)$, and use them to localize the reflectors.

Heavy clutter arises in applications of imaging through foliage or the turbulent atmosphere, in nondestructive testing 
of materials, and so on. It has received much attention lately, specially in the context of imaging with passive arrays of receivers which are either near the imaging region or are separated from it by a non scattering medium \cite{GPST15}. In these problems the waves  emitted from remote sources travel through clutter before reaching the receivers and  the small reflectors.
Due to the favorable placement of the receivers, the clutter effects can be suppressed by computing the cross-correlations of the 
recordings and using appropriate time windowing \cite{bakulin,GPST15}. The images formed with such cross-correlations  are as good as if there
were no clutter, as shown with analysis and numerical simulations in  
\cite{GPST15}.  

In many applications it is not possible to place receiver arrays near the imaging region, or behind the heavy clutter. 
For example, in nondestructive testing, the measurements are necessarily confined to the surface of the tested body, and 
the small reflectors (defects) are buried deep inside, as we assume in this paper. 
The suppression of  clutter backscatter  is much more challenging in this case, and requires 
carefully designed data filters.  

A filter  of waves backscattered by  a randomly layered medium was  proposed and studied in  \cite{LAY_PAP2}. It is  efficient, but since it relies on the layered structure it does not generalize to other clutter. 
The  filter  in \cite{Aubry-Derode09a,Aubry-Derode09b,Aubry-Derode09c}  seeks to separate single from 
multiple scattering waves by performing a rotation of the response matrix followed by a projection.
It uses that when the array aperture  is small with respect to the distance to the small reflectors, the single scattering part of  $\bP(t)$ {\em i.e.}, the direct arrivals from the small reflectors, is approximately a Hankel matrix. After the rotation, 
which involves discarding  a large part of $\bP(t)$, the filtering is carried out by a projection on the space of certain rank one matrices.
The detection method in  \cite{Aubry-Derode09a,Aubry-Derode09b,Aubry-Derode09c} requires measurements of  the response matrix  from  a  part of the medium that does not contain the small reflectors.

The detection and filtering method proposed in this paper  is an extension of that in  \cite{BPT-LCT}. 
It analyzes the response matrix $\bP(t)$ in sequentially refined time windows,  using the singular value decomposition (SVD) of the local-cosine transform (LCT) of $\bP(t)$. The point is that in time windows that contain only clutter backscatter, $\bP(t)$ 
resolved over frequencies is  a "noise" matrix\footnote{The quote stands for the fact that clutter backscatter does not give a usual noise matrix
with identically distributed and uncorrelated entries, such as Gaussian.}.  Its SVD analysis 
reveals that  the larger singular values  are clustered together, and have similar behavior across frequencies. In the windows that contain echoes from the small scatterers, $\bP(t)$ 
is a perturbation of a noise matrix, and detection can be carried out by seeking singular values that are significantly larger than the others
across frequencies. 
The success of the detection depends on the strength of the perturbation relative to noise. This improves as we  refine the time windows. 
However, there is a trade-off. If the windows are too small, they cannot capture the arrival of the echoes from
the small reflectors at all the receivers in the array.  The arrival times vary across the array, and the window selection must take this into account.
The adaptive time-frequency algorithm in \cite{BPT-LCT} is designed to address this trade-off.

An analysis of the adaptive time-frequency algorithm in \cite{BPT-LCT} was carried out in 
\cite{ABPT} in the case of randomly layered media, but the method applies to general clutter.  Here we extend the algorithm so that 
it also selects the direction of arrival of the echoes from the small reflectors. This leads to improved data filtering and better resolution of the 
images obtained with any coherent method. 
We illustrate this using both the coherent interferometric imaging method \cite{BPT-05,BPT-ADA, BGPT_11} and the Kirchhoff migration method \cite{BIONDI,BLEISTEIN}.

The paper is organized as follows: In section \ref{sec:general} we formulate the problem. In section 
\ref{sec:clutter} we  illustrate with numerical simulations the difficulty of imaging in  heavy clutter. 
In section \ref{sect:alg} we present our detection and imaging algorithm. We review 
its first step from \cite{BPT-LCT} in section \ref{sec:lct}, and describe in detail the  new step for direction of arrival detection and filtering in section \ref{sec:angle}.  The performance of the algorithm is illustrated in section \ref{sec:numerics} using  numerical simulations carried out in a setup
relevant to non-destructive testing. We end with a summary in section \ref{sec:sum}

\section{Formulation of the problem}
\label{sec:general}
The array gathers the response matrix $\bP(t)$ with entries 
$P(t,\vx_r,\vx_s)$  by emitting pulses $f(t)$ from $\vx_s$ for
$s=1,\ldots,N$, and recording the scattered waves at
the receiver locations  $\vx_r$ for $r=1,\ldots,N$. The measurements  are modeled by the 
 solution of the wave equation 
\begin{eqnarray}
  \dsp \frac{1}{v^2(\vx)} \frac{\partial^2 P(t,\vx,\vx_s)}{\partial t^2} -  
  \Delta  P(t,\vx,\vx_s) &=& f(t) \delta(\vx -\vx_s), 
\qquad  \vx  = (\bx,z) \in \mathbb{R}^d,   \label{eq:pwaves}
 \end{eqnarray}
for $d \ge 2$ and time $t > 0$, with initial conditions
\begin{eqnarray}
  P(0,\vx) = 0,    \quad \dsp \frac{\partial P(0,\vx)}{\partial t} &=& 0. \label{eq:pwavesi} 
\end{eqnarray}
Here we introduced the system of coordinates with range axis $z$ in the 
direction of propagation of the waves, pointing from the array to the reflectors that we wish to image, and cross-range $\bx $ in the  plane $\mathbb{R}^{d-1}$ orthogonal to it.   

We model the emitted pulse as 
\[
f(t) = e^{-i \om_o t} f_{B}(t),
\]
where $\om_o$ is the carrier frequency and $f_B$ is a 
function with Fourier transform $\widehat f_B$ supported in the interval $(-\pi B,\pi B)$, where $B$ is the 
bandwidth. Then,
\begin{equation}
\widehat f(\om) = \dsp \int_{-\infty}^\infty e^{i (\om-\om_o) t} f_{B}(t) d
t = \widehat f_{B}(\om-\om_o),
\label{fft}
\end{equation}
is supported at frequencies  $\om \in (\om_o-\pi B, \om_o + \pi B)$. 

If the small reflectors are penetrable inclusions, we can model them and the clutter 
by  $v(\vx)$ in (\ref{eq:pwaves}), satisfying 
\begin{equation}
\label{fluct}
\dsp 
\frac{1}{v^2(\vx)} = \frac{1}{c^2}\left[1+\varepsilon
  \mu(\vx)  + 
  \rho(\vx) \right].
\end{equation}
Here $c$ is the constant reference speed and 
$\rho(\vx)$ is the reflectivity of the   inclusions, supported in the union of the disjoint domains $\Omega_m$, centered at 
points $\vy_m$,  for $m = 1, \ldots, M$.  The inclusions are round and small, meaning 
that their volumes $|\Omega_m|$  satisfy 
$
|\Omega_m|^{1/d} < \lambda_o,
$
where $\lambda_o = 2 \pi c/\omega_o$ is the central wavelength. However, they
have a much larger reflectivity  than the heterogeneities in the cluttered medium.
This is why we can hope to  image them. 

If the small reflectors are impenetrable, 
they are modeled with boundary conditions 
at $\partial \Omega_m$. In the simulations they 
are soft scatterers, so 
\begin{equation}
P(t,\vx,\vx_s) = 0, \quad \vx \in \partial \Omega_m, ~ ~ m = 1, \ldots, M,
\end{equation}
and the wave speed $v(\vx)$ satisfies 
\begin{equation}
\label{fluct1}
\dsp 
\frac{1}{v^2(\vx)} = \frac{1}{c^2}\left[1+\varepsilon
  \mu(\vx) \right].
\end{equation}

The clutter is a conglomerate of small and weak heterogeneities, which are impossible to know in detail. They introduce
uncertainty in the wave propagation model which translates into uncertainty of the waves measured at the array. This impedes 
the imaging process. 
We model the uncertainty of $v(\vx)$ with the mean zero random process $\mu$, which is assumed 
statistically homogeneous, bounded almost surely, with integrable autocorrelation
\[
\mathcal{C}(\vx) = \mathbb{E} [ \mu(\vx + \vx') \mu(\vx')],
\]
where $\mathbb{E}$ denotes expectation. We normalize the process by 
$
\mathcal{C}({\bf 0}) = 1, $ so $\eps \ll 1$ scales the small amplitude of the fluctuations. 

In imaging we probe a single heterogeneous medium, corresponding to one realization of the process $\mu$.
Any heterogeneity in this medium is a weak scatterer when compared with the reflectors that we wish to image,
as modeled by $\eps \ll 1$. However, there are many heterogeneities and their cumulative scattering effects 
add up over long distances of propagation of the waves. This cumulative scattering is responsible for the 
strong reverberations registered at the array,  the heavy clutter 
backscatter. 

The  detection problem seeks to identify the time and direction of arrival of the single scattered waves 
at the reflector locations $\vy_m$, for $m = 1, \ldots, M$. The goal of filtering is to suppress the heavy clutter backscatter
and emphasize these direct arrivals, so that better estimates of $\{\vy_m\}_{1 \le m \le M}$ can be obtained with coherent imaging 
methods such as coherent interferometry (CINT) \cite{BPT-ADA, BGPT_11} or Kirchhoff migration (KM) \cite{BIONDI,BLEISTEIN}.  

The KM imaging function is 
\begin{align}
\IKM(\vy) &= \sum_{r = 1}^{N} \sum_{s = 1}^{N}
 P\left(\tau(\vx_s,\vy) + \tau(\vy,\vx_r),\vx_r, \vx_s  
\right) \nonumber \\
&= \sum_{r = 1}^{N} \sum_{s = 1}^{N} \int_{-\infty}^\infty \frac{d \om}{2 \pi} \, 
\widehat P(\om,\vx_r,\vx_s)
 \exp{\left\{ -i \om \left[ \tau(\vx_s,\vy) + \tau(\vy,\vx_r) \right] \right\}}, 
\label{eq:KMom}
\end{align}
where $\vy$ are the search points  in the imaging region. It adds the entries of the response matrix delayed by  
the travel time  from the sources to the imaging point and then back to the receivers. The 
travel times are calculated in the reference medium, at wave speed $c$, 
\begin{equation}
\tau(\vx,\vy) = {|\vx-\vy|}/{c},
\label{eq:tau}
\end{equation}
and the evaluation of $P(t,\vx_r,\vx_s)$ at the round trip travel time 
$\tau(\vx_s,\vy) + \tau(\vy,\vx_r)$ is called backpropagation to $\vy$. 
The estimates of the reflector locations are the peaks of $\IKM(\vy)$. The direct arrivals from the reflectors add constructively at points  $\vy \in \{\vy_1, \ldots, \vy_M\}$,  and the KM imaging  method works well when 
the clutter backscatter is weak.

The CINT imaging function is given by 
\begin{eqnarray}
\ICINT(\vy)&=&\int_{-\infty}^\infty \frac{d \om}{2 \pi}  \int\limits_{|\omega-\omega'|\leq
\Omega_d} \hspace{-0.2in}  \frac{d\omega'}{2 \pi} 
 \sum_{r,r' \in \mathscr{S}_d(\om+\om') }  \sum_{s,s' \in \mathscr{S}_d(\om+\om') } \hspace{-0.1in}
\widehat{P}(\om,\vx_r,\vx_s)\overline{\widehat{P}(\om',\vx_{r'},\vx_{s'})}
\nonumber \\ && \vspace{0.1in} \exp\left\{-i\omega
[\tau(\vx_r,\vy)+\tau(\vy,\vx_s)] + i  \omega' [\tau(\vx_{r'},\vy) +
\tau(\vy,\vx_{s'}]\right\}.
\label{cint}
\end{eqnarray} 
It also uses backpropagation to $\vy$ 
via travel time delays,  but it does not sum directly the measurements. It sums 
their local cross-correlations, calculated at nearby frequencies $\om$ and $\om'$ satisfying 
$|\om-\om'| \le \Omega_d$, and at nearby sources and receivers, with indexes in the frequency dependent
sets 
\[
\mathscr{S}(\om+\om') = \left\{r,r' = 1, \ldots, N, ~ ~ |\vx_r - \vx_{r'}| \le X_d\left(\frac{\om+\om'}{2} \right) \right\}.
\]
Here $\Omega_d$ and $X_d$ are the decoherence frequency and length. They define the frequency
and sensor location offsets over which the waves scattered in clutter decorrelate statistically. They play 
an important role in the statistical stabilization of the CINT imaging function, and can be obtained 
adaptively during the image formation as explained in detail in 
\cite{BPT-ADA, BGPT_11,borcea2007asymptotics}.  CINT can mitigate moderate clutter backscatter. Explicitly, 
it can image at distances that  do not exceed a few transport mean free paths in the cluttered medium. 
In this paper we consider stronger clutter backscatter,  which cannot be handled by CINT alone, as 
shown with 
numerical simulations in the 
next section.

 \section{Illustration of heavy clutter effects on imaging}  
\label{sec:clutter}
\begin{figure}[H]
\subfigure[The imaging problem setup.]{
\includegraphics[height=4cm]{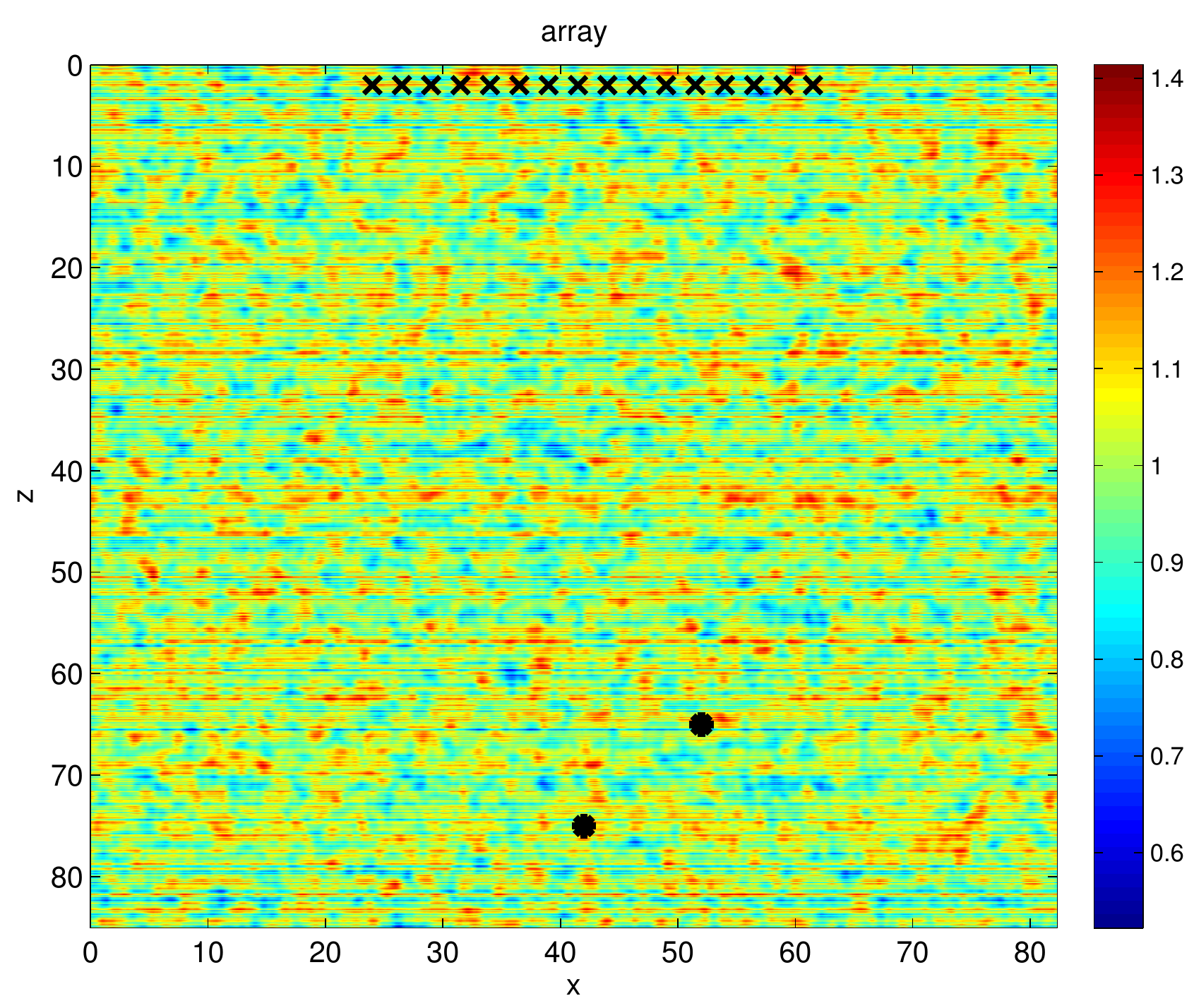}}
\subfigure[Time traces]{
\raisebox{0.5cm}{
\includegraphics[width=7.5cm]{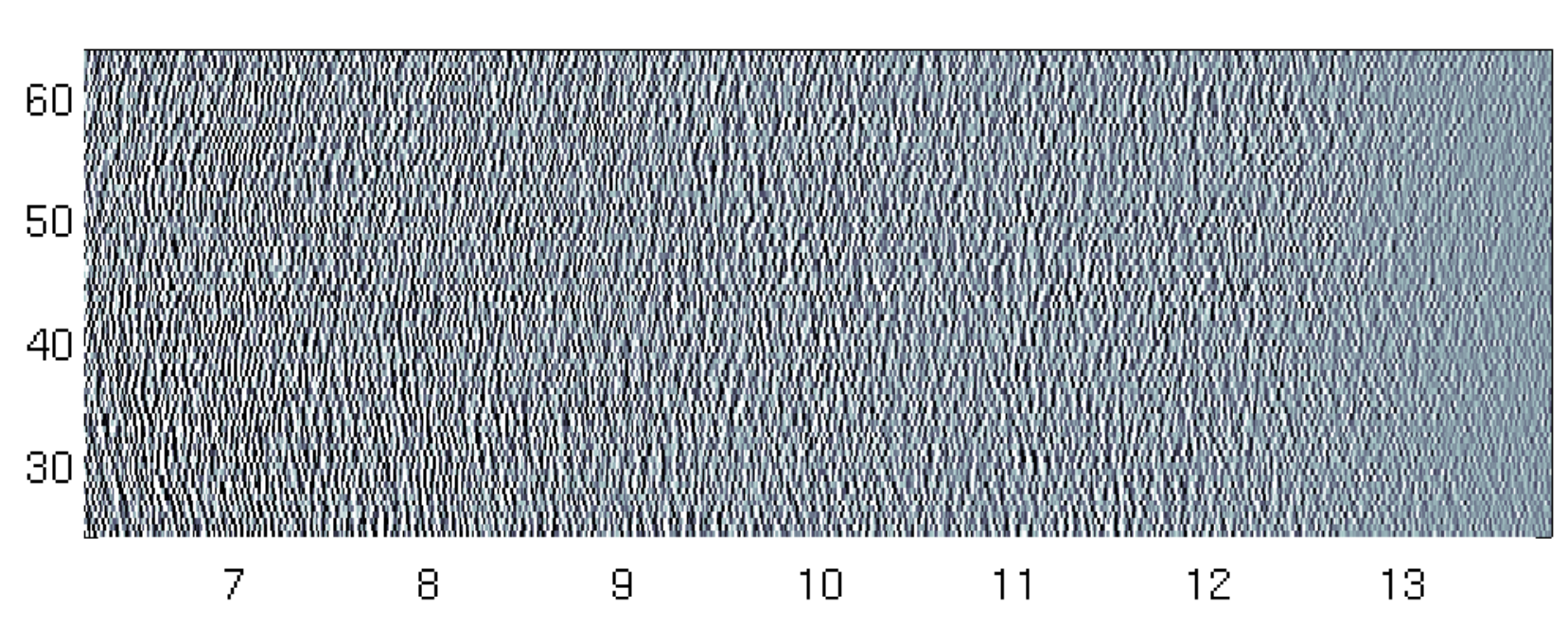}}}
\caption{(a) Two small, sound soft reflectors  embedded in a strongly scattering
  medium. The array is on the top. The velocity of the medium
  fluctuates around the constant $c=1$Km/s. The fluctuations are shown with colors.  The horizontal axis
  is cross-range and the vertical axis  is range,  in units of
  $\la_o$. (b) The display of $P(t,\vx_r,\vx_s)$ as a function of time on the abscissa and $x_r$ on the ordinate, for the source at $\vx_s = (44 \la_o,2 \lambda_o)$.}
\label{fig:setup}
\end{figure}
To illustrate how clutter impedes imaging, we present here the results of a numerical simulation 
in two dimensions, in  the setup depicted   on the left in Figure \ref{fig:setup}. There 
are two small reflectors to image, shown  with the black dots.  They are  modeled as 
 sound 
soft disks of radius $\lambda_o/4$, centered at $\vy_1 = (42\lambda_o,75 \lambda_o)$ and $\vy_2 = (52 \lambda_o,65 \lambda_o)$. 
The array is linear, and consists 
of $N = 80$ transducers. The range axis is orthogonal to it, 
and points downward in the figure. The transducer locations are 
\[
\vx_r=\left( x_r,2\lambda_o\right), \quad x_r = 24\lambda_o+(r-1)\frac{\lambda_o}{2}, \quad r = 1, \ldots, 80,
\]
so the array has aperture $a \approx 40 \la_o$, which is about half the range of the reflectors.

The  clutter is a realization of 
\begin{equation}
\label{eq:mu_c}
\mu(\vx) =\frac{1}{\sqrt{2}} \left[ \mu_{i}(\vx)  + \mu_{l} \Big(z\Big)\right],
\end{equation}
where $\mu_i$ and $\mu_l$ are mean zero, statistically homogeneous random processes.
The first models an isotropic random medium with autocorrelation 
\begin{equation}
\label{eq:mu_i}
\mathbb{E}[\mu_i(\vx)\mu_i(\vx')]= \left(1+\frac{|\vx-\vx'|}{\ell}\right)
e^{-\frac{|\vx-\vx'|}{\ell}},
\end{equation}
and correlation length $\ell = \la_o/4$. The second models a randomly layered 
medium with autocorrelation 
\begin{equation}
\label{eq:mu_l}
\mathbb{E}[\mu_l(z)\mu_l(z')]=\left(1+\frac{|z-z'|}{\ell_z}\right)
e^{-\frac{|z-z'|}{\ell_z}},
\end{equation}
and correlation length $\ell_z = \la_o/50$. The amplitude scale  of the fluctuations $\mu(\vx)$ is $\eps = 0.1$, and the 
actual wave speed   $v(\vx)$ used in the simulation is shown with colors in Figure \ref{fig:setup}. 

The simulation parameters are typical for an  ultrasonic non-destructive testing experiment \cite{Aubry-Derode09a}. 
The array probes the medium with Ricker pulses, which are second derivatives of a Gaussian, with central frequency 
 $\om_o /(2 \pi) = 10$MHz and standard deviation $10$MHz.  The reference velocity is $c = 1$Km/s, so $\la_o = 0.1$mm. All the lengths in Figure \ref{fig:setup} are scaled by $\lambda_o$.

The array response matrix $\bP(t)$ is obtained by solving numerically the wave
equation (\ref{eq:pwaves}) in $\mathbb{R}^2$, using  the perfectly matched
absorbing layer technique \cite{Berenger94PML}. The numerical method uses a finite
element discretization in space of \eqref{eq:pwaves}, written as a 
first order hyperbolic system \cite{Bec1,Bec2}. The discretization in time is
with  standard finite differences. 

We display on the right in Figure \ref{fig:setup} the recordings $P(t,\vx_r,\vx_s)$ for $r = 1, \ldots, 80$ and 
$s = 41$. Borrowing  terminology from the seismic literature, we call the recordings time traces. The direct arrivals from the two 
sound soft disks are weak and cannot be seen because they are dominated by the clutter backscattered waves, which arrive before and after them.
\begin{figure}[H]
\centerline{\subfigure[KM]{\includegraphics[width=0.3\textwidth]{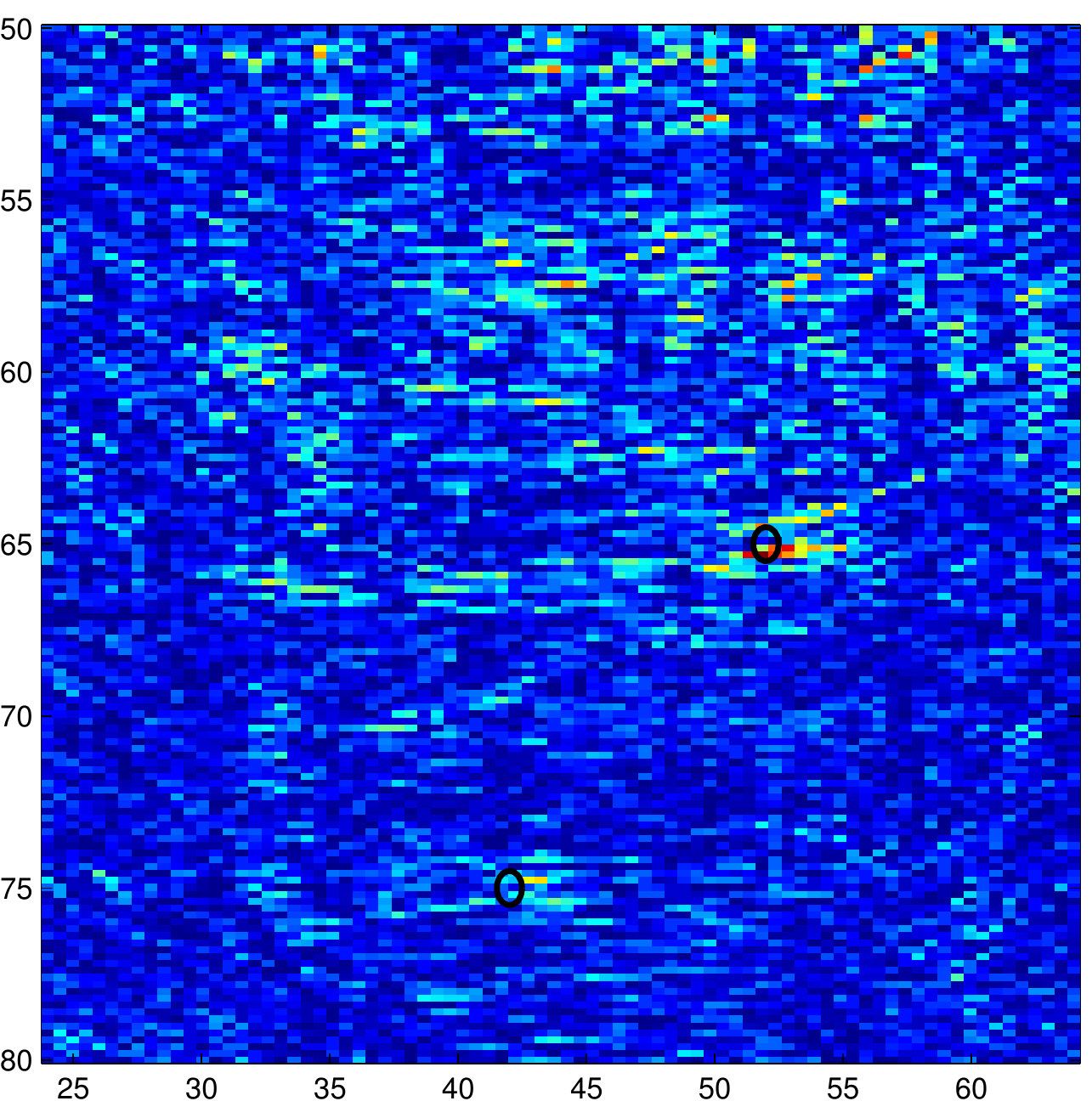}}
\subfigure[CINT]{\includegraphics[width=0.3\textwidth]{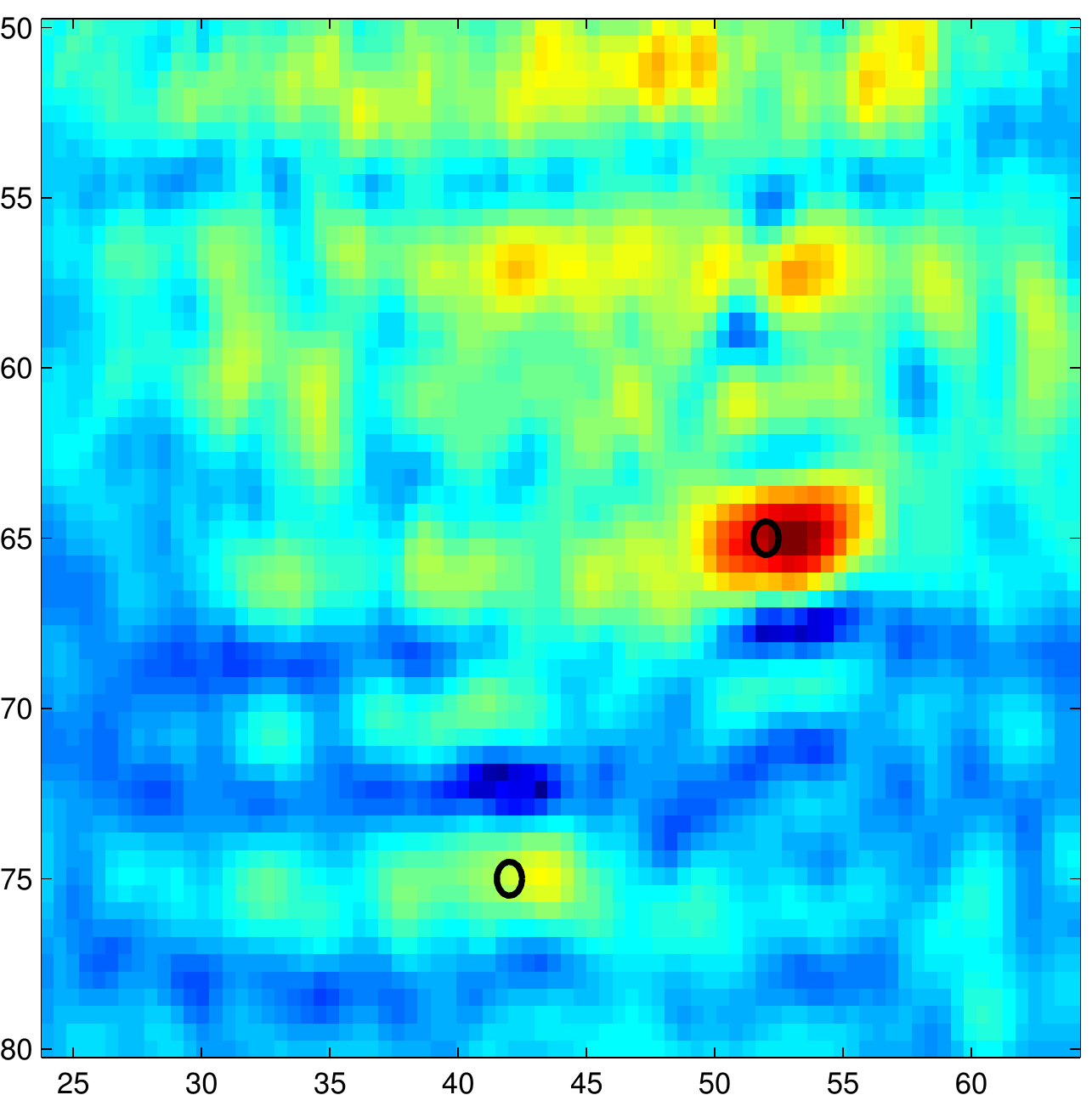}}}
\caption{Kirchhoff  migration (a) and CINT (b) images obtained for the data
  shown on Figure \ref{fig:setup}(b). The full array response matrix is
  used to obtain these images and not just the central illumination
  data. The true location of the scatterers is shown with black circles.}
\label{fig:imaging}
\end{figure}

The KM and CINT images are shown in Figure
\ref{fig:imaging}, 
where the two sound soft reflectors  are indicated  with black circles. We note that  both images 
have peaks near the locations of the reflectors. In particular, CINT produces a strong peak at the 
reflector that is closer to the array.  However, there are many other peaks,  which are stronger than
the peak at the second reflector. The algorithm described in the next section is designed to mitigate 
the clutter backscatter, and therefore improve the quality of the images. 

\section{Detection and filtering of clutter backscatter}
\label{sect:alg}
Our method of detection of the arrival of the weak echoes from the small reflectors, and of filtering the unwanted clutter backscatter, 
consists of two main parts, outlined here.  

The first part is as in \cite{BPT-LCT}, and we review it in section \ref{sec:lct}.
It analyzes the response matrix $\bP(t)$ in sequentially refined time windows using the discrete local cosine transform (LCT). Time windowing is useful because over the entire duration of the recordings the energy carried by the clutter backscatter (the "noise") overwhelms that of the useful echoes (the "signal").
As the time windows containing these echoes become smaller and smaller, the signal to noise ratio (SNR) improves. The LCT
allows a systematic window refinement and search, using a tree structure. 
It decomposes the entries of $\bP(t)$ over frequencies, locally in each window, and the 
detection  is done  by tracking the behavior of the leading singular values of the transformed matrix. 
Data filtering consists of zeroing the LCT coefficients in the windows where no signal is detected, projecting 
on the subspace spanned by the singular vectors of the distinguished singular values, 
and then inverting the LCT transform. We call the filtered response matrix $\bP^{^{TF}}\hspace{-0.04in}(t)$, where the index 
TF stands for time filtering. 

The second part of the method is the main contribution of this paper, and is described in section \ref{sec:angle}. 
It seeks to detect the direction of arrival of the desired echoes from the small reflectors, in addition to the arrival time, and to 
improve the data filter for better resolution of the images.
To do so, we begin with the backpropagation of $\bP^{^{TF}}\hspace{-0.04in}(t)$
using travel time delays from the array transducers to a reference point in the imaging region.  
This reference point is  at range equal to half the center time of the selected time window, 
multiplied by the wave speed $c$. If more  time windows are selected, than the backpropagation  is done for each of them. 
The purpose of the backpropagation is to remove the large phase of the entries of 
$\bP^{^{TF}}\hspace{-0.04in}(t)$,  so that we can better analyze the data 
around the detected arrival time.

We call the backpropagated matrix $\bP^{^{BP}}\hspace{-0.04in}(t)$ and its Fourier transform $\widehat{\bP}^{^{BP}}\hspace{-0.04in}(\omega)$.
Its entries are indexed by the receivers located at $\vx_r$ and sources at $\vx_s$, and we rotate it next  by representing it in the center and difference coordinates   $(\vx_s+\vx_r)/2$ and $\vx_r-\vx_s$. The motivation for the 
rotation is similar to that in \cite{Aubry-Derode09a,Aubry-Derode09b,Aubry-Derode09c}. Assuming that the array aperture is small
with respect to the range of the reflectors,  we may use the paraxial approximation to model the  direct arrivals from the small reflectors, the useful (coherent)  part of the data.  After the backpropagation, this  part  is approximately  
independent of the difference $\vx_r-\vx_s$, which is why it is advantageous to rotate  $\widehat{\bP}^{^{BP}}\hspace{-0.04in}(\omega)$.  

We denote the rotation by $\mathcal{R}$, and 
to suppress the remaining
clutter backscatter, we calculate the best approximation of  $\mathcal{R}\widehat{\bP}^{^{BP}}\hspace{-0.04in}(\omega)$ by a rank 
one matrix which is independent of $\vx_r-\vx_s$. The approximation is with respect to the Frobenius norm, and the 
result is denoted by $\mathcal{R} \widehat{\bP}^{^{AF}}\hspace{-0.04in}(\omega)$.  

To determine the direction of arrival of the coherent echoes {\em i.e.}, the wave vectors associated with the single scattered waves, 
we decompose  $\mathcal{R} \widehat{\bP}^{^{AF}}\hspace{-0.04in}(\omega)$ in plane waves using Fourier transforms.  The detection 
amounts to seeking maxima of the Fourier coefficients (the plane wave amplitudes), and the filtering is done by careful tapering over the other directions. The output of the algorithm is the inverse Fourier transform of the result,  rotated back to the coordinates $\vx_r$ and $\vx_s$. This is the filtered data to be used in the image formation.

\subsection{Adaptive time-frequency detection and filtering}
\label{sec:lct}
We review here the steps of the algorithm introduced in \cite{BPT-LCT}. 
The  input  is the array response matrix $\bP(t)$, for time $t \in [0,T]$  sampled in uniform $N_T$ time steps, where $N_T$ equals an integer power of $2$. The LCT decomposition \cite{Mallat} is done in  time windows arranged in a binary tree structure. 
At each level $l = 0,  \ldots, N_l$, 
the recording window $[T_o,T]$ is divided in $2^l$ windows, of size $\Delta_l = (T-T_o)/2^{l}$. The minimum size of the time windows is determined by the user defined maximum tree level $N_l$. 

Let us index the nodes of the tree by $(j,l)$, with $j =0,  \ldots, 2^l-1$ and $l = 0, \ldots, N_l$. 
Each  node is associated with the subspace spanned by the orthonormal bases
\bequ \label{eq:Fj}
\mathscr{B}_j^l=\left\{ \sqrt{\frac{2}{\Delta_l}}
  \chi \Big(\frac{t-t_j^l}{\Delta_l}\Big) \cos[\omega_n^l(t-t_j^l)], ~ ~ n \in \mathbb{Z}^+\right\},
\eequ
with discrete times  $t_j^l = T_o+j \Delta_l$ and frequencies 
$
\om_n^l = {\pi \left( n + 1/2 \right)}/{\Delta_l}$  of the decomposition in the smooth windows $\chi$. 
For any $l$, the union over $j$ of the bases $\mathscr{B}_j^l$  forms  an orthonormal basis of $L^2[T_o,T]$, and  at the next tree level
the span of $\mathscr{B}_j^l$ is split in two orthogonal subspaces, with bases $\mathscr{B}_{2j}^{l+1}$ and $\mathscr{B}_{2j+1}^{l+1}$.  We 
refer to \cite{Mallat} for details\footnote{
In the simulations the basis  (\ref{eq:Fj}) is discretized 
at the $N_T$ points $t$ of the interval $[T_o,T]$, and the
frequencies  $\omega_n^l$ sample the { same} bandwidth $\left( 0,
  {\pi}{N_T}/{T} \right), $ in steps $\pi/\Delta_l$, that increase
with the tree level $l$. The implementation
uses the Wavelab 850 MATLAB package \cite{Wavelab850} with window $\chi$ option 
"Sine".}.

\vspace{0.1in}
The steps of the time-frequency detection and fitering algorithm are: 
\begin{enumerate}
\itemsep 0.1in
\item {\em Computation of the discrete LCT transform of 
the array response matrix on a binary tree with maximum level $N_l$. 
This gives the $N \times N$ matrices
\begin{equation}
\widehat \bP^l{(t_j^l,\omega_n^l)} = \left\{ \widehat
P^l(t_j^l,\omega_n^l,\vx_r,\vx_s) \right\}_{r,s = 1,\ldots, N},
\label{eq:coeff}
\end{equation}
for $j = 0,1, \ldots, 2^l-1$ and $l = 0, \ldots, N_l$, with entries 
\begin{equation}
  \dsp \widehat{P}^l(t_j^l,\omega_n^l,\vx_r,\vx_s) = \int dt \, 
  P(t,\vx_r,\vx_s) \sqrt{\frac{2}{\Delta_l}} \chi\Big( \frac{t-t_j^l}{\Delta_l}\Big) 
  \cos[\omega_n^l (t-t_j^l)].
\label{eq:coeff2}
\end{equation}
}
\item {\em Calculate the singular value decomposition of  $\widehat\bP^l{(t_j^l,\omega_n^l)}$. Let 
 $\sigma^{l,j}_q(\omega_n^l)$ be the singular values,
for $q = 1, \ldots N$.  }
\item {\em Choose the frequency band $\mathcal{B} \in (0,\pi N_T/T)$ and the number $\bar{q}$ of largest 
singular values to be used in the detection.    }
\item {\em Detect the time window of interest as follows: \\ 
For $l=0:N_l$
\begin{enumerate}
\item[ ] Decide if there is at least one window indexed by $(j,l)$, where the largest singular values
are distinguished from the others across the frequencies in $\mathcal{B}$. 
If yes, let $l_o = l$ and $j_\star^{l_o}  = j$ and stop.\\
\end{enumerate}

\vspace{-0.1in}
\noindent For $l = l_0+1 :N_l$
 \begin{enumerate}
\item[ ] Let $j \in \{ 2j_\star^{l-1}, 2j_\star^{l-1}+1\}$ and decide in which of the two windows the largest singular values
are better separated from the rest.  Call the decision $j_\star^l = j$. 
 If the selection is ambiguous, set $l = l-1$ and stop. 
 \end{enumerate}
}
\item {\em Let the chosen time window be indexed by $(j_\star^l,l)$. Set to zero the 
LCT coefficients in all other windows at level $l$. This is equivalent to multiplying \eqref{eq:coeff2} with  the Kronecker $\delta_{j,j_\star^l}$.}

\item {\em Project
$ \delta_{j,j_\star^l}\widehat \bP^l(t_{j}^l,\om_n^l)$ on the subspace of low rank matrices
with singular vectors corresponding to the distinguished top singular
values. The projection is done for frequencies $\om_n \in \mathcal{B}$.  All other coefficients are set to zero. 
}

\item {\em The output of the algorithm is the filtered response matrix $\bP^{^{TF}}\hspace{-0.02in}(t)$ obtained 
with the inverse LCT of the entries of the matrix obtained at step (6).
}
\end{enumerate}
\subsection*{Remarks} 
The  number $\bar{q}$ of singular values at step (3) should be larger than the number $M$ of reflectors that we wish to image. 
We also should have enough measurements, meaning that $N \gg \bar{q} > M$. The bandwidth ${\mathcal B}$ is the 
part of the frequency support of the probing pulse over which the reflectors are detectable. This ${\mathcal B}$ depends on 
the clutter, but in  general it is at the lower frequencies that the detection is easier. 

The details on how the algorithm searches for the distinguishable, leading singular values are given in \cite{BPT-LCT}. Note that 
at step (4) we search first from the bottom to the top of the tree.  At the root level $l = 0$, the data is expected to be dominated 
by the clutter backscatter, so $\widehat \bP^0(T_o,\omega_n^0)$ are like noise matrices. This is illustrated in Figure \ref{fig:lct-time} and the numerical simulations 
in \cite{BPT-LCT} by the fact that all singular values $\sigma_q^{0,0}(\om_n^0)$ are clustered together across the frequencies. 
There is no distinguished or significant singular value. When the window sizes become small enough, the SNR in the 
windows that contain the useful echoes from the reflectors (the signal) improves, and  the largest singular values become well separated from the others. This is the level $l_o$ at step (4). The second part of the search at step (4) refines sequentially the windows of interest until the selection becomes ambiguous. 

The filters at steps (5) and (6) are for suppressing the clutter backscatter. First, they remove all the arrivals outside the selected
time window and then, they project the result on the subspace spanned by the singular vectors corresponding to the distinguishable 
singular values. 

\subsection{Direction of arrival detection and filtering}
\label{sec:angle}
The filtered array response matrix  $\bP^{^{TF}}\hspace{-0.04in}(t)$ given by the first part of the algorithm is localized in
a small time window  which contains the
echoes from the reflectors that we wish to image. Here we explain how we can detect the direction of arrival of these 
echoes and how we can improve the data filtering. 

 \begin{figure}[H]
    \begin{minipage}{1\textwidth}
      \centering
      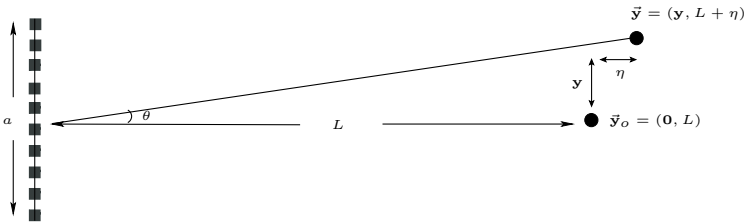
    \end{minipage}
    \caption{Illustration of an imaging setup with array
      aperture $a$ that is small compared to the range $L$. The reference point $\vy_o$ is determined from 
      the center time of a selected window and is along the range axis originating at the center of the array.} 
    \label{fig:array}
  \end{figure} 
Suppose that a small reflector  at $\vy$ is detected in the selected time window centered at $t_o$. Its distance to the center 
of the array, the origin of coordinates,  
is approximately $L = c/(2t_o)$, and we define the 
reference point   $\vy_o = ({\bf 0},L)$. 
As illustrated in Figure   \ref{fig:array}, $\vy_o$ is offset from $\vy$ by $\by$ in the cross-range plane and $\eta$ in range,
meaning that $\vy = (\by,L+\eta)$. 

The useful echoes for imaging, which are single scattered 
at $\vy$,  have the deterministic phase $k ( |\vy-\vx_s|+|\vy-\vx_r|)$ for the source receiver pair $(s,r)$, 
where $k = \om/c$ is the wavenumber. 
Assuming that the array is planar, with small aperture with respect to the range $L$, we let $\vx_r = (\bx_r,0)$ and obtain with
the paraxial 
approximation that
\begin{align}
|\vy-\vx_s|+|\vy-\vx_r| &= \sqrt{(L+\eta)^2 + |\bx_r-\by|^2} + \sqrt{(L+\eta)^2 + |\bx_s-\by|^2} \nonumber \\
&\approx 2 (L+\eta) + \frac{|\bx_r - \by|^2 + |\bx_s-\by|^2}{2 L} \nonumber \\
&= 2(L+\eta) + \frac{|\bar{\bx}_{rs}|^2}{L} + \frac{|\widetilde{\bx}_{rs}|^2}{4 L} + \frac{|\by^2|}{L} - \frac{2 \bar{\bx}_{rs} \cdot \by}{L},
\label{eq:parax1}
\end{align}
where $\bar{\bx}_{rs} = (\bx_r + \bx_s)$ and $\widetilde \bx_{rs} = \bx_r-\bx_s.$
For the reference point $\vy_o$ we have similarly 
\begin{align}
|\vy_o-\vx_s|+|\vy_o-\vx_r| &= \sqrt{L^2 + |\bx_r|^2} + \sqrt{L^2 + |\bx_s|^2} \nonumber \\
&\approx  2L + \frac{|\bar{\bx}_{rs}|^2}{L} + \frac{|\widetilde{\bx}_{rs}|^2}{4 L},
\label{eq:parax3}
\end{align}
so when backpropagating the filtered data $\bP^{^{TF}}\hspace{-0.04in}(t)$ to $\vy_o$ we achieve the following 
phase reduction of the direct arrivals in the selected time window
\begin{equation}
\label{eq:parax4}
k ( |\vy-\vx_s|+|\vy-\vx_r| -   |\vy_o-\vx_s|+|\vy_o-\vx_r|) \approx k \left(2 \eta + \frac{|\by^2|}{L} - \frac{2 \bar{\bx}_{rs} \cdot \by}{L}\right).
\end{equation}
The observation that these reduced phases are independent of the difference coordinates $\widetilde \bx_{rs}$ leads to
the detection and filtering algorithm described below.

The algorithm can be used for three dimensional problems, but to avoid cumbersome index notation we present it here
in two dimensions, for a linear array. The extension to three dimensions requires a modification of the indexing in the rotation operation at step (2) below.
 The steps of the algorithm are:

\begin{enumerate}
\itemsep 0.1in 
\item{\em For the selected time window, centered at $t_o$, define the $N \times N$ input matrix 
\begin{equation}
\itbf{P}^{^{IN}}\hspace{-0.04in}(t) = \itbf{P}^{^{TF}}\hspace{-0.04in}(t-t_o),
\label{eq:defPo}
\end{equation}
and 
Fourier transform it with respect to time $t$,
    \begin{equation}
    \widehat{\bP}^{^{IN}}\hspace{-0.04in}(\omega) = \int_{-\infty}^{\infty} e^{i \om t} \bP^{^{IN}}\hspace{-0.04in}(t) dt  = e^{i \om t_o} \int_{-\infty}^\infty e^{i \omega(t-t_o)} 
    \itbf{P}^{^{TF}}\hspace{-0.04in}(t-t_o) dt.
    \label{eq:defPo1}
    \end{equation}
Denote the entries of this matrix by 
$\widehat{P}^{^{IN}}\hspace{-0.04in}(\omega,x_r,x_s)$,  since for the linear array $\vx_r = (x_r,0)$, with $r = 1, \ldots, N$.
 
Backpropagate  $\widehat{\bP}^{^{IN}}\hspace{-0.04in}(\omega)$ to the test point $\vy_o$, and denote the  resulting  matrix by 
$\widehat{\bP}^{^{BP}}\hspace{-0.02in}(\omega)$, with entries  defined by
  \begin{equation}\widehat{P}^{^{BP}}\hspace{-0.02in}(\omega,x_r,x_s)  =   \widehat{P}_{o}(\omega,x_r,x_s) e^{-i \om t_o - i k
        (|\vy_o-\vx_s|+|\vy_o-\vx_r|)}, \quad r,s = 1, \ldots, N.
        \label{eq:BP}
   \end{equation}
}

\item {\em  Rotate $\widehat{\bP}^{^{BP}}\hspace{-0.04in}(\omega)$ by ninety degrees, to form a larger $(2N-1) \times (2N-1)$ 
 matrix, with entries indexed by the center and difference coordinates 
\[
\bar{x}_{rs} = \frac{x_r+x_s}{2}, \quad \widetilde{x}_{rs} = x_r-x_s.
\]
The rotation is done with the following  commands:

\begin{lstlisting}[mathescape]
$\itbf{H} = \widehat{\bP}^{^{BP}}\hspace{-0.04in}(\omega)$ 
${\mathcal R} \itbf{H}  = zeros(2N-1,2N-1)$
For  i=1:N 
   For  j=1: N 
          i1=i+j-1 
          i2=i-j-(1-N)+1
          $\mathcal{R}$H(i1,i2)= H(j,i)
   end
end
\end{lstlisting}
The rotated matrix has a rhombus structure as illustrated in Figure \ref{fig:rhombus}. The 
diagonals in $\itbf{H}$, which correspond to 
constant source receiver offsets $\widetilde{x}_{rs}$, form the columns of  $\mathcal{R}\itbf{H}$. The anti-diagonals of 
$\itbf{H}$, which correspond to common midpoints  $\bar{x}_{rs}$, form the rows of $\mathcal{R}\itbf{H}$. 
The 
resulting matrix is $\mathcal{R}\widehat{\bP}^{^{BP}}\hspace{-0.04in}(\omega)$.}

\begin{figure}[H]
      \centering
      \includegraphics[width=0.5\textwidth]{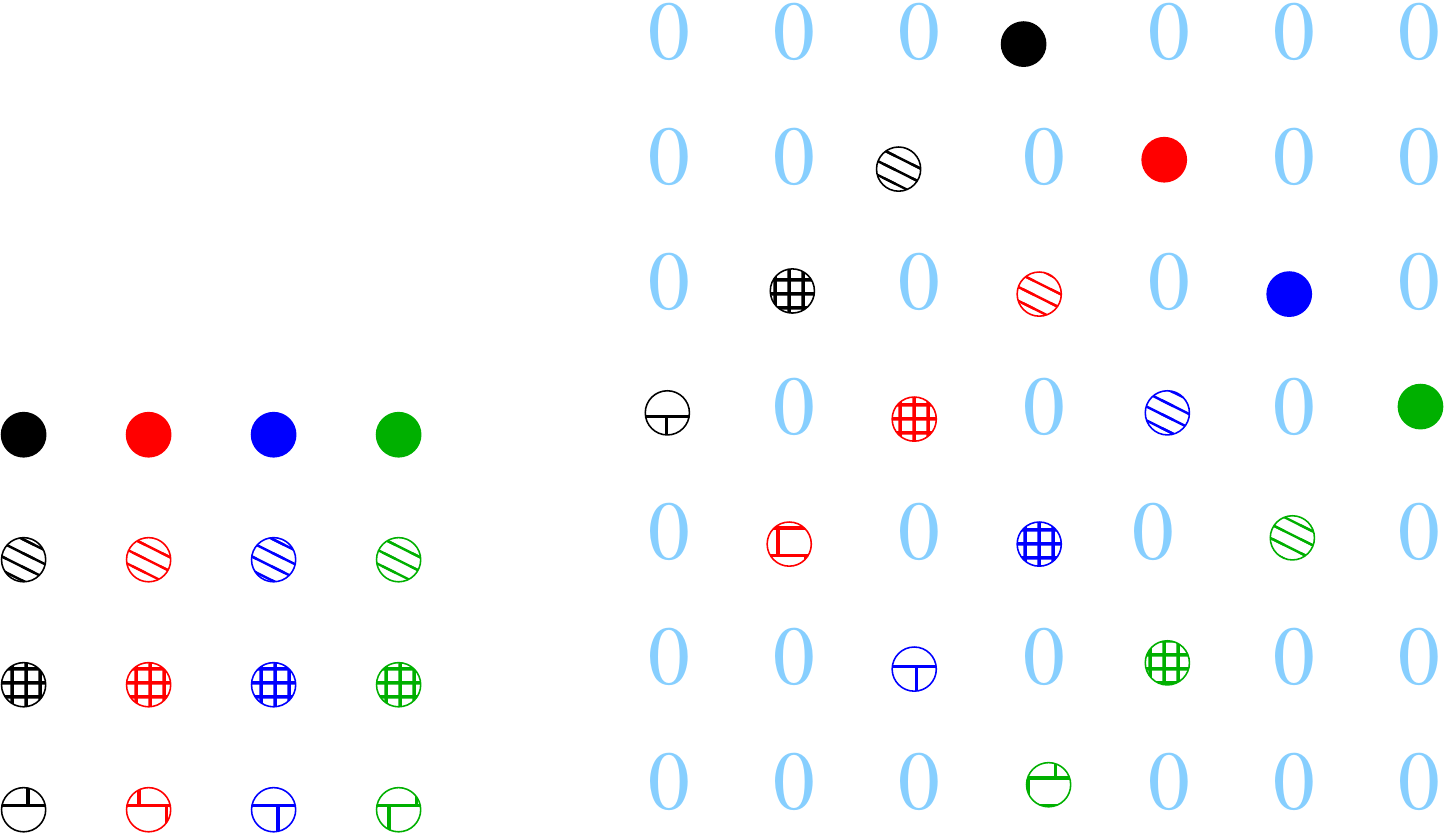}
    \caption{Illustration of a square $4 \times 4$ matrix $\itbf{H} $ on the left and its rotation ${\mathcal R} \itbf{H}$ on the right.   }
    \label{fig:rhombus}
  \end{figure}
  
\item {\em We know from equation  \eqref{eq:parax4} that the desired, coherent part of the matrix calculated at step (2)  should be independent of the 
source receiver offsets $\widetilde{x}_{rs}$. Therefore, we calculate the best approximation of ${\mathcal R} \widehat{\bP}^{^{BP}}\hspace{-0.02in}(\omega)$ by a matrix $\mathcal{R} \widehat{\bP}^{^{AF}}\hspace{-0.04in}(\omega)$ with identical columns $\widehat{\itbf{p}}(\omega)$, restricted to the support  of  ${\mathcal R} \widehat{\bP}^{^{BP}}\hspace{-0.02in}(\omega)$, {\em i.e.}, the non-zero elements of the rhombus. 
Let $\mathscr{S}$ be the set of indexes $(i,j)$ in this support, for $i,j = 1, \ldots, 2N-1$ enumerating the center and difference locations, 
renamed henceforth $\bar{x}_i$ and $\widetilde{x}_j$, and define
the indicator function  
\[
1_{_\mathscr{S}}(i,j) = \left\{ \begin{array}{ll}1, \quad & (i,j) \in \mathscr{S} \\
0, &\mbox{otherwise} \end{array} \right. .
\]
Then, $ \widehat{\itbf{p}}(\om)$ is the $2N-1$ column vector with entries $p(\om,\bar{x}_i)$ that minimize
\[
\sum_{i,j=1}^{2N-1} 1_{_{\mathscr{S}}}(i,j) \left| {\mathcal R} \widehat{P}^{^{BP}}\hspace{-0.02in}(\omega,\bar{x}_i, \widetilde{x}_j) - \widehat{p}(\omega,\bar{x}_i)\right|^2.
\]
We obtain that 
\begin{equation}
\widehat{p}(\omega,\bar{x}_i) = \frac{1}{n_{i}} \sum_{j=1}^{2 N - 1} {\mathcal R} \widehat{P}^{^{BP}}\hspace{-0.02in}(\omega,\bar{x}_i, \widetilde{x}_j), \quad 
n_i = \sum_{j=1}^{2N-1} 1_{_\mathscr{S}}(i,j),  \label{eq:FrobAp}
\end{equation}
where $n_i$ is 
is the number of non zero entries of the $i-$th row in the set $\mathscr{S}$. 
Here we used that $\mathcal{R} \widehat{\bP}^{^{BP}}\hspace{-0.04in}(\omega)$ is supported in $\mathscr{S}$. 

The approximation $\mathcal{R} \widehat{\bP}^{^{AF}}\hspace{-0.04in}(\omega)$ is the  matrix with entries
\begin{equation}
\mathcal{R} \widehat{P}^{^{AF}}\hspace{-0.04in}(\omega,\bar{x}_i, \widetilde{x}_j) = 1_{_\mathscr{S}}(i,j) \widehat{p}(\om,\bar{x}_i), 
\quad i,j = 1, \ldots, 2N-1.
\label{eq:FrobAp1}
\end{equation}
}

\item  {\em Take the Fourier transform  of \eqref{eq:FrobAp} with respect to $\bar{x}$. This is 
a plane wave decomposition, with wave vector samples $\kappa_i$, the dual variable to $\bar{x}_i$.  The transformed vector  is 
\begin{equation}
\label{eq:PLANEWAVE}
\breve{\itbf{p}}(\om) = \left(\breve{p}(\om,\kappa_i)\right)_{i=1, \ldots, 2N-1}.
\end{equation} 
}
\item {\em The detection of the direction of arrival of the direct echoes from the reflectors amounts to seeking maxima 
of \eqref{eq:PLANEWAVE} that are above a user defined tolerance. If there is one dominant reflector  for the selected time window, 
we expect a single maximum, denoted by $\kappa^\star(\om)$. For multiple 
reflectors we may have multiple maxima. If they are well separated, we use them one at a time.}
\item {\em To filter the residual unwanted clutter backscatter, we taper off the arrivals from the directions that are different 
than the selected $\kappa^\star(\om)$ at step (5).  The taper function is determined by the aperture $a$ of the array, which defines 
the resolution of order $a/L$ in the plane wave decomposition. 

In theory, the taper should be  a sinc function, due to the support of the entries of 
$\widehat{\itbf{p}}(\om)$ in the interval $[-a/2,a/2]$. Since we are interested only 
in the vicinity of the peak wave vector, we taper using a Gaussian centered at $\kappa^\star(\om)$, with  standard deviation 
$\beta$ determined by minimizing the least squares error between $ \breve{p}(\om,\kappa_i)$ and 
$\breve{p}(\om,\kappa_i) \exp[-(\kappa_i-\kappa_i^\star)^2/ (2 \beta^2)]$ in the vicinity of $\kappa^\star$, where $\breve{p}(\om,\kappa_i)$
drops up to half 
of its peak value.

We denote the tapered vector by $\breve{\itbf{p}}^{^{DoA}}\hspace{-0.04in}(\omega)$, with index DoA standing for 
direction of arrival. Its entries are defined by 
\begin{equation}
\breve{{p}}^{^{DoA}}\hspace{-0.04in}(\omega,\kappa_i) =  \breve{p}(\om,\kappa_i) 
e^{-\frac{[\kappa_i-\kappa^\star(\omega)]^2}{2 \beta^2}}, \quad i= 1, \ldots, 2N-1.
\label{eq:FrobAp3}
\end{equation}
} 
\item {\em Compute the inverse Fourier transform (with respect to $\kappa$)  of the tapered vector \eqref{eq:FrobAp3}. Its 
entries  are 
\[
\widehat{{p}}^{^{\,DoA}}\hspace{-0.04in}(\omega,\bar{x}_i) \sim \widehat{p}(\omega,\bar{x}_i) \star_{\bar{x}} e^{i 
\kappa^\star(\om) \bar{x}_i-\frac{\beta^2 \bar{x}_i^2}{2}},
\] 
where $\star_{\bar{x}}$ denotes convolution and $\sim$ denotes equal, up to a multiplicative constant. The phase in the right hand side of this equation carries the direction of arrival selected at step (6). 

}
\item {\em Define the filtered, rotated matrix $\mathcal{R} \widehat{\itbf{P}}^{^{DoA}}\hspace{-0.04in}(\omega)$, with entries 
\begin{equation}
\mathcal{R} \widehat{P}^{^{DoA}}\hspace{-0.04in}(\omega,\bar{x}_i, \widetilde{x}_j) = 1_{_\mathscr{S}}(i,j) \widehat{p}^{^{\,DoA}}\hspace{-0.04in}(\om,\bar{x}_i), \quad i,j = 1, \ldots, 2N-1.
\label{eq:FrobAp3p}
\end{equation}
Rotate it back to obtain the $N \times N$ matrix $\widehat{\itbf{P}}^{^{DoA}}\hspace{-0.04in}(\omega)$ with entries $\widehat{P}^{^{DoA}}\hspace{-0.04in}(\omega,x_r,x_s)$, for $r,s = 1, \ldots, N$.} 
\item {\em Undo the back propagation at step (1), equation \eqref{eq:BP}, by multiplying ~\\$\widehat{P}^{^{DoA}}\hspace{-0.04in}(\omega,x_r,x_s)$  with $\exp\left[i \om t_o + i k (|\vy_o-\vx_s| + |\vy_o - \vx_r|)\right]$, for $r,s = 1, \ldots, N$.}
\item {\em  The output of the algorithm is the inverse Fourier transform in time of the matrix calculated at step (9). We call it $\bP^{^{OUT}}\hspace{-0.04in}(t)$.}
\end{enumerate}

\subsection*{Remarks}

Equation \eqref{eq:parax4}, which states that after the backpropagation the direct echoes from the sought-after reflectors  carry phases that are independent of the source-receiver offset location, is used by the algorithm in two ways:  First, it rotates at step (2) the backpropagated data matrix 
 to the center and difference system of coordinates $\bar{x}_i$, and $\widetilde{x}_j$, and then approximates at step (3) the result by the closest matrix with identical columns,  
 independent of the offset coordinates $\widetilde{x}_{j}$.  Second, it Fourier transforms the result with respect to the center coordinates 
 $\bar{x}_{i}$, to determine at steps (4) and (5) the wave vector $\kappa^\star(\om)$ corresponding to the desired direct echoes. Equation \eqref{eq:parax4}  says that this should be approximately 
 $
 \kappa^\star(\om) \approx 2 k y/L.
 $
 The algorithm then suppresses the returns with wave vectors away from $\kappa^\star(\om)$  at step (6). The remaining steps (7)-(9) undo the rotation and the Fourier transform to return to 
 the source-receiver coordinates and the time domain in which the array response matrix is represented.

As stated before, the algorithm applies to three dimensions, with the only difference being in the indexing in the rotation 
operation at step (2), and the search for the direction of arrival at step (5) in two dimensions instead of one.

\section{Numerical simulations}
\label{sec:numerics}
 We begin in section \ref{sec:illustration} with the illustration of the 
direction of arrival detection and filtering algorithm, for the setup considered in section \ref{sec:clutter}. Then we present 
in section \ref{sec:numdata} imaging results for two nearby reflectors in three location arrangements and three 
different types of clutter.

\begin{figure}[H]
\begin{minipage}{1.\textwidth}
\centering
\includegraphics[width=0.32\textwidth]{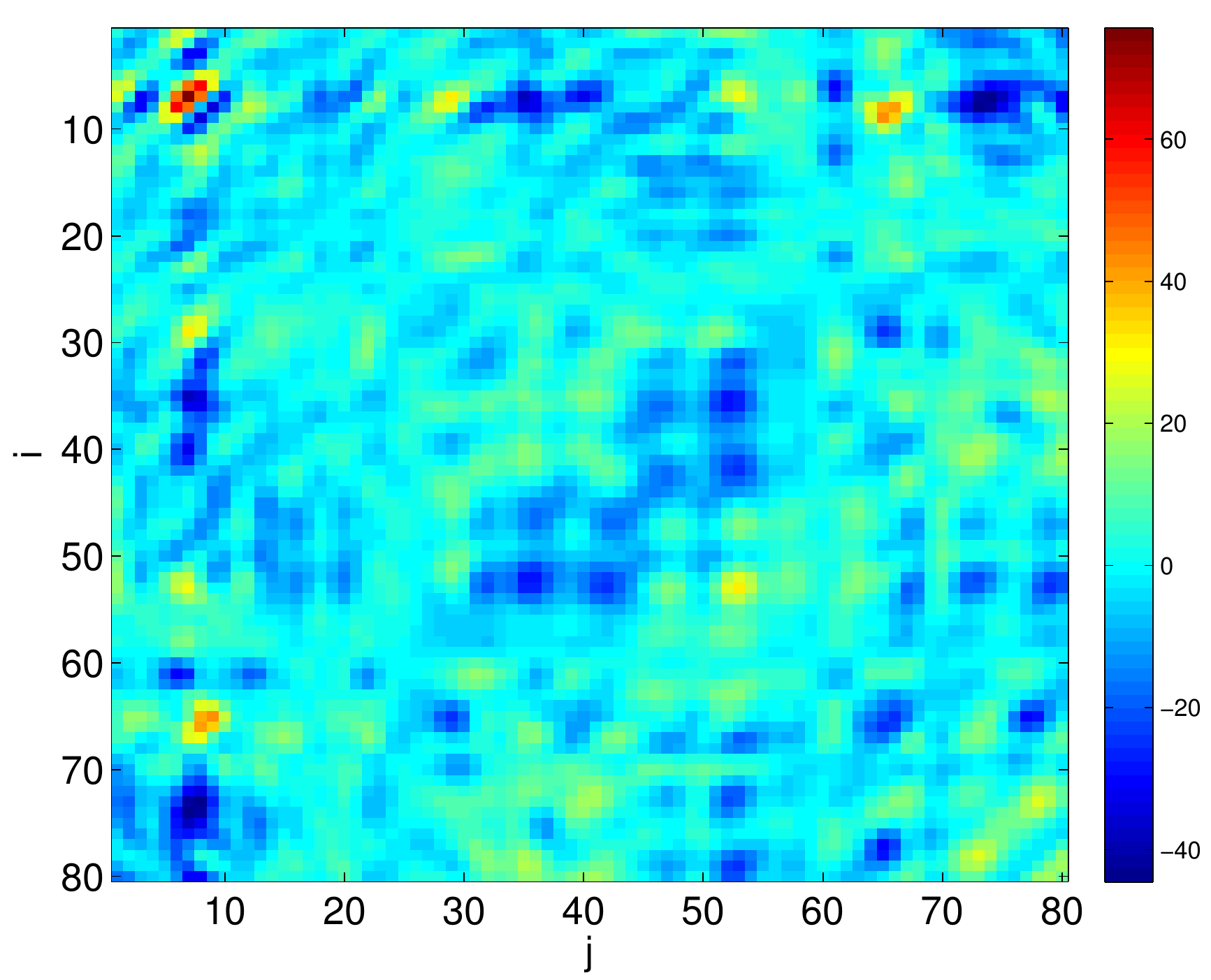}
\hspace{-0.1in}\includegraphics[width=0.32\textwidth]{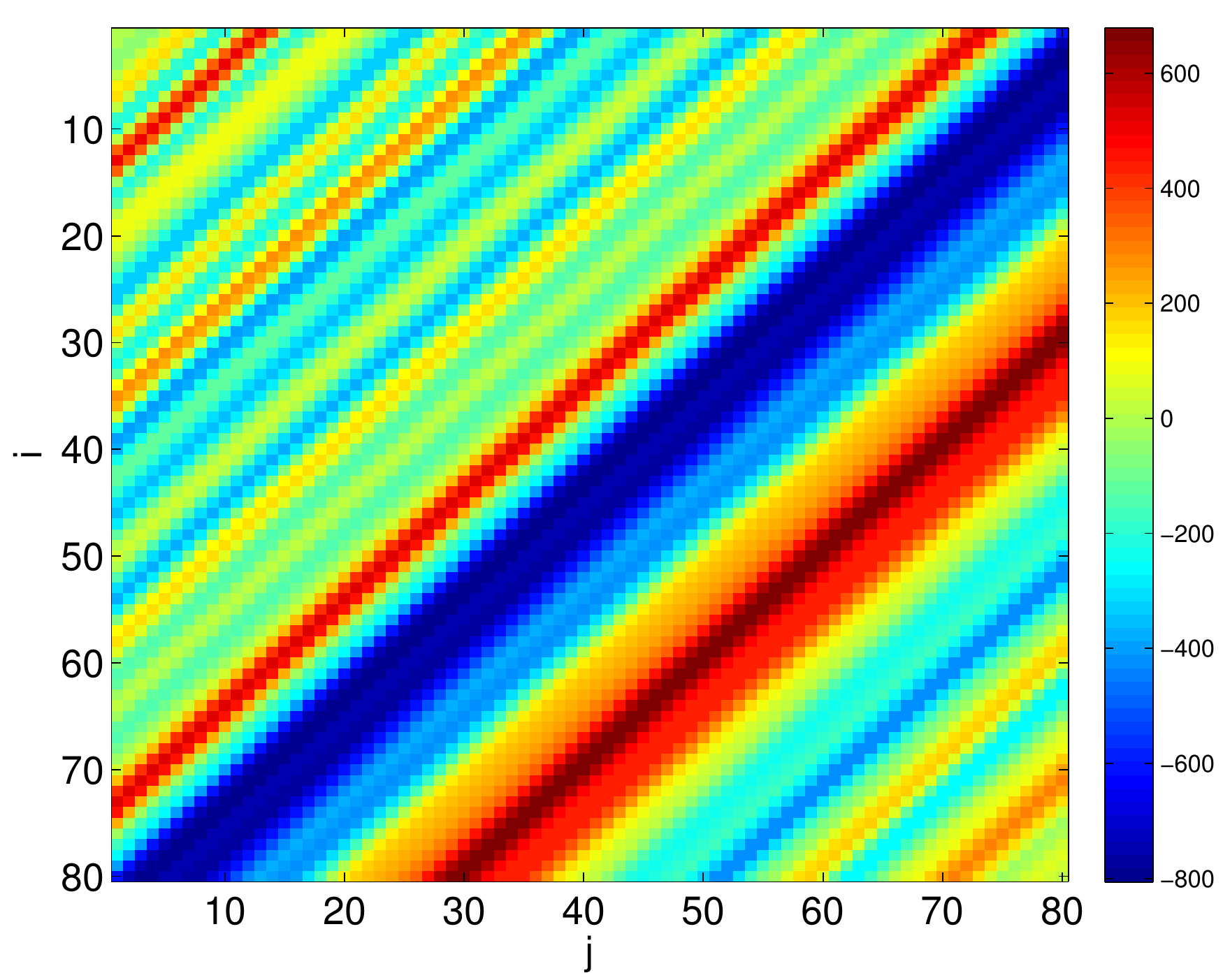}
\hspace{-0.1in}\includegraphics[width=0.32\textwidth]{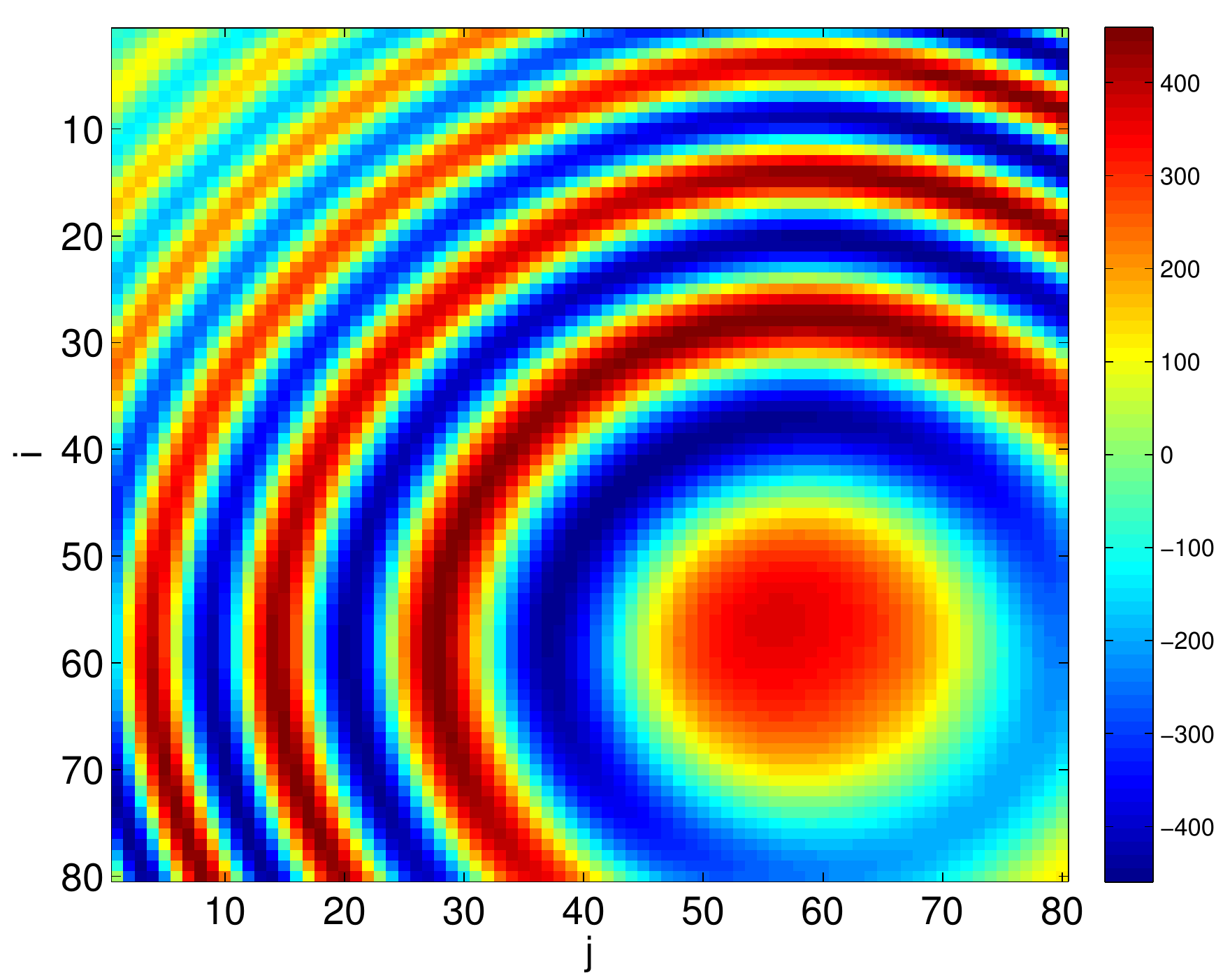}
\end{minipage}~\\
\caption{From left to right: the real part of $\widehat{\bP}^{^{BP}}(\omega)$,
  its  approximation  by the Hankel matrix $\widehat{\bP}^{^{AF}}(\omega)$
   and the filtered matrix $\widehat{\bP}^{^{OUT}}\hspace{-0.04in}(\omega)$. The axes are the indexes of the transducers and 
    $\omega/(2\pi) = 5$MHz.} 
\label{fig:matrixC1}
\end{figure}

\subsection{Illustration of  direction of arrival detection and filtering}
\label{sec:illustration}
The numerical simulations in this section are for the setup illustrated in Figure \ref{fig:setup} and described in detail  in section \ref{sec:clutter}. We focus attention on the reflector that is closer to the array, and illustrate 
how the algorithm introduced in section \ref{sec:angle} detects the arrival of the direct echoes from it and removes
the clutter backscatter. 

Since there are $N = 80$ transducers in the simulation, the array response matrix $\bP(t)$ is of size $80 \times 80$, 
and the time recordings are $N_T=2^{13}$ for the duration $T-T_o=13.7748$$\mu$s. We start the recordings at 
time $T_o=6.2252$$\mu$s
and end them at time $T=20$$\mu$s.
The selection of the time window containing the arrival of the direct echoes is done as explained in section \ref{sec:lct}
and described in detail in \cite{BPT-LCT}. It identifies the window at the level $l = 4$ of the LCT tree, indexed by $j_\star^4 = 7,$ and centered 
at time \[
t_o = T_o + (j_\star^4+1/2) \Delta_4, \quad \Delta_4 = (T-T_o)/2^4.
\] 

\begin{figure}[H]
\begin{minipage}{1.\textwidth}
\centering
\includegraphics[width=0.32\textwidth]{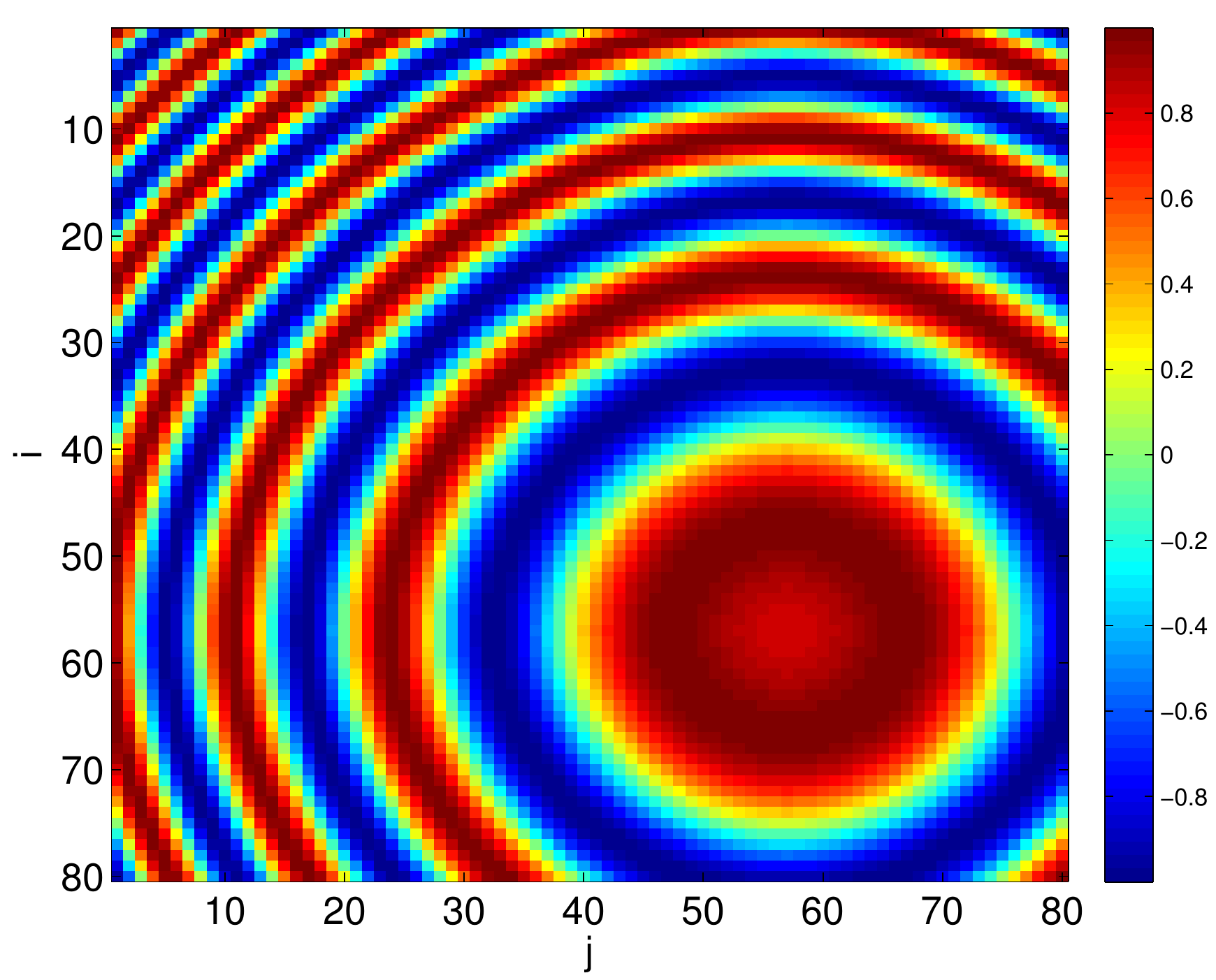}
\hspace{-0.1in}\includegraphics[width=0.32\textwidth]{./MatrixKa_10C2TSC3_Level4_window7_Pband70_Neigs2_freq70_mul1_average}
\hspace{-0.1in}\includegraphics[width=0.32\textwidth]{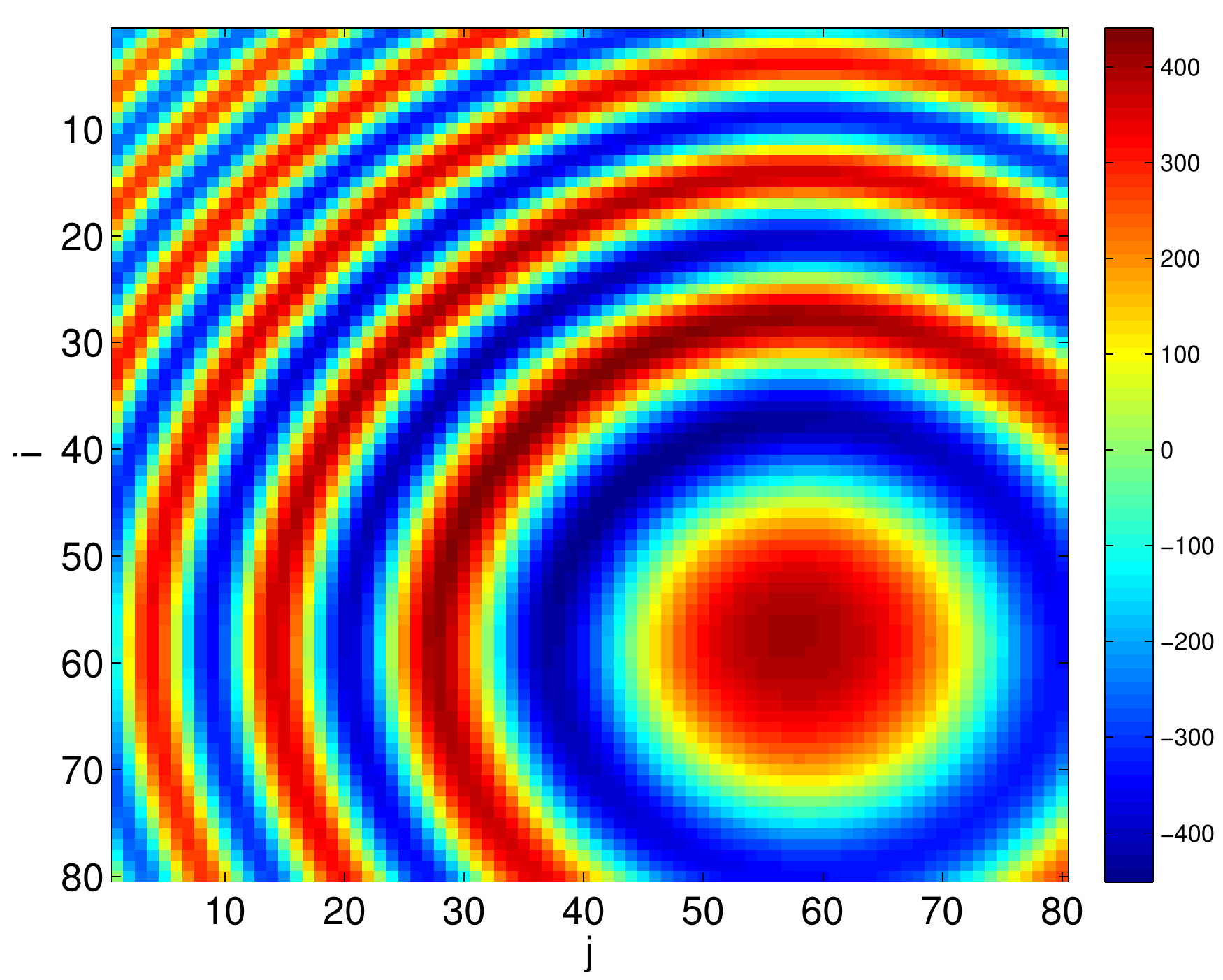}
\end{minipage}~\\
\caption{From left to right: the real part of the ideal 
  response matrix $\widehat{\bP}^{^{HOM}}\hspace{-0.04in}(\om)$ in the homogeneous medium,  the filtered matrix $\widehat{\bP}^{^{OUT}}\hspace{-0.04in}(\omega)$, and its rank one approximation. The axes are the indexes of the transducers and 
    $\omega/(2\pi) = 5$MHz.} 
\label{fig:matrixC1TH}
\end{figure}

\begin{figure}[H]
\begin{minipage}{1\textwidth}
\centering
\includegraphics[width=0.3\textwidth]{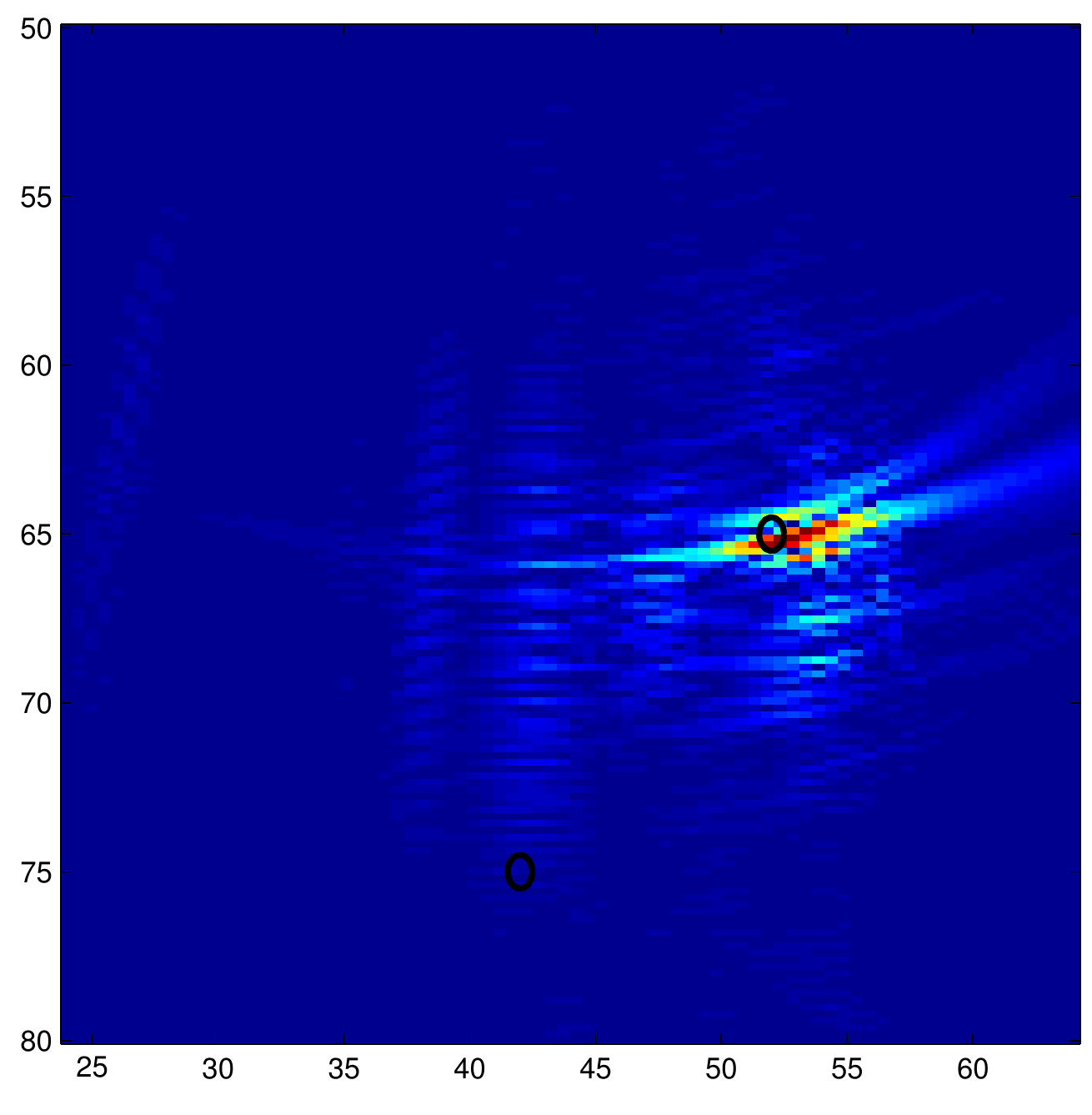}
\includegraphics[width=0.3\textwidth]{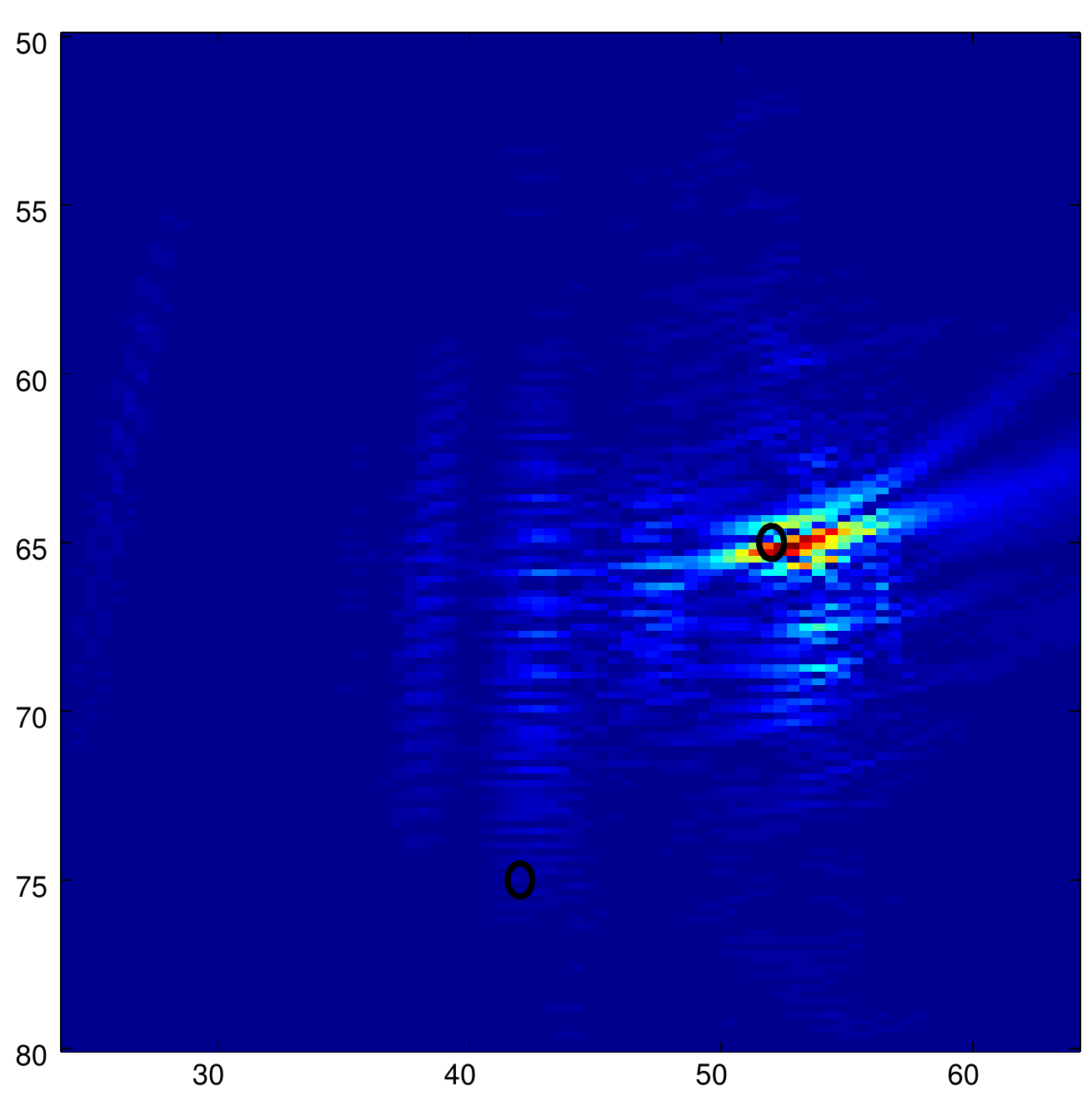}
\end{minipage}~\\
\caption{Kirchhoff migration images formed with the filtered data matrix ${\bP}^{^{OUT}}(t)$    (left) and the inverse Fourier transform 
of its rank one approximation (right). They are almost  the same. The abscissa is cross-range in units of $\la_o$ and the ordinate 
is range in units of $\la_o$.} 
\label{fig:imDORT} 
\end{figure}

The direction of arrival detection and filtering begins with the matrix $\bP^{^{IN}}\hspace{-0.04in}(t)$ defined in equation \eqref{eq:defPo1}. We Fourier transform 
it and backpropagate it to the reference point $\vy_o = (0,c/(2 t_o))$ using equation \eqref{eq:BP}, and display in the left plot of Figure \ref{fig:matrixC1} the real part of the resulting
matrix  $\widehat{\bP}^{^{BP}}\hspace{-0.04in}(\omega)$, at frequency $\om/(2 \pi) = 5$MHz.
In the middle plot we display its approximation $\widehat{\bP}^{^{AF}}\hspace{-0.04in}(\omega)$ obtained by 
rotating the matrix defined in  equation \eqref{eq:FrobAp1} to the system of coordinates corresponding to the source and receiver locations. 
This is a Hankel matrix by construction. In the right plot we display the filtered  matrix 
$\widehat{\bP}^{^{OUT}}\hspace{-0.04in}(\omega)$, the Fourier transform of the output matrix at step (9) of the algorithm. 
We compare it in Figure \ref{fig:matrixC1TH} with the ideal array response matrix in the homogeneous medium, which has rank one 
and entries 
\[
\widehat P^{^{HOM}}\hspace{-0.04in}(\om,\vx_r,\vx_s) = \frac{e^{i k \left(|\vx_r-\vy) + |\vx_s-\vy|\right)}}{16 \pi^2 |\vx_r-\vy||\vx_s-\vy|} .
\]
We note that the matrices are quite close, so the filtering algorithm works well. In the right plot 
of Figure \ref{fig:matrixC1TH} we display the rank one approximation of $\widehat{\bP}^{^{OUT}}\hspace{-0.04in}(\omega)$, 
the matrix with the leading left and right singular vectors and singular value of $\widehat{\bP}^{^{OUT}}\hspace{-0.04in}(\omega)$.
The improvement is slight, and has little effect on the images displayed in Figure \ref{fig:imDORT}.  Comparing these images with those 
in Figure \ref{fig:imaging} we note the dramatic improvement brought by the  filtering algorithm.

  \begin{figure}[H]
    \begin{minipage}{\textwidth}
      \centering
      \includegraphics[width=0.45\textwidth]{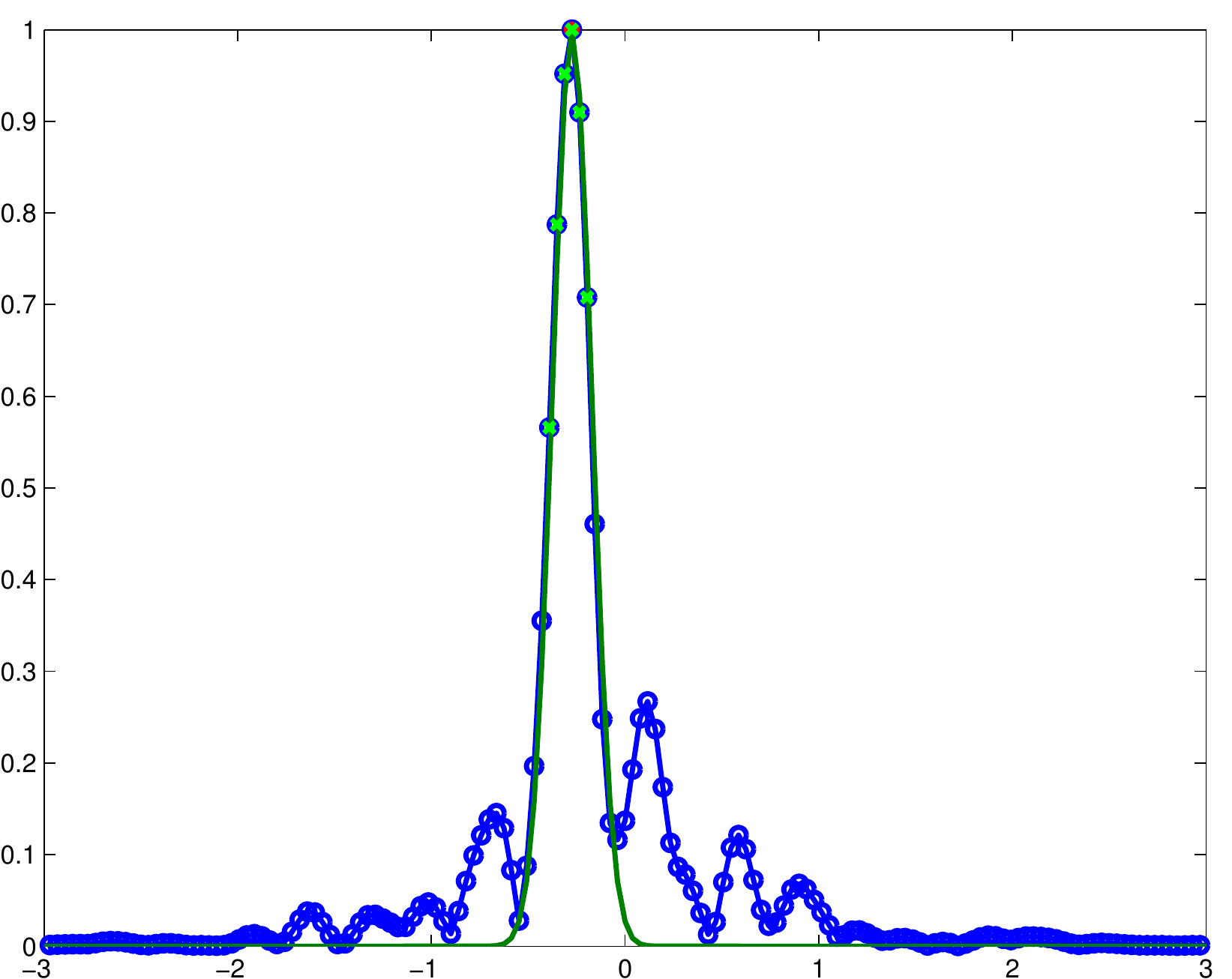}
    \end{minipage}
    \caption{We display the absolute value of the entries of $\breve{\itbf{p}}(\omega)$ defined in equation \eqref{eq:PLANEWAVE} (blue line) 
    and the Gaussian taper in green. The
      points used to determine the least squares fit by the  Gaussian at step (6)  are shown with
      green stars. The abscissa  is scaled by the wavenumber $k$ so the peak corresponds to $2 y /L$.
       }
    \label{fig_adapt1}
  \end{figure}
To illustrate the use of the rotation and approximation at step (3) of the algorithm, equation \eqref{eq:FrobAp} in particular, 
we plot in Figure \ref{fig_adapt1} the entries of the vector $\breve{\itbf{p}}(\omega)$ as a function of $\kappa$. We note that 
there is a clear peak, corresponding to the arrival of the coherent  echoes from the reflector at $\vy$, that is fitted with the Gaussian taper shown in green. 
 
\subsection{Imaging two reflectors in different geometrical configurations and types of clutter}
\label{sec:numdata}
We assess in this section the performance of the direction of arrival filtering algorithm for different 
geometric configurations of two reflectors and different cluttered
media, as illustrated in Figures \ref{fig:configsC} and \ref{fig:config3}. We begin with the numerical setup in section
\ref{sec:numdata} and then we show the results in the following sections.
\subsubsection{Description of the numerical setup}
\label{sec1:numdata}

\begin{figure}[H]
\begin{minipage}{1\textwidth}
\centering
\includegraphics[width=0.31\textwidth]{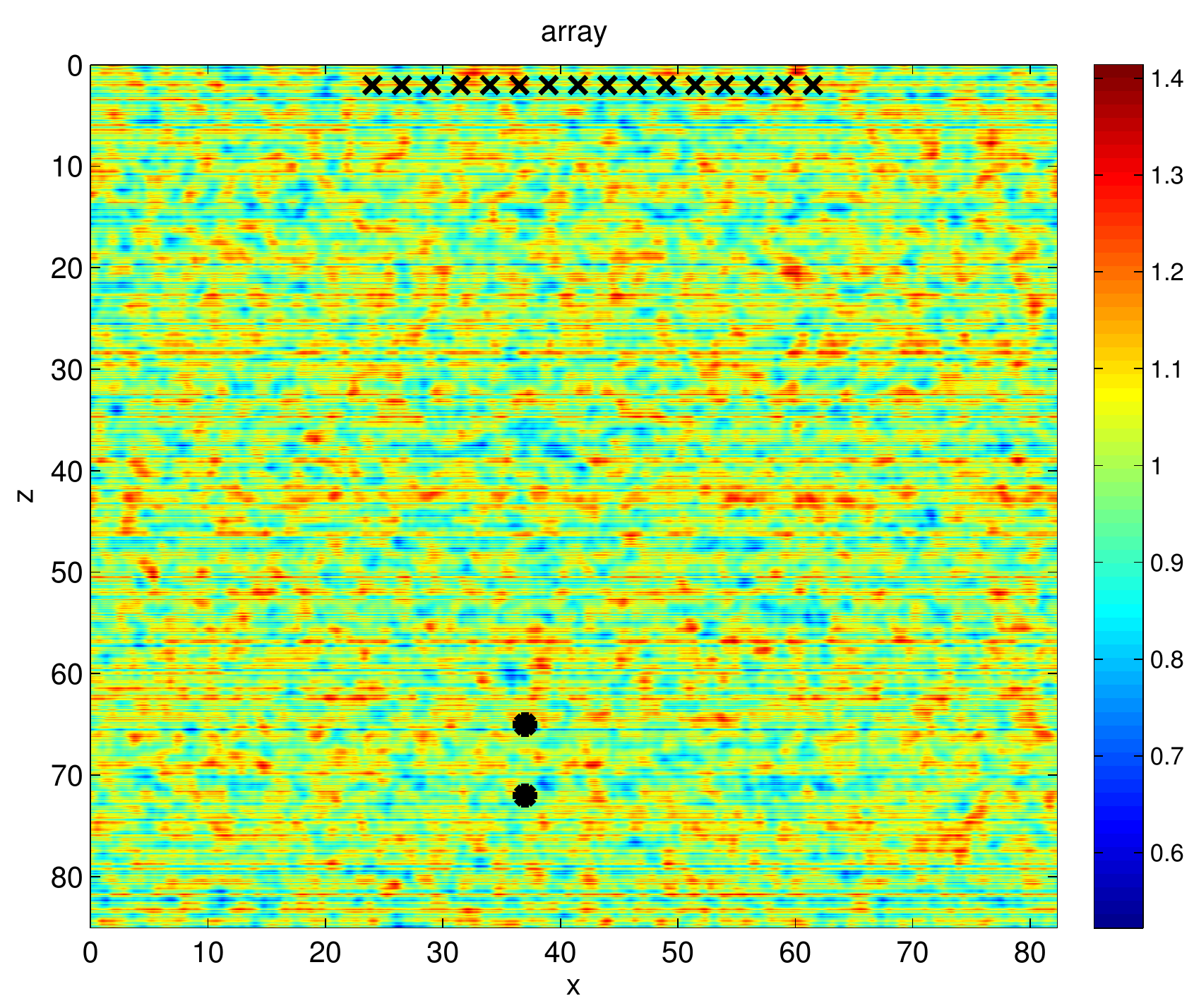}
\includegraphics[width=0.31\textwidth]{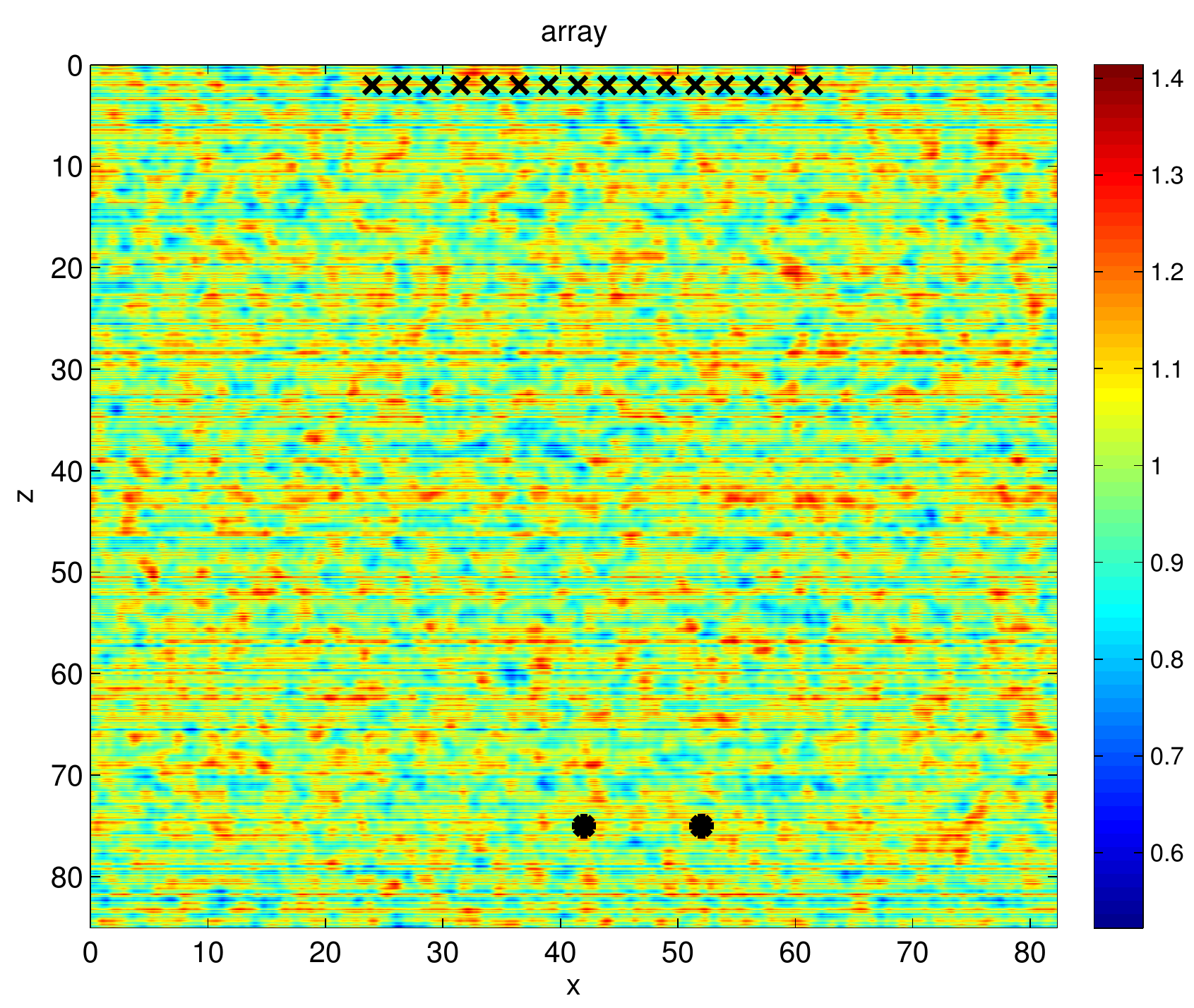}
\includegraphics[width=0.31\textwidth]{./random10CT2C3}
\end{minipage}
\caption{Combined cluttered medium modeled by the process $\mu$ in equation \eqref{eq:mu_c}. 
Three configurations of two reflectors. We
  call them configurations 1, 2 and 3 from left to right.}
\label{fig:configsC}
\end{figure}

\begin{figure}[H]
\begin{minipage}{1\textwidth}
\centering
\includegraphics[width=0.31\textwidth]{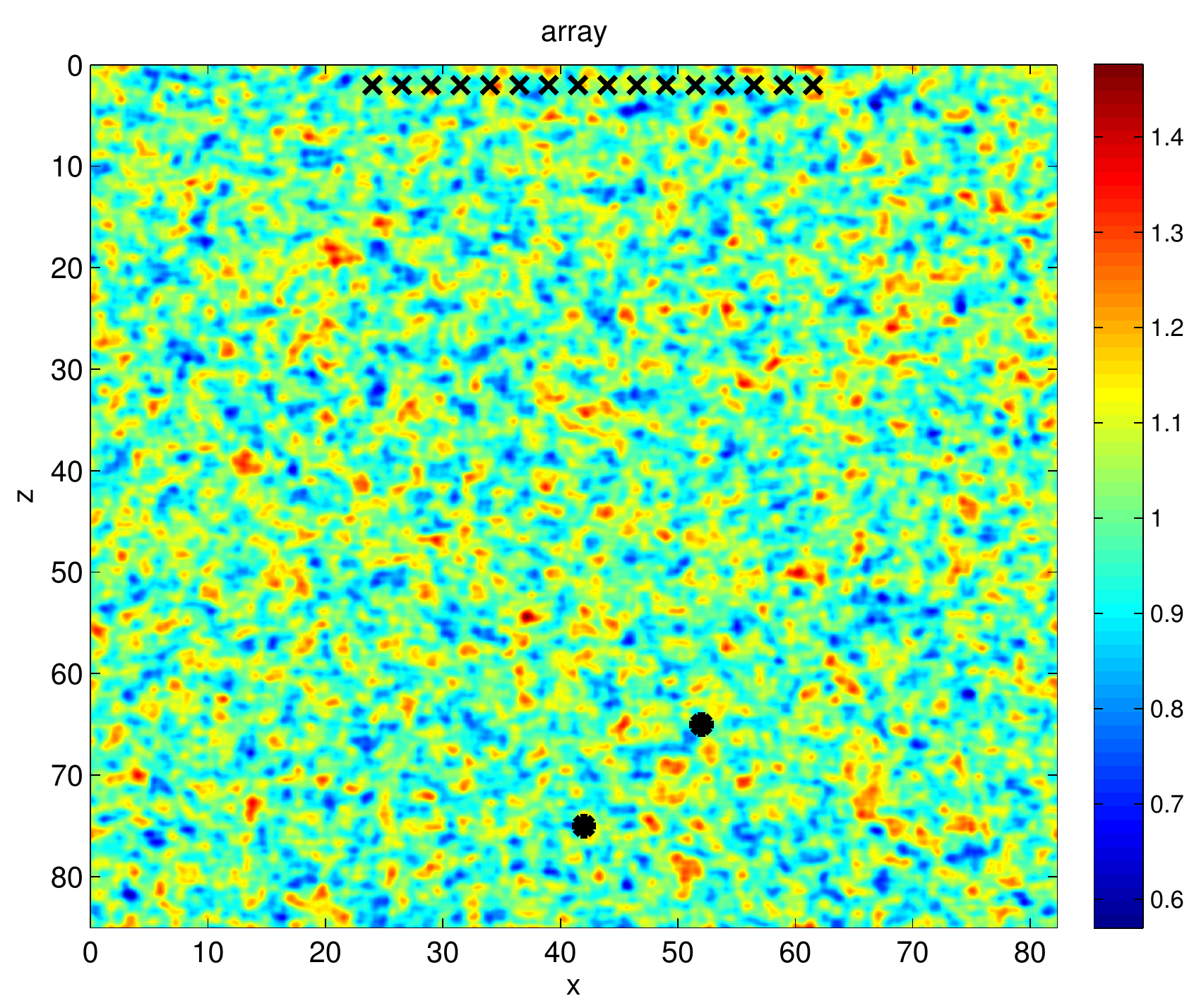}
\includegraphics[width=0.31\textwidth]{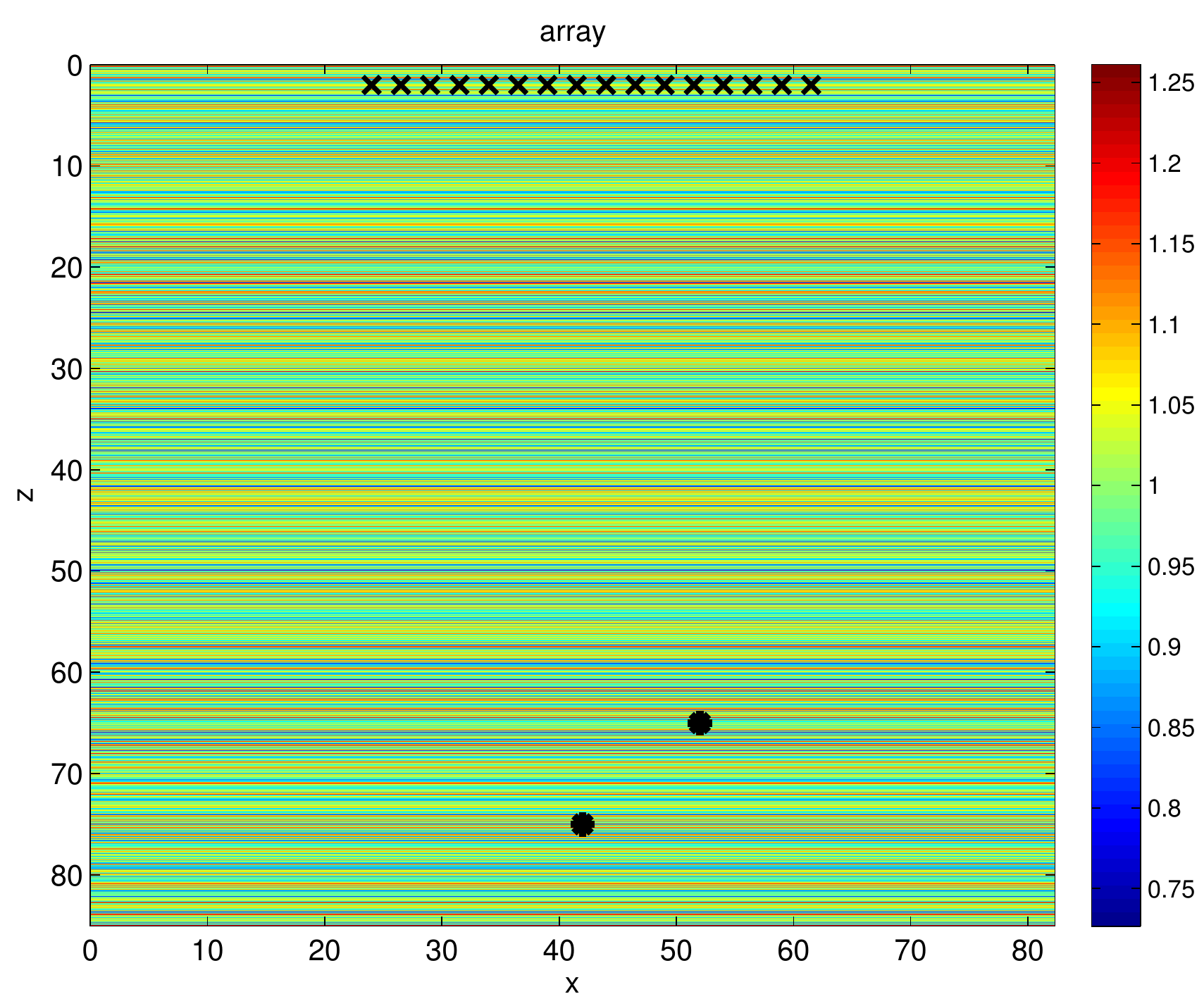}
\includegraphics[width=0.31\textwidth]{./random10CT2C3}
\end{minipage}
\caption{The third configuration of reflectors in three different
  cluttered media. Isotropic on the left, layered in the middle and
  combined on the right.}
\label{fig:config3}
\end{figure}

We consider three geometrical arrangements  of two reflectors which are 
offset either in  range, in cross-range or both directions, as illustrated in Figure 
\ref{fig:configsC}. The reflectors are sound soft disks of radius $\la_o/4$, located at $\vy_1=(37\la_o,
65\la_o)$, $\vy_2=(37\la_o,72\la_o)$ for configuration 1, at 
$\vy_1=(42\la_o, 75\la_o)$, $\vy_2=(52\la_o,75\la_o)$ for
configuration 2,  and at $\vy_1=(42\la_o, 75\la_o)$, $\vy_2=(52\la_o,65\la_o)$ 
for the third configuration. 

We test  the direction of arrival filtering algorithm  in  three different types of clutter modeled by the isotropic random process
$\mu_i$ in equation \eqref{eq:mu_i}, the layered one $\mu_l$ in equation \eqref{eq:mu_l}, 
and the combined  $\mu$ in equation \eqref{eq:mu_c}. In all cases, the smooth part of the speed is constant $c=
1$km/s, and the fluctuations are generated with random Fourier series.
In the isotropic medium the standard deviation of the fluctuations is $\varepsilon =0.1$ and
the correlation length is $\ell=\lambda_o/4$. For
the layered medium $\varepsilon =0.17$ and   $\ell=\lambda_o/50$.  
For the combined medium  \eqref{eq:mu_c} the standard deviation is $\varepsilon =0.1$.
We display in Figure \ref{fig:config3} the realizations
of the wave speed  used in the 
simulations.

The array is linear, as described in section \ref{sec:clutter}, and gathers the response matrix by probing 
the medium with one source at a time, emitting the same Ricker pulse. The receivers record the echoes in the time window
$[T_o,T]$ with $T_o=6.2252$$\mu$s and  $T=20$$\mu$s. The time discretization is with  $N_T=2^{13}$ steps. 



\subsubsection{Imaging results for the three reflector configurations}
\label{sec:numimage}
\begin{figure}[H]
\begin{minipage}{1\textwidth}
\centering
\includegraphics[width=0.28\textwidth]{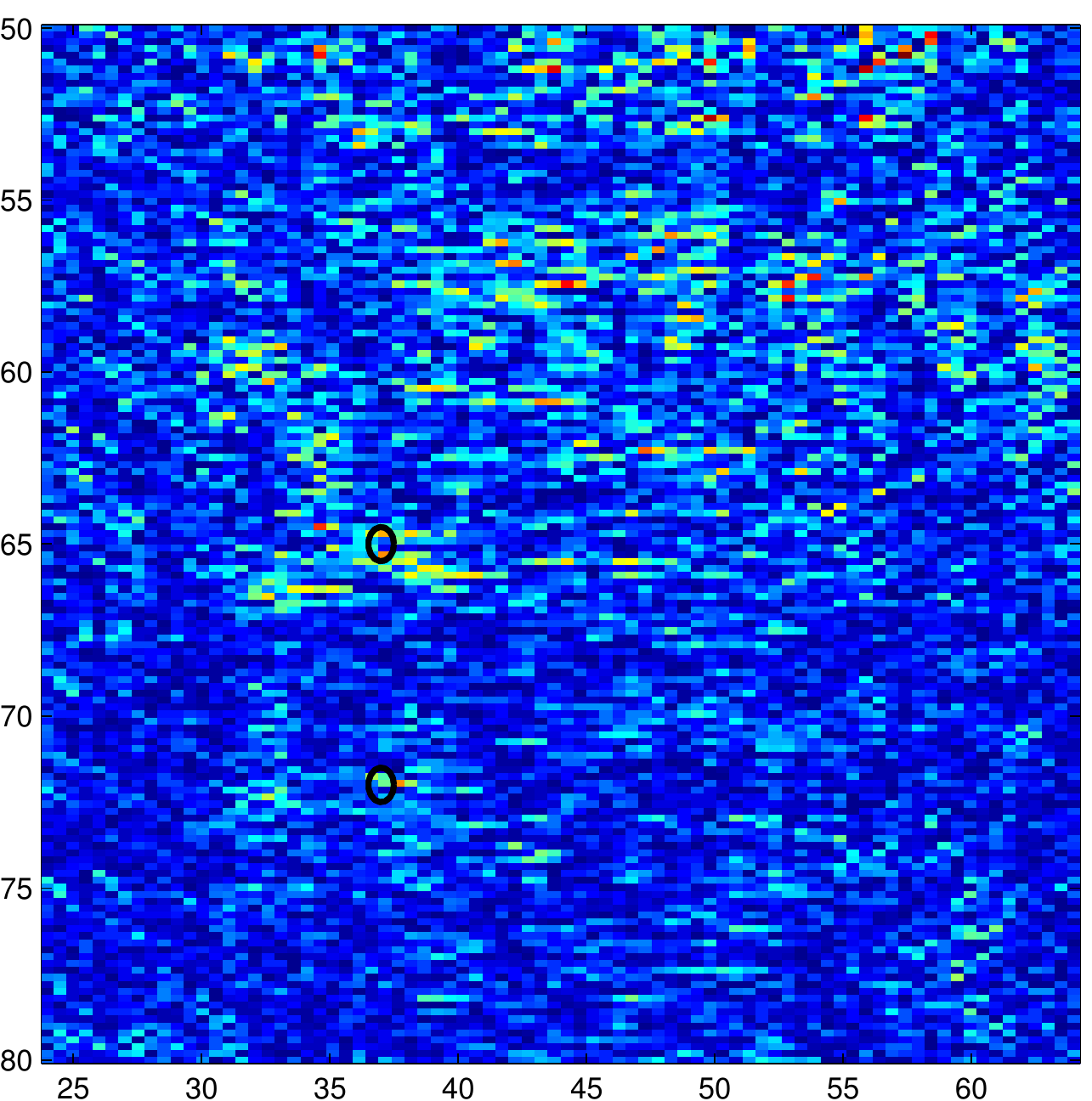}
\includegraphics[width=0.28\textwidth]{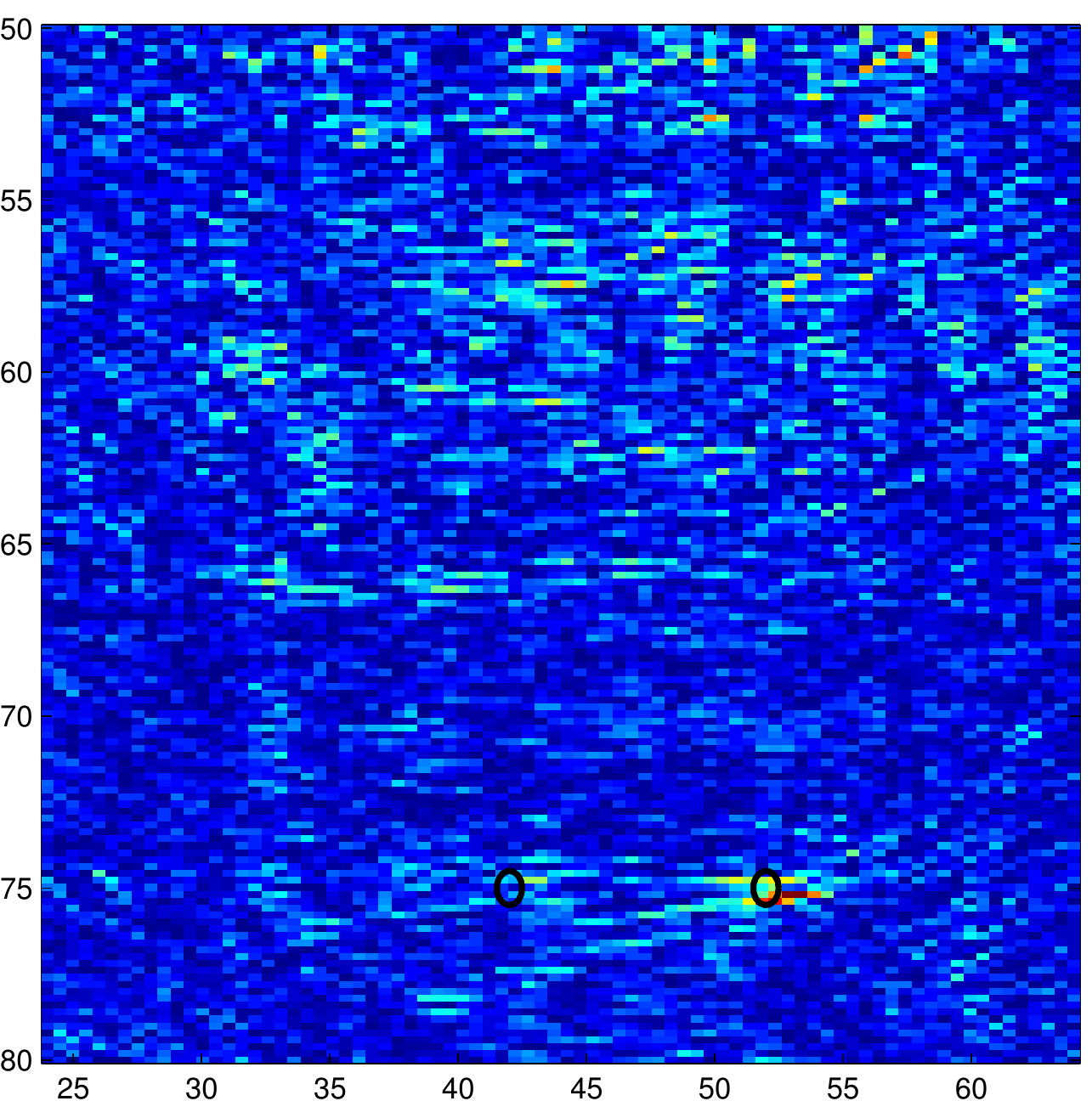}
\includegraphics[width=0.28\textwidth]{./Image_10C2TSC3_Level0_window0}
\end{minipage}
\caption{KM images for the three 
  configurations and clutter shown in Figure \ref{fig:configsC}. The images are  formed with the unfiltered response matrix $\bP(t)$. The abscissa is 
  cross-range in units of $\la_o$ and the ordinate is range in units of $\la_o$. The reflectors are indicated with the black circles.}
\label{fig:KMinitCall}
\end{figure}

We present here imaging results in the clutter modeled by the process $\mu$ in 
equation \eqref{eq:mu_c}, for the three geometric arrangements of the reflectors. 
We begin with Figure \ref{fig:KMinitCall}, where we display the KM images formed with the 
unfiltered response matrix $\bP(t)$.  Due to the strong clutter,
the images are noisy and difficult to 
interpret. Repeated simulations, in different realizations of $\mu$, also show that the images
change dramatically, and unpredictably. We do not show the CINT images because 
they are also not useful in this strong clutter, as illustrated in Figure \ref{fig:imaging}.

The first part of the filtering algorithm selects the time windows that contain the direct arrivals 
from the reflectors, at level $l = 4$ in the LCT tree. They are indexed by $j_\star^4 = 7$ and 
$j_\star^4 = 9$ in configurations 1 and 3 and by $j_\star^4 = 9$ in configuration 2, where a single window is selected.
This is because in the second configuration the reflectors are at the same range location. 

\begin{figure}[H]
  \begin{minipage}{\textwidth}
 \centering
 $j_\star^0 = 1$ \hspace*{3cm}  $j_\star^4 = 9$ \hspace*{3cm} $j_\star^4 = 15$ \\
  \includegraphics[width=0.3\textwidth]{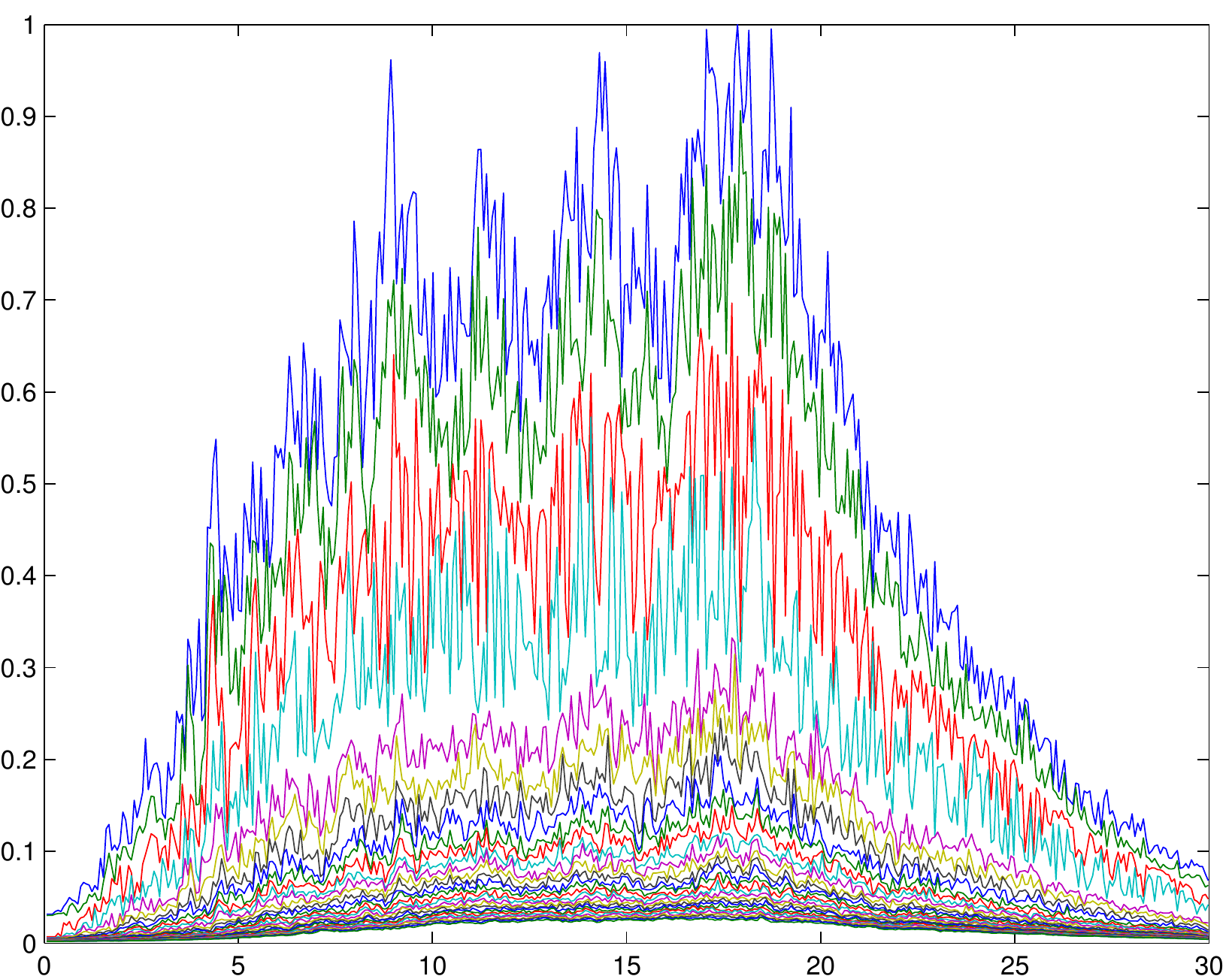}
 \includegraphics[width=0.3\textwidth]{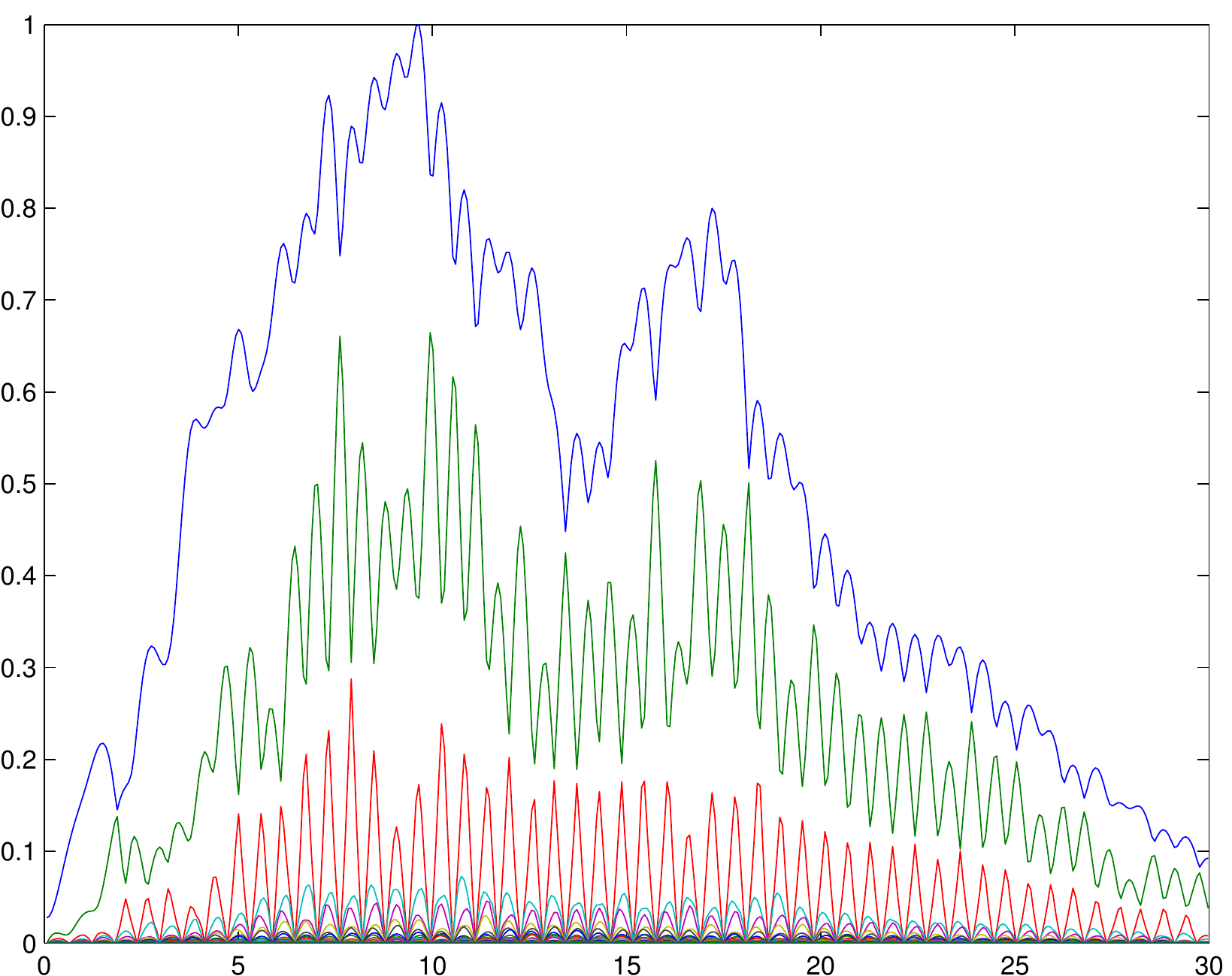}
 \includegraphics[width=0.3\textwidth]{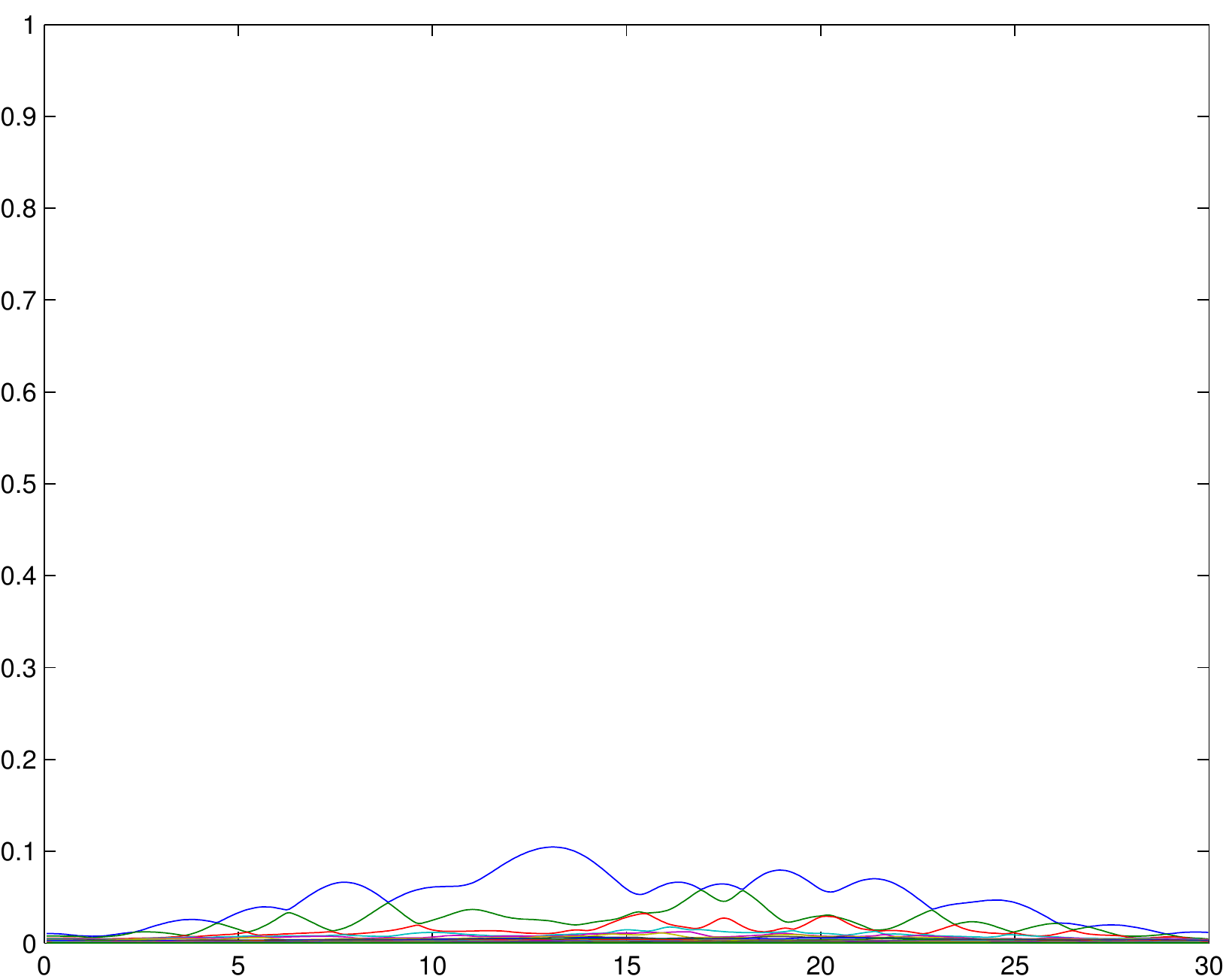}
    \end{minipage}
    \caption{The singular values of the array response matrix as a function of frequency for configuration 2 in the combined medium. 
    The left plot is at root level $l=0$ of the LCT tree i.e., for the entire duration of the recordings. The other plots are at 
    the level $l = 4$ of the tree, corresponding to $2^4$ time windows in the recording interval.  The middle plot is for the 
    selected window indexed by $j_\star^4 = 9$ which contains direct echoes from one of the reflectors. The right plot 
    is for the unselected window indexed by $j_\star^4 = 15$ which contains only clutter backscatter.    
    The plots are normalized  by dividing the singular values at each tree level  $l$ by the 
    maximum one over all windows and frequencies at level $l$. }
    \label{fig:lct-time}
  \end{figure}

To illustrate the benefit of time
windowing, we plot in 
Figure \ref{fig:lct-time} the singular values of the array response matrix for the second configuration,  at level 0 of the LCT tree, and then at level 4 
for windows  $j_\star^4 = 9$  and $j_\star^4 = 15$. The first window is selected by the algorithm as containing echoes from a reflector, and the second 
contains just clutter backscatter. We note that it is more difficult to distinguish the singular values  at the root level, 
because they  are clustered together. In the selected time window there is a clear separation of the larger singular values, signaling 
the arrival of the coherent echoes. In the last window, containing the clutter backscatter, the singular values are smaller and clustered together. 

\begin{figure}[H]
    \begin{minipage}{\textwidth}
      \centering
 \includegraphics[width=0.45\textwidth]{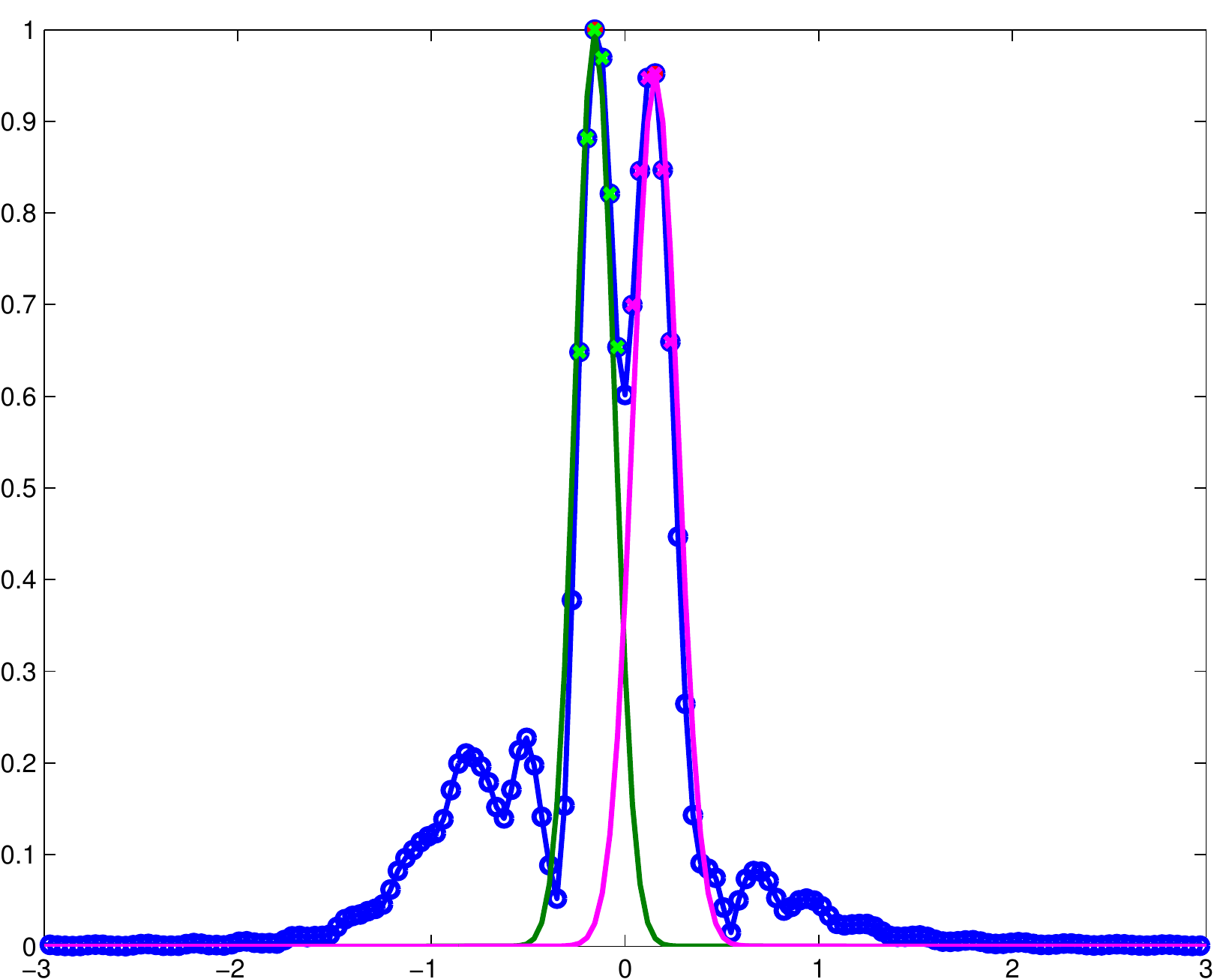}
    \end{minipage}
    \caption{Absolute value of the entries of $\breve{\itbf{p}}(\omega)$ defined in equation \eqref{eq:PLANEWAVE} (blue line) 
    and the Gaussian tapers in green and pink. The
      points used to determine the least squares fit by the  Gaussian at step (6) of the algorithm are shown with
      green and pink stars. The abscissa is scaled by the wavenumber $k$.}
    \label{fig_adapt1p}
  \end{figure}

The input matrix 
$\bP^{^{IN}}\hspace{-0.04in}(t)$ of the detection of the direction of arrival and filtering algorithm introduced 
in section \ref{sec:angle}  is calculated using equation \eqref{eq:defPo}, for each selected window centered at 
$t_o = T_o+(j_\star^4 +1/2) \Delta_4$,  with $\Delta_4 = (T-T_o)/2^4$.  The images displayed below are formed with the 
filtered matrix $\bP^{^{OUT}}\hspace{-0.04in}(t)$, the output of the algorithm. The direction of arrival selection
at step (5) is as easy as in Figure \ref{fig_adapt1} in configurations 1 and 3, because the selected time 
windows contain the direct echoes form a single reflector. In the second configuration the echoes from both 
reflectors arrive in the selected window, and the plot of the vector $\breve{\itbf{b}}(\omega)$ calculated 
at step (4) is shown in Figure \ref{fig_adapt1p}. There are two peaks, corresponding to the direction of each 
reflector. 

\begin{figure}[H]
\begin{minipage}{1\textwidth}
\centering
\includegraphics[width=0.25\textwidth]{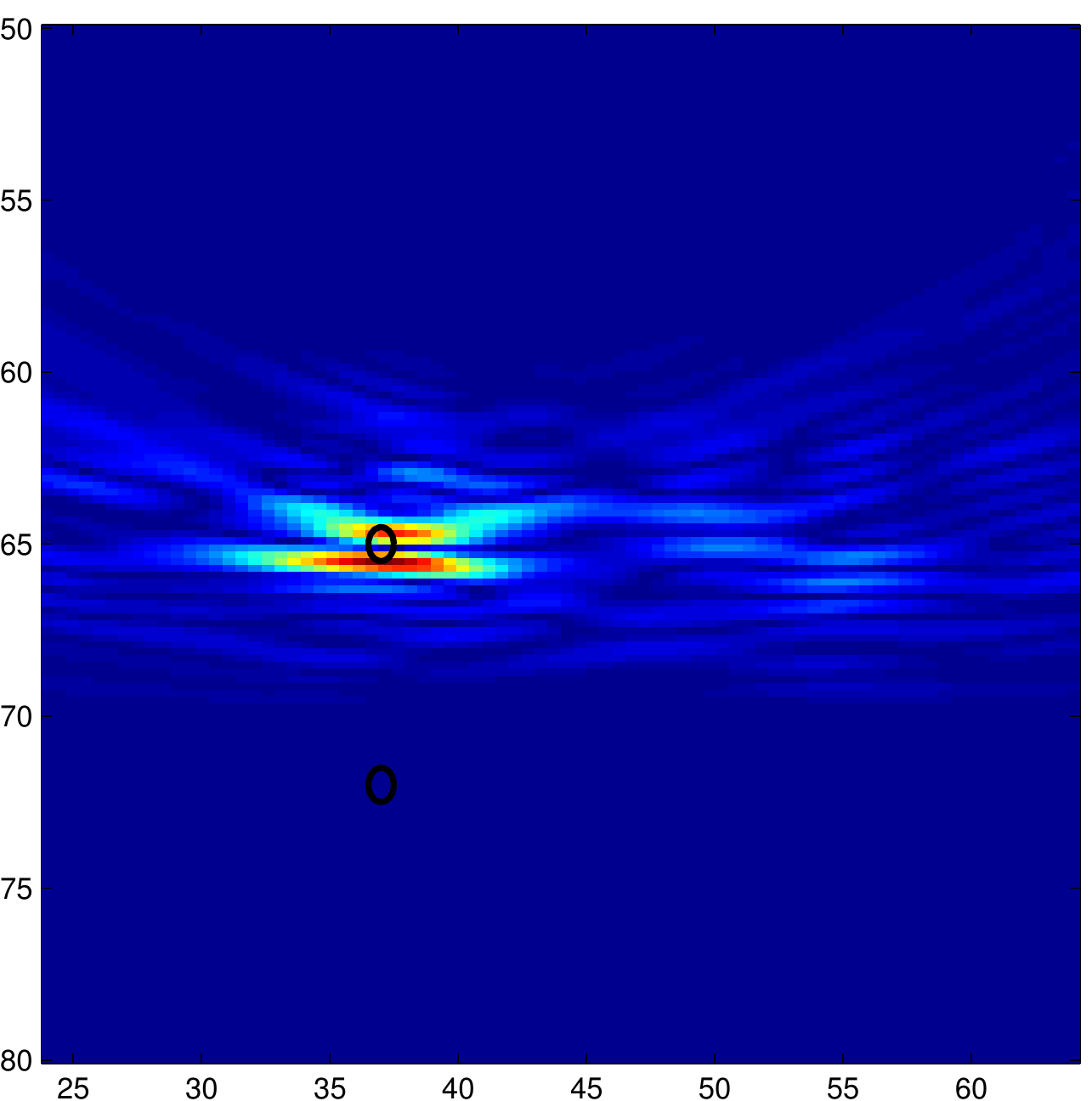}
 \includegraphics[width=0.25\textwidth]{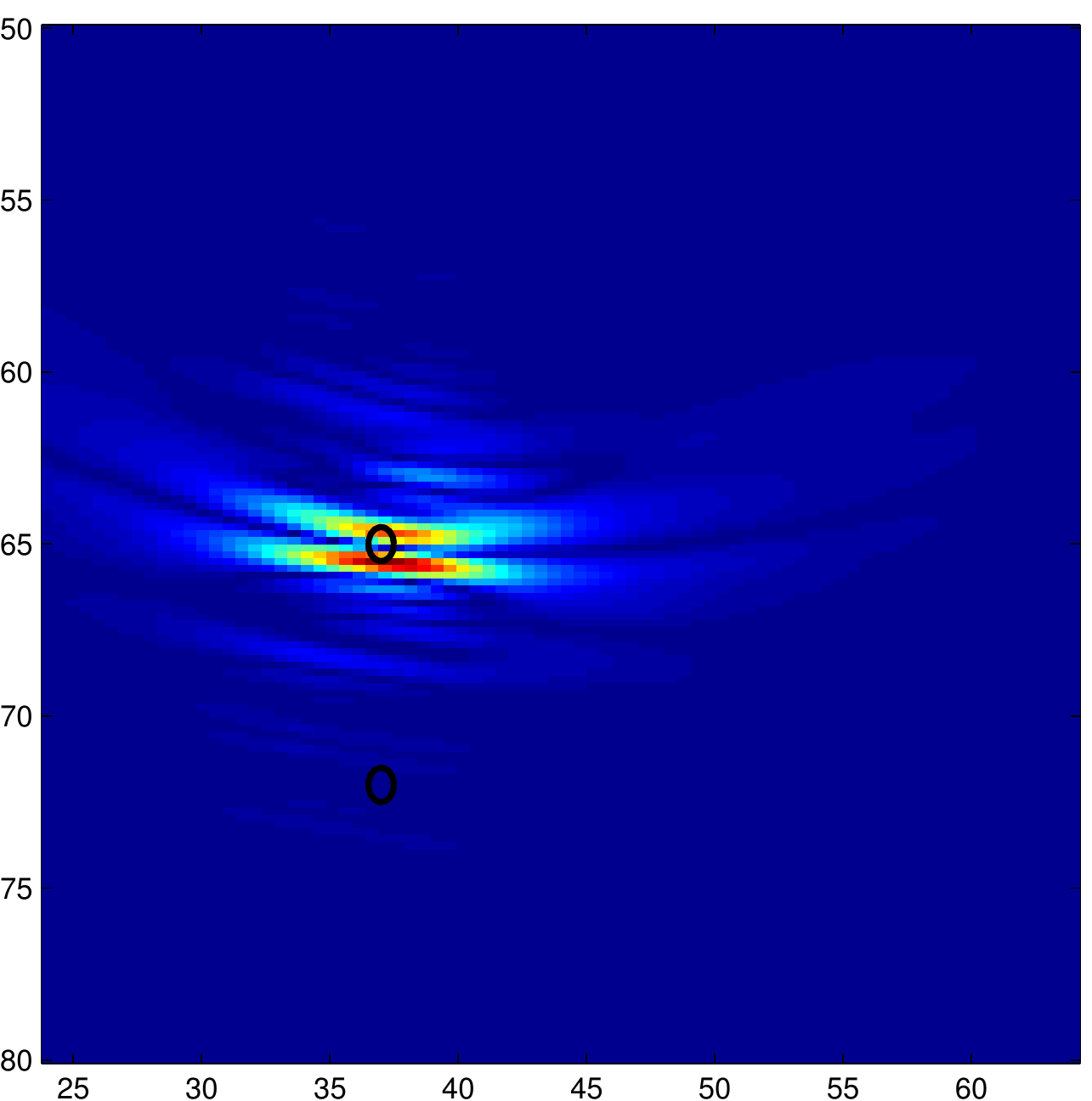}
 \includegraphics[width=0.25\textwidth]{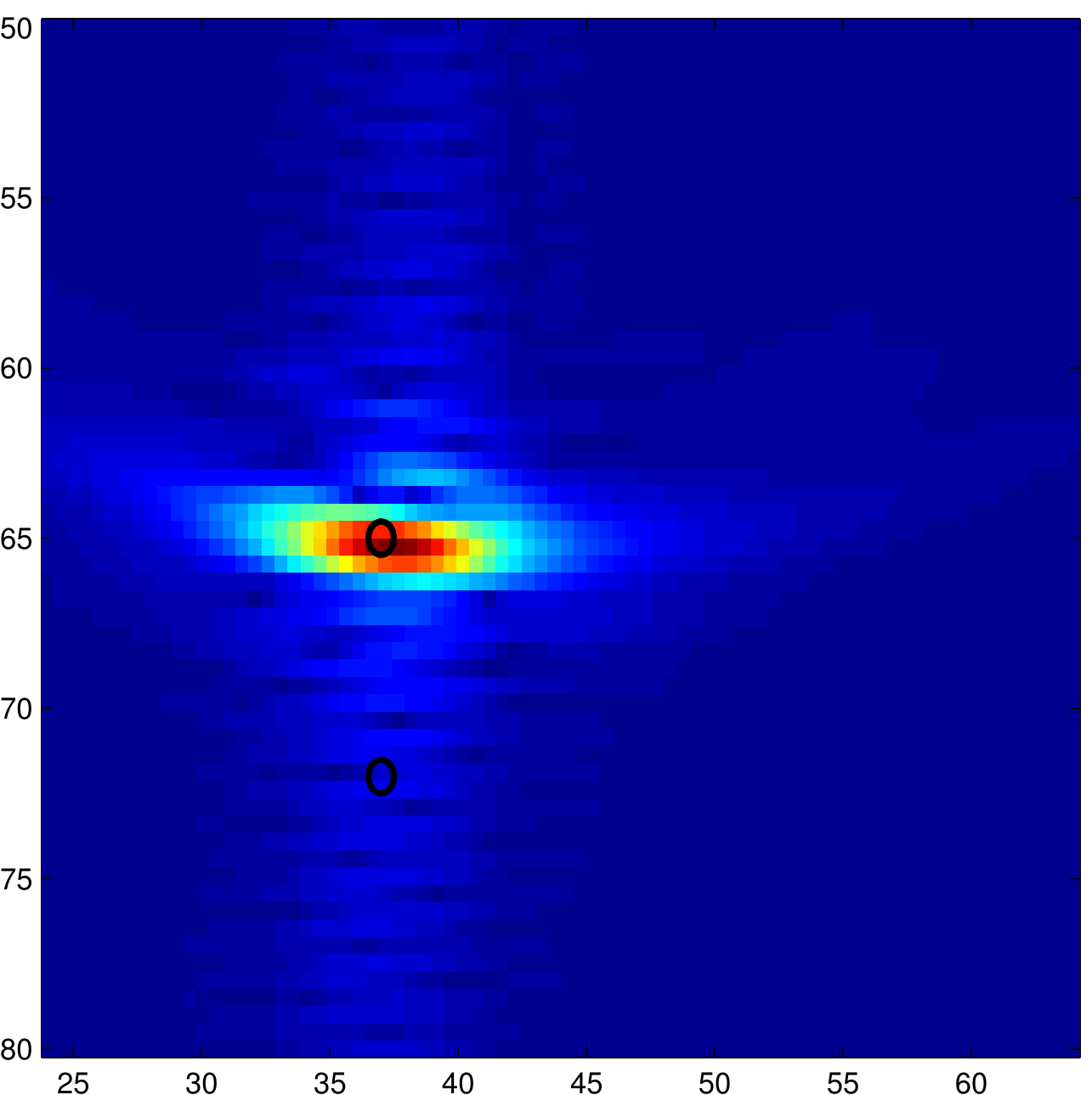}
\end{minipage}~\\
\begin{minipage}{1\textwidth}
\centering
\includegraphics[width=0.25\textwidth]{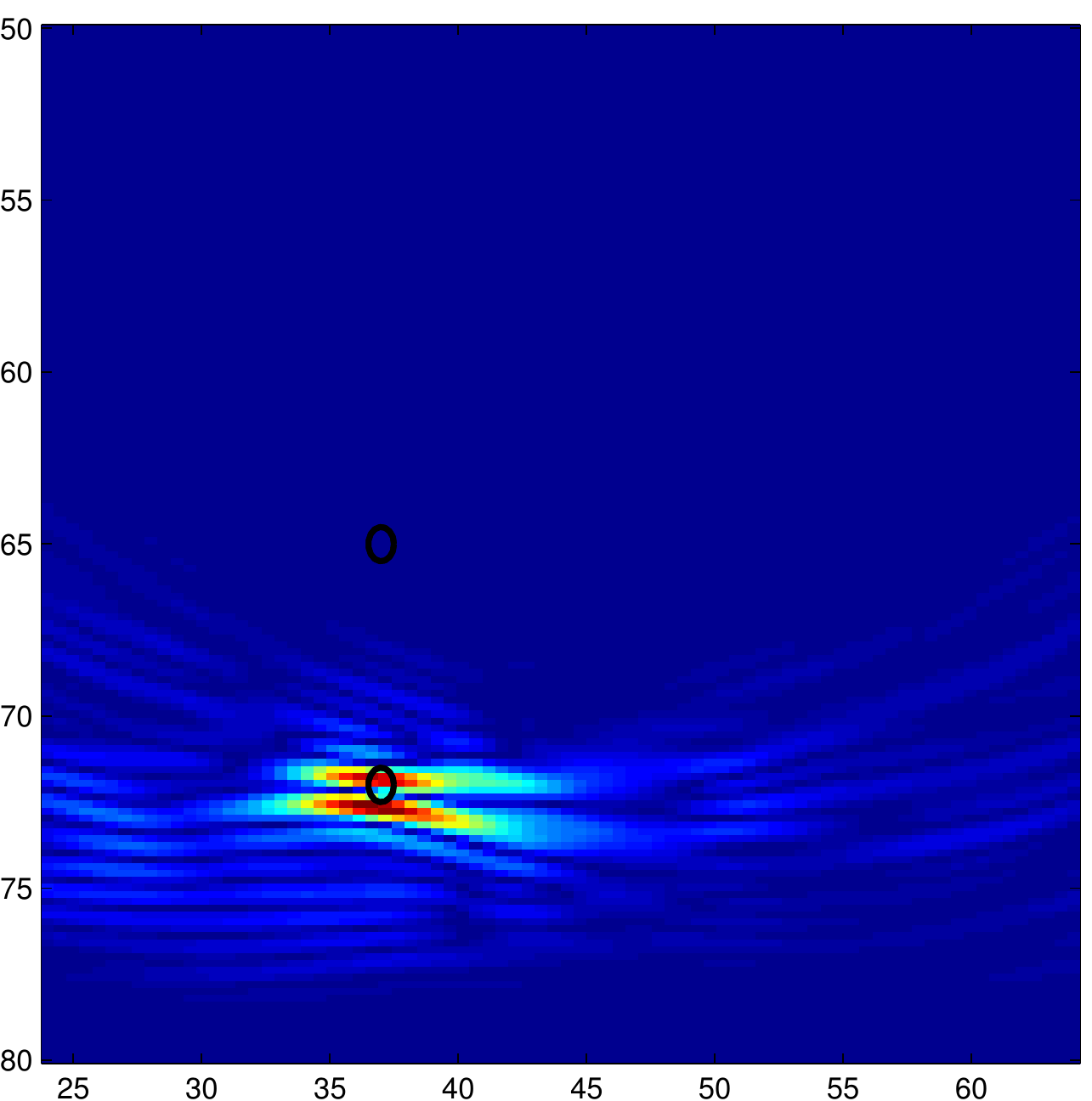}
 \includegraphics[width=0.25\textwidth]{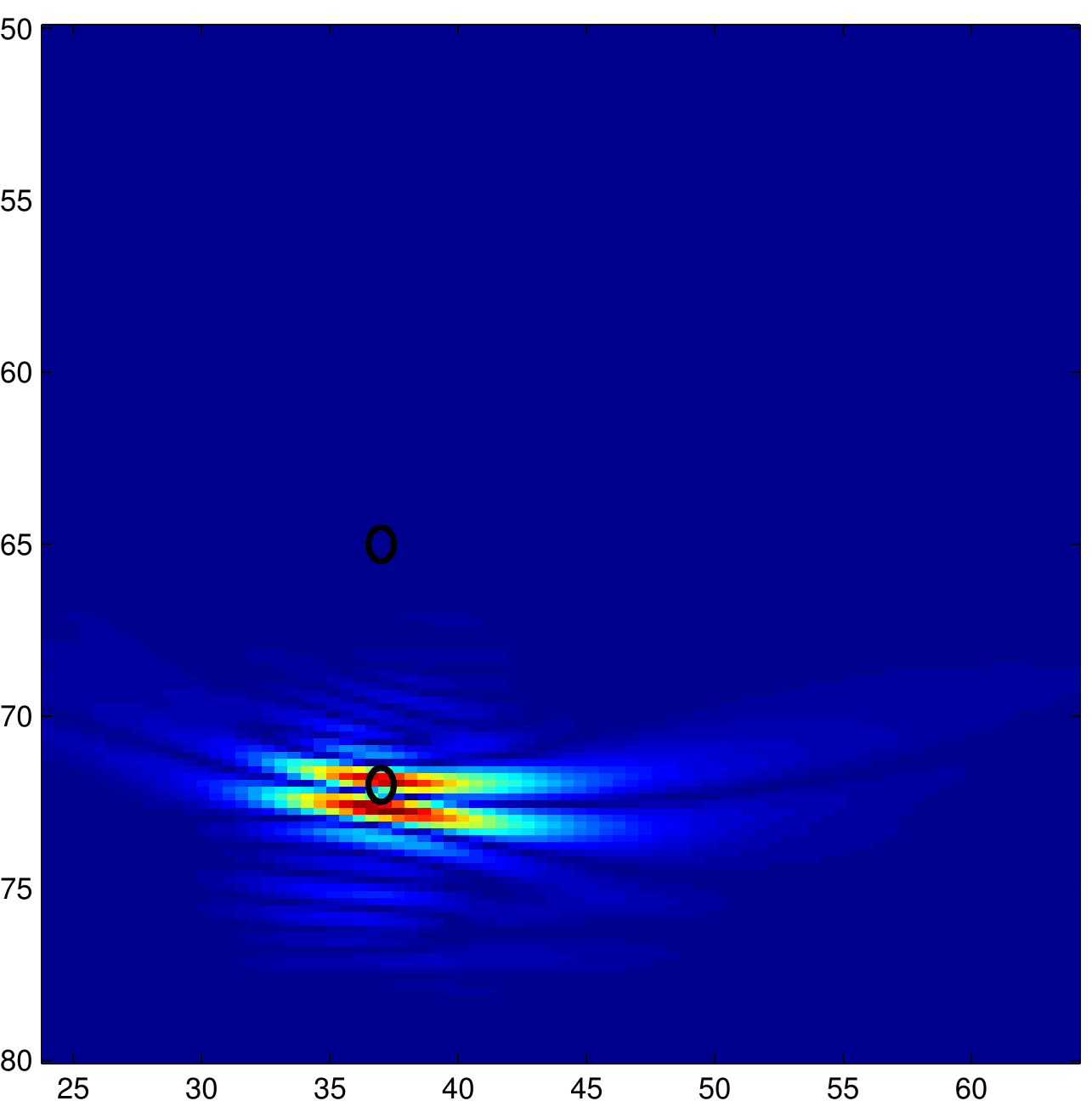}
 \includegraphics[width=0.25\textwidth]{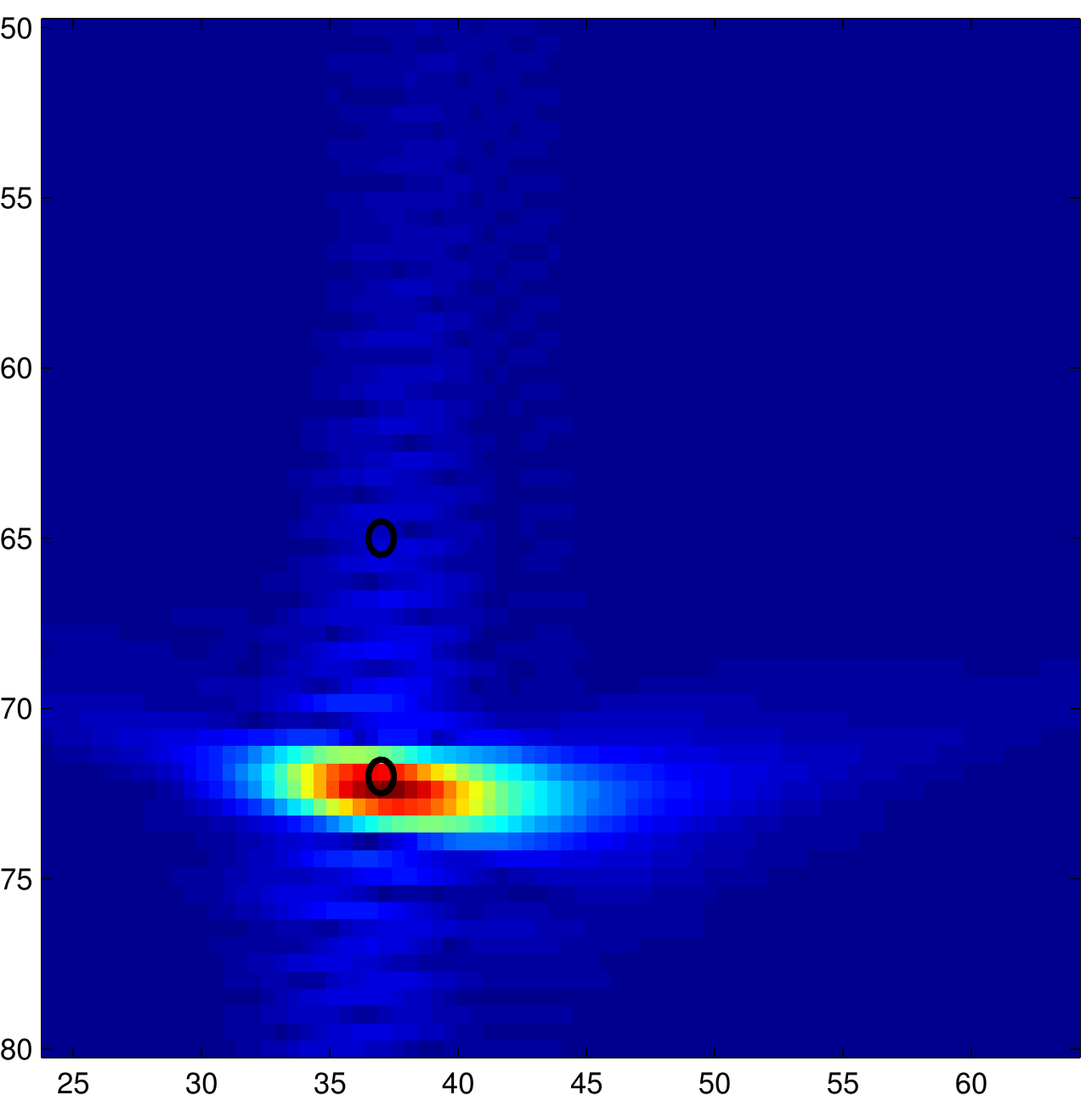}
\end{minipage}
\caption{Imaging results for configuration 1 and the clutter shown in Figure \ref{fig:configsC}. Top row the images obtained with the data in the first selected time window 
and bottom row in the second selected time window. Left column:  KM images formed with $\bP^{^{IN}}\hspace{-0.04in}(t)$.
Center column: KM images formed with $\bP^{^{OUT}}\hspace{-0.04in}(t)$. Right column: CINT images formed with $\bP^{^{OUT}}\hspace{-0.04in}(t)$.  The abscissa is 
  cross-range in units of $\la_o$ and the ordinate is range in units of $\la_o$.}
\label{fig:imagC1all}
\end{figure}
\begin{figure}[H]
\begin{minipage}{1\textwidth}
\centering
\includegraphics[width=0.25\textwidth]{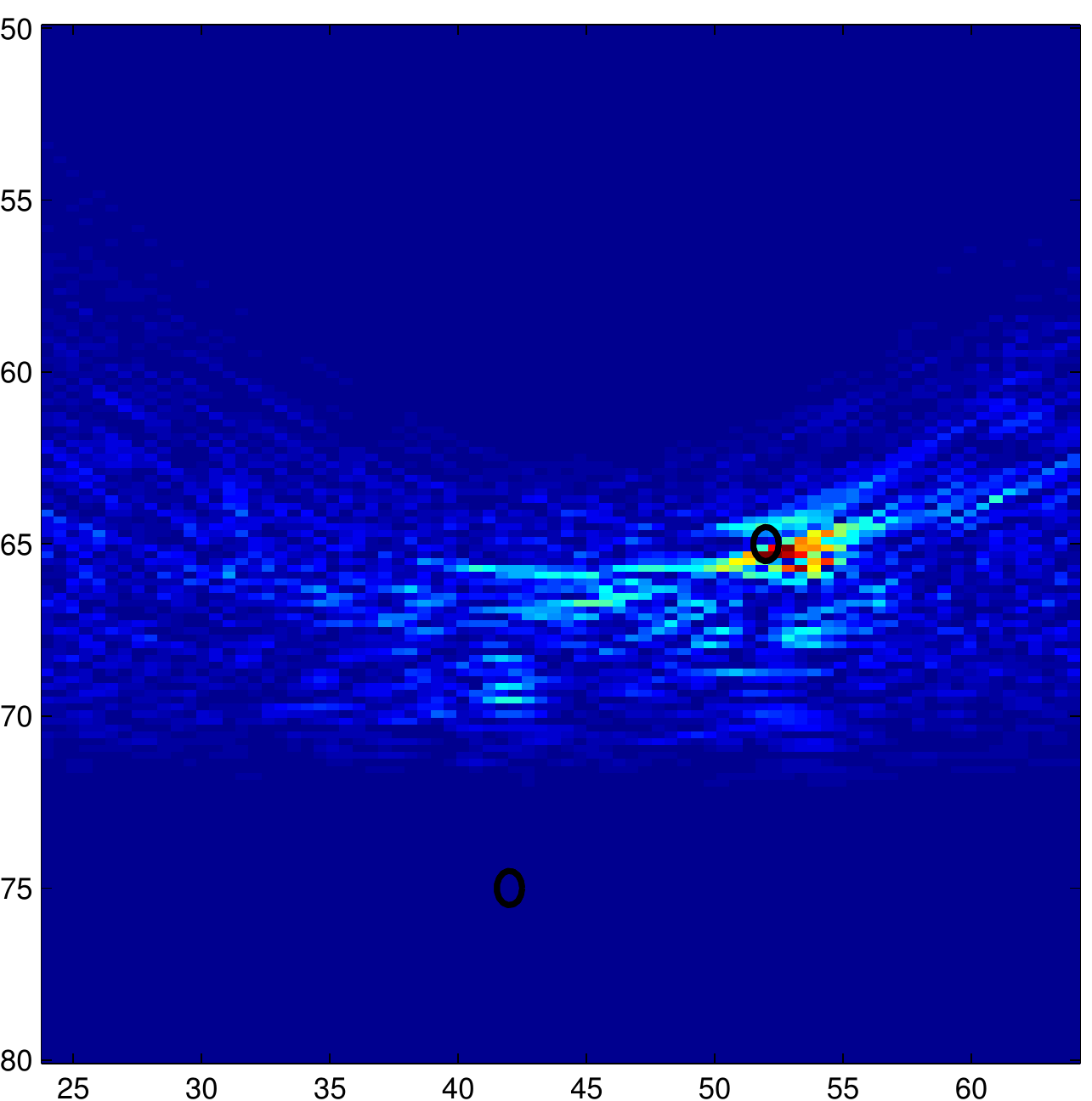}
 \includegraphics[width=0.25\textwidth]{./ImageDA_10C2TSC3_Level4_window7_Pband70_Neigs2_mul1_adaptive_average}
 \includegraphics[width=0.25\textwidth]{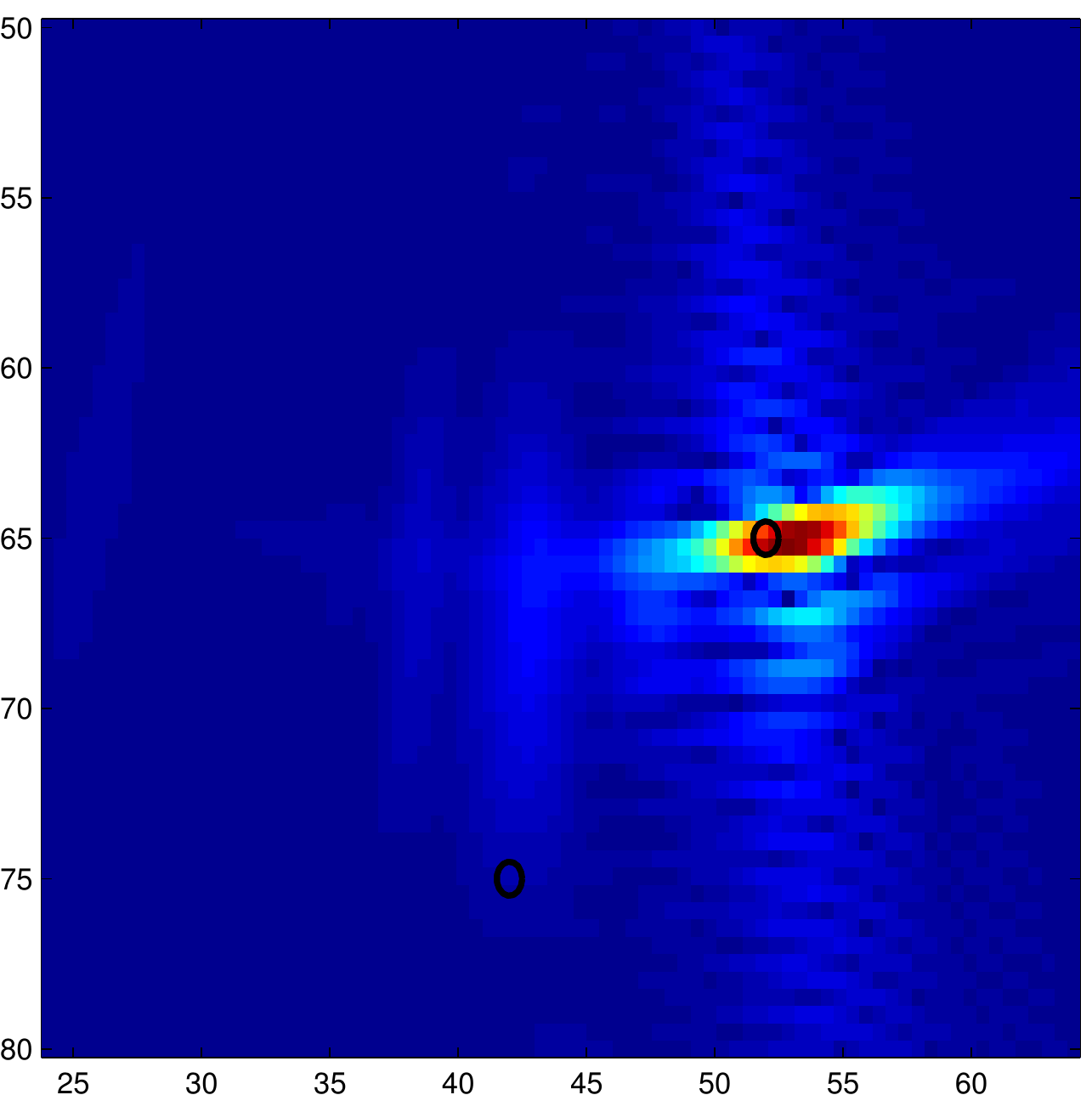}
\end{minipage}~\\
\begin{minipage}{1\textwidth}
\centering
\includegraphics[width=0.25\textwidth]{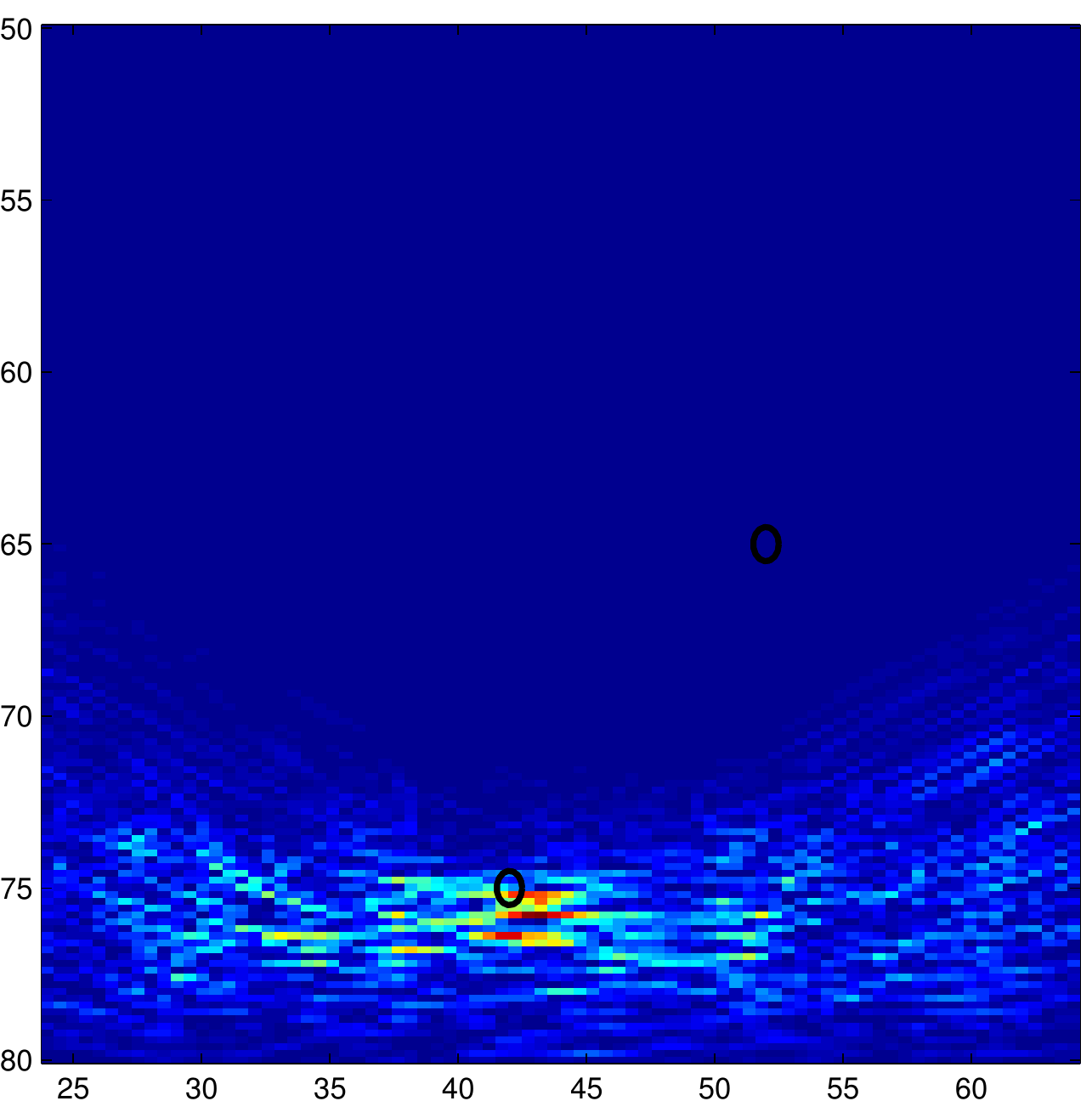}
 \includegraphics[width=0.25\textwidth]{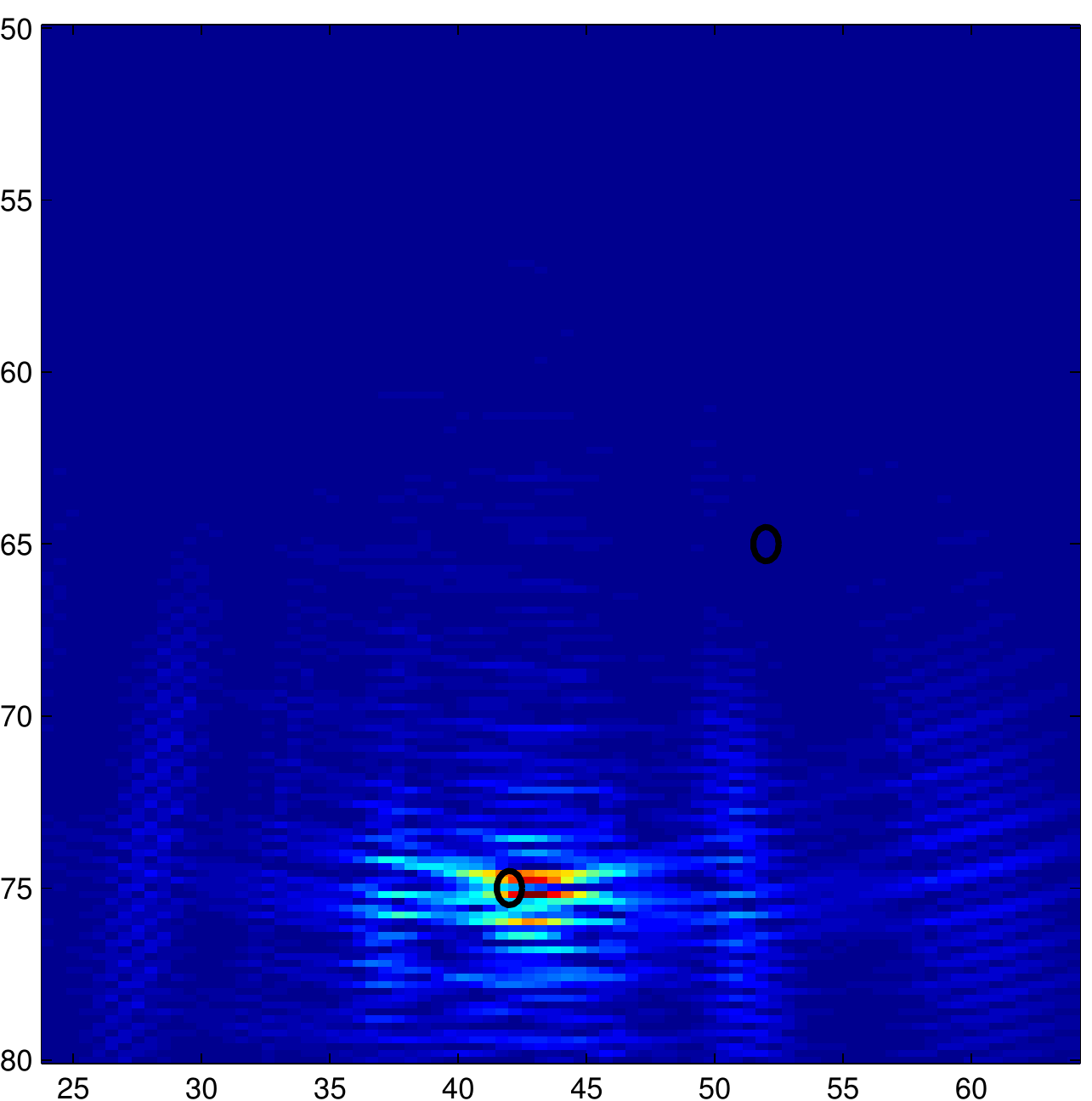}
 \includegraphics[width=0.25\textwidth]{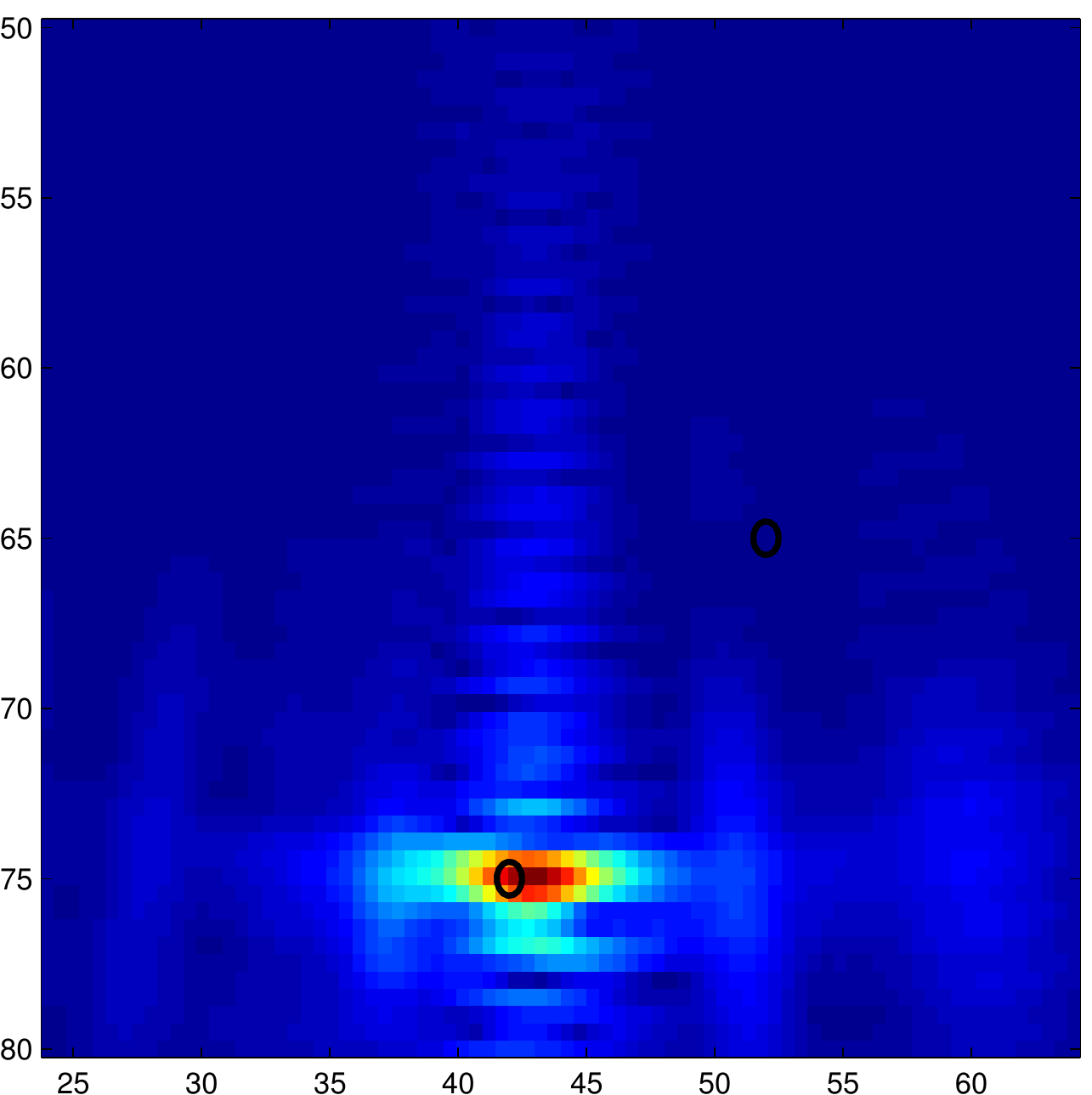}
\end{minipage}~\\
\caption{Imaging results for configuration 3 and the clutter shown in Figure \ref{fig:configsC}. Top row the images obtained with the data in the first selected time window 
and bottom row in the second selected time window. Left column:  KM images formed with $\bP^{^{IN}}\hspace{-0.04in}(t)$.
Center column: KM images formed with $\bP^{^{OUT}}\hspace{-0.04in}(t)$. Right column: CINT images formed with $\bP^{^{OUT}}\hspace{-0.04in}(t)$.  The abscissa is 
  cross-range in units of $\la_o$ and the ordinate is range in units of $\la_o$.}
\label{fig:imagC3all}
\end{figure}

The imaging results for configurations 1 and 3 are shown in  Figures \ref{fig:imagC1all} and \ref{fig:imagC3all}, 
and are a significant improvement over those in Figure \ref{fig:KMinitCall}. 
Because the direct echoes from the reflectors are well separated in time and are captured in two different windows, we image one reflector at a time. The images are good even before the filtering over the direction of arrival, but  this filtering sharpens the 
focusing of the images, specially in the third configuration. 

\begin{figure}[H]
\begin{minipage}{1\textwidth}
\centering
\includegraphics[width=0.25\textwidth]{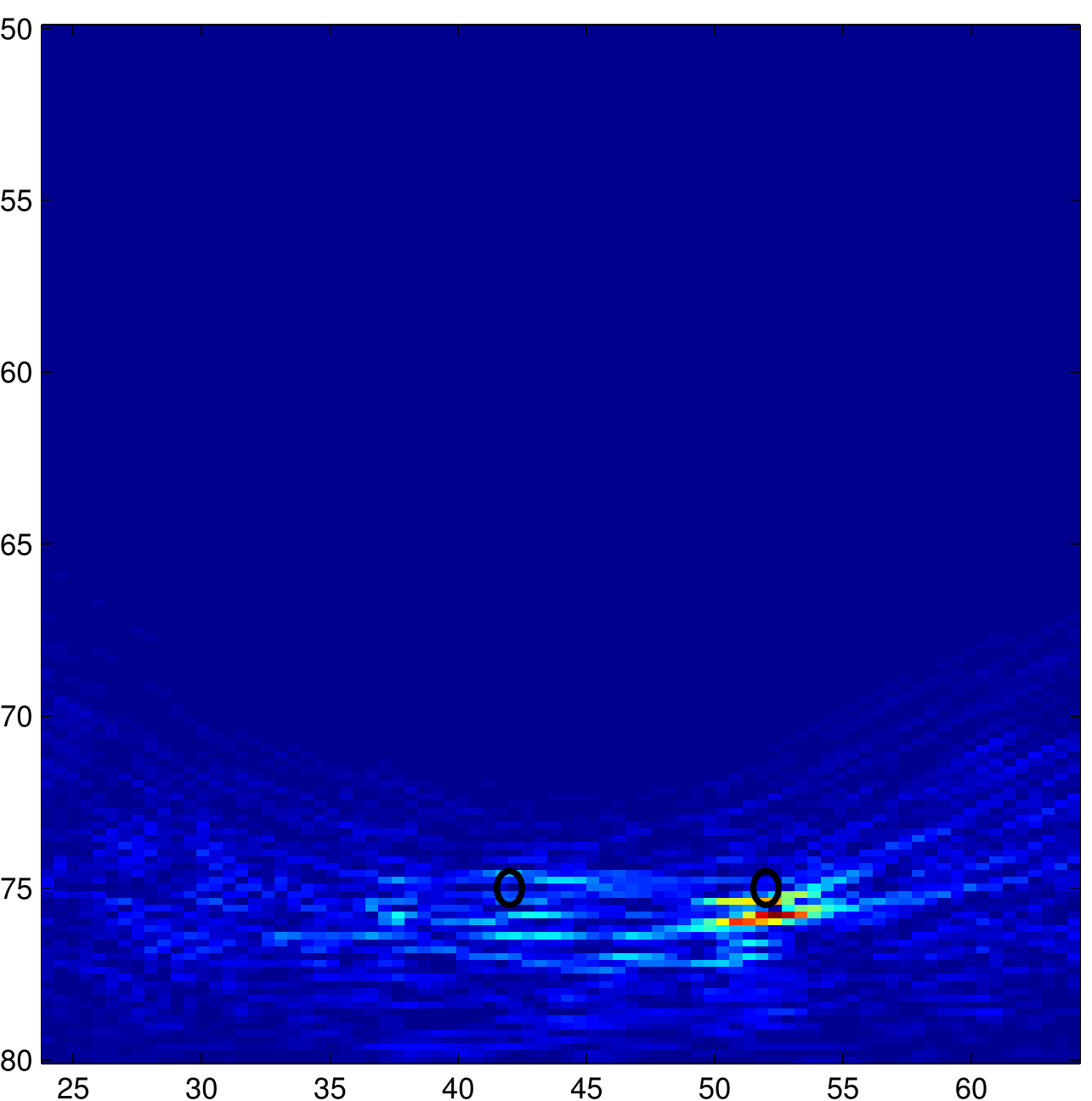}
\includegraphics[width=0.25\textwidth]{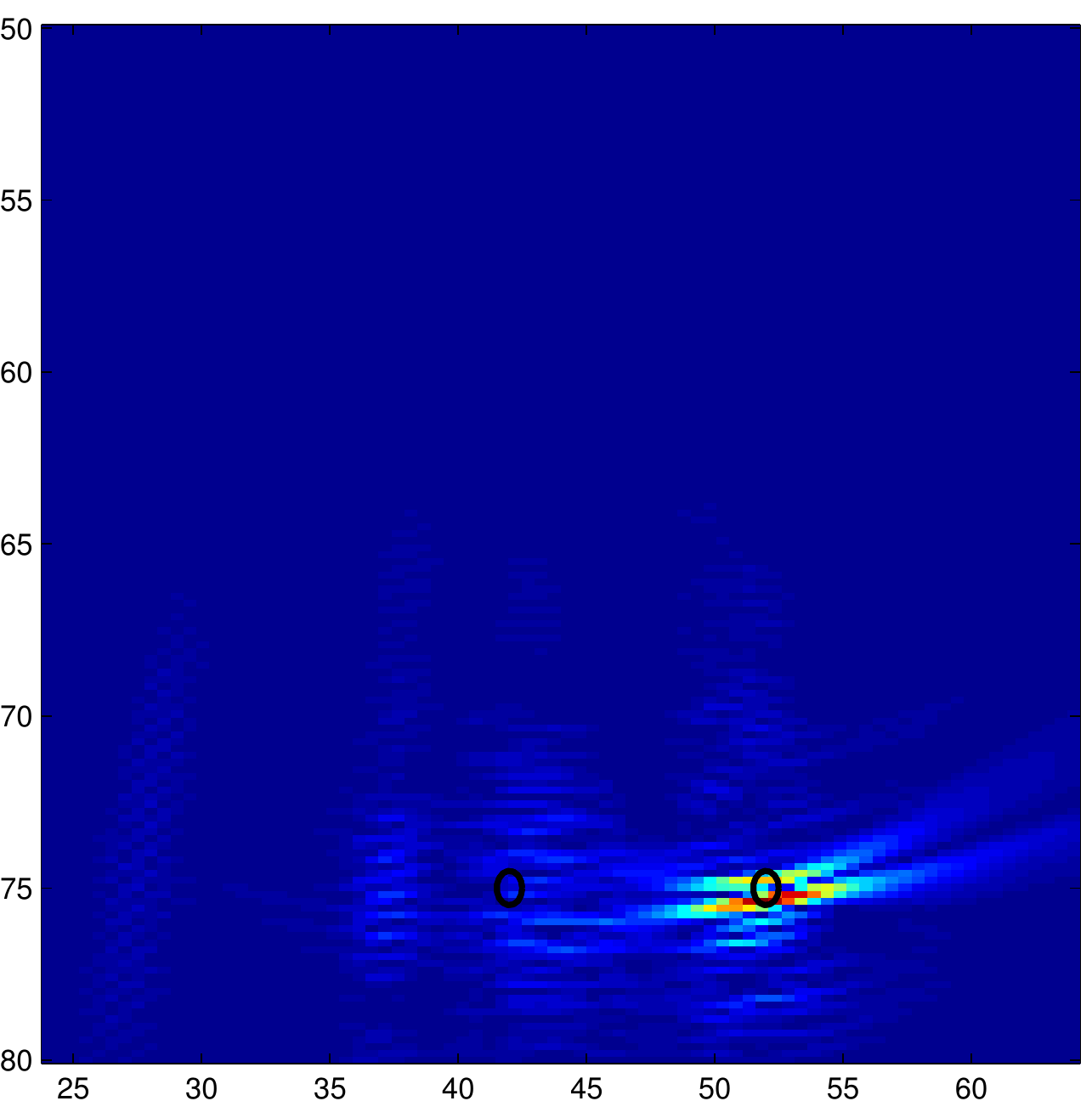}
\includegraphics[width=0.25\textwidth]{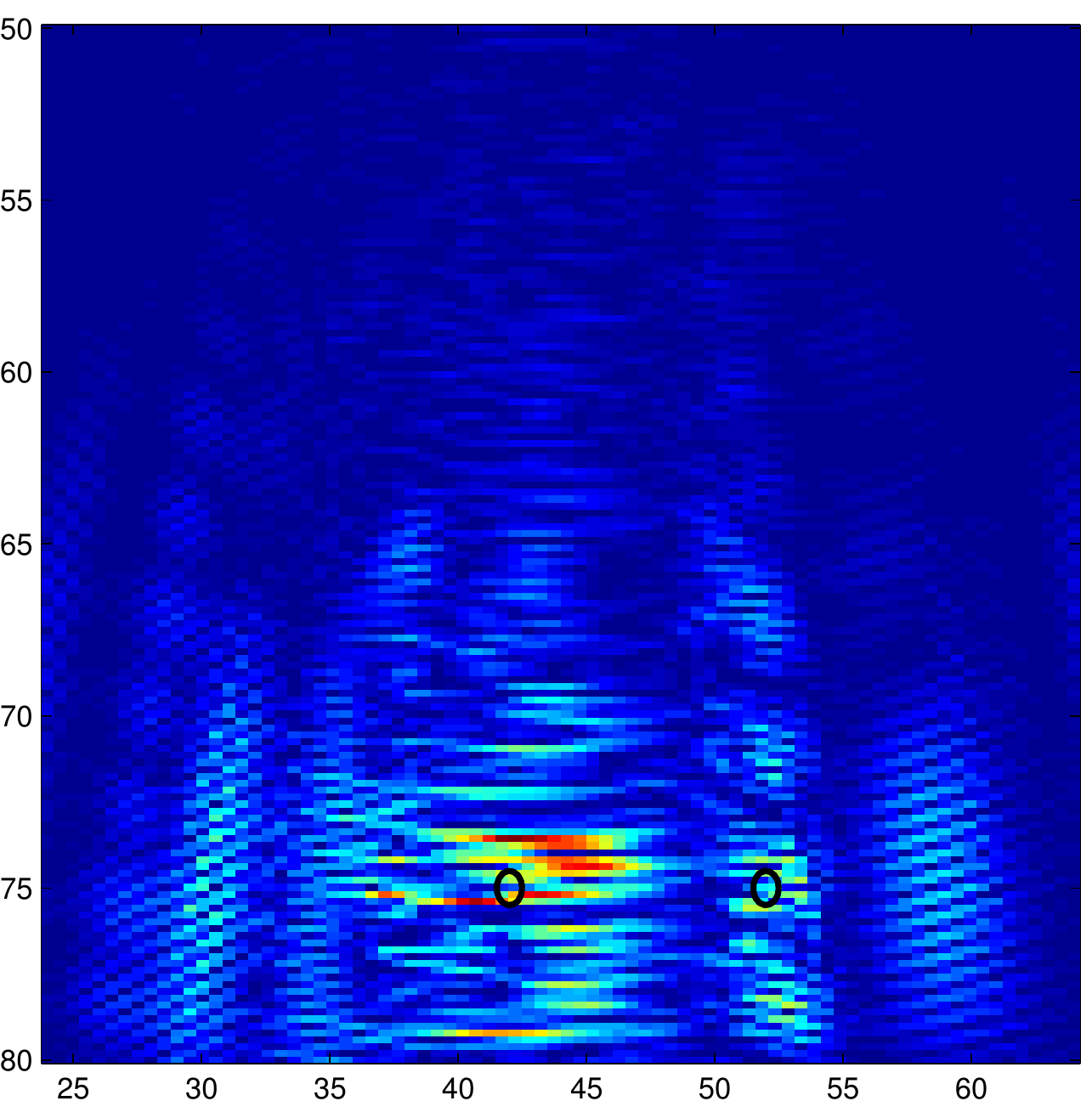}
~\\
\hspace*{0.25\textwidth}
 \includegraphics[width=0.25\textwidth]{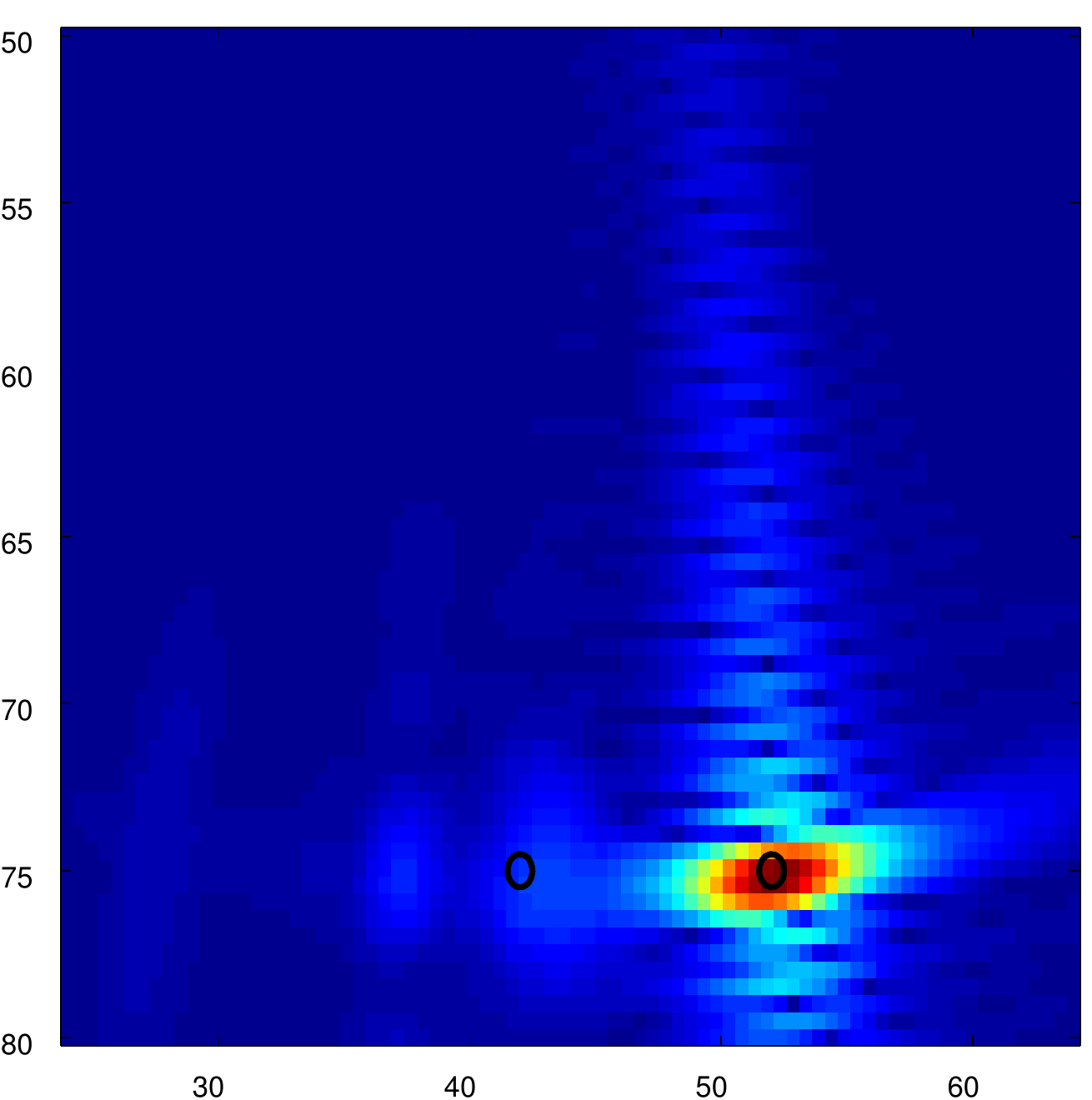}
 \includegraphics[width=0.25\textwidth]{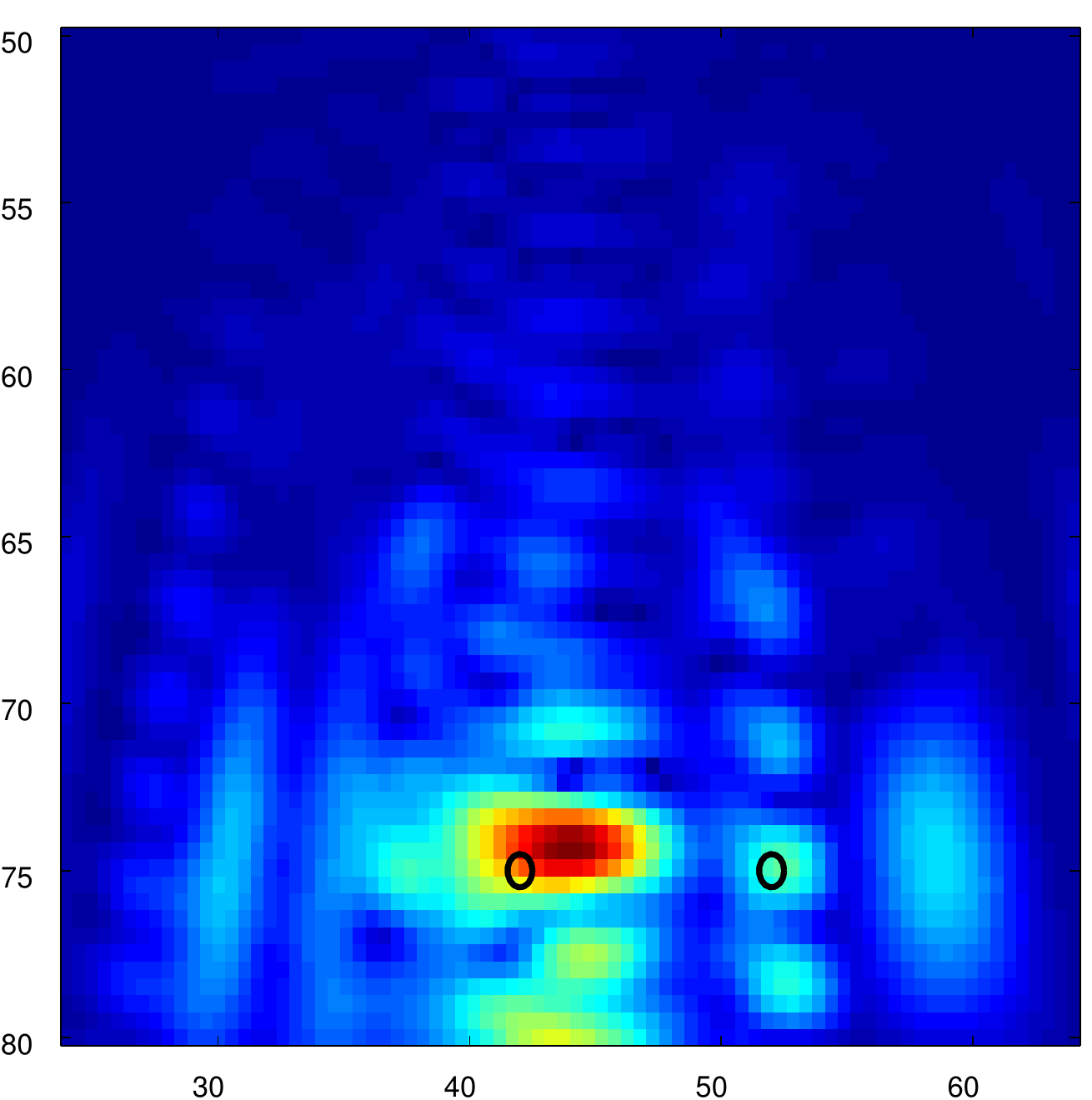}
\end{minipage}~\\
\caption{Imaging results for configuration 2 and the clutter shown in Figure \ref{fig:configsC}. There is a single selected 
time window containing the echoes from both reflectors. Left column:  KM image formed with $\bP^{^{IN}}\hspace{-0.04in}(t)$.
Top row middle and right column: KM  images formed with the filtered matrices $\bP^{^{OUT}}\hspace{-0.04in}(t)$ for the two selected 
arrival directions in Figure \ref{fig_adapt1p}. Bottom row: CINT images formed with $\bP^{^{OUT}}\hspace{-0.04in}(t)$ for the two selected 
arrival directions.  The abscissa is 
  cross-range in units of $\la_o$ and the ordinate is range in units of $\la_o$.}
\label{fig:imagC2all}
\end{figure}

The  direction of arrival  filtering is important in the second configuration, where the echoes from both reflectors 
arrive in the same time window.  Without it, only one reflector can be seen in the left image in Figure \ref{fig:imagC2all}.
The CINT method performs better than KM, as it mitigates the reverberations between the reflectors and the medium 
in their vicinity, as seen in the plots in the second row of the figure. 



%
\subsubsection{Imaging results for different types of clutter}

\begin{figure}[H]
\begin{minipage}{1\textwidth}
\centering
\includegraphics[width=0.25\textwidth]{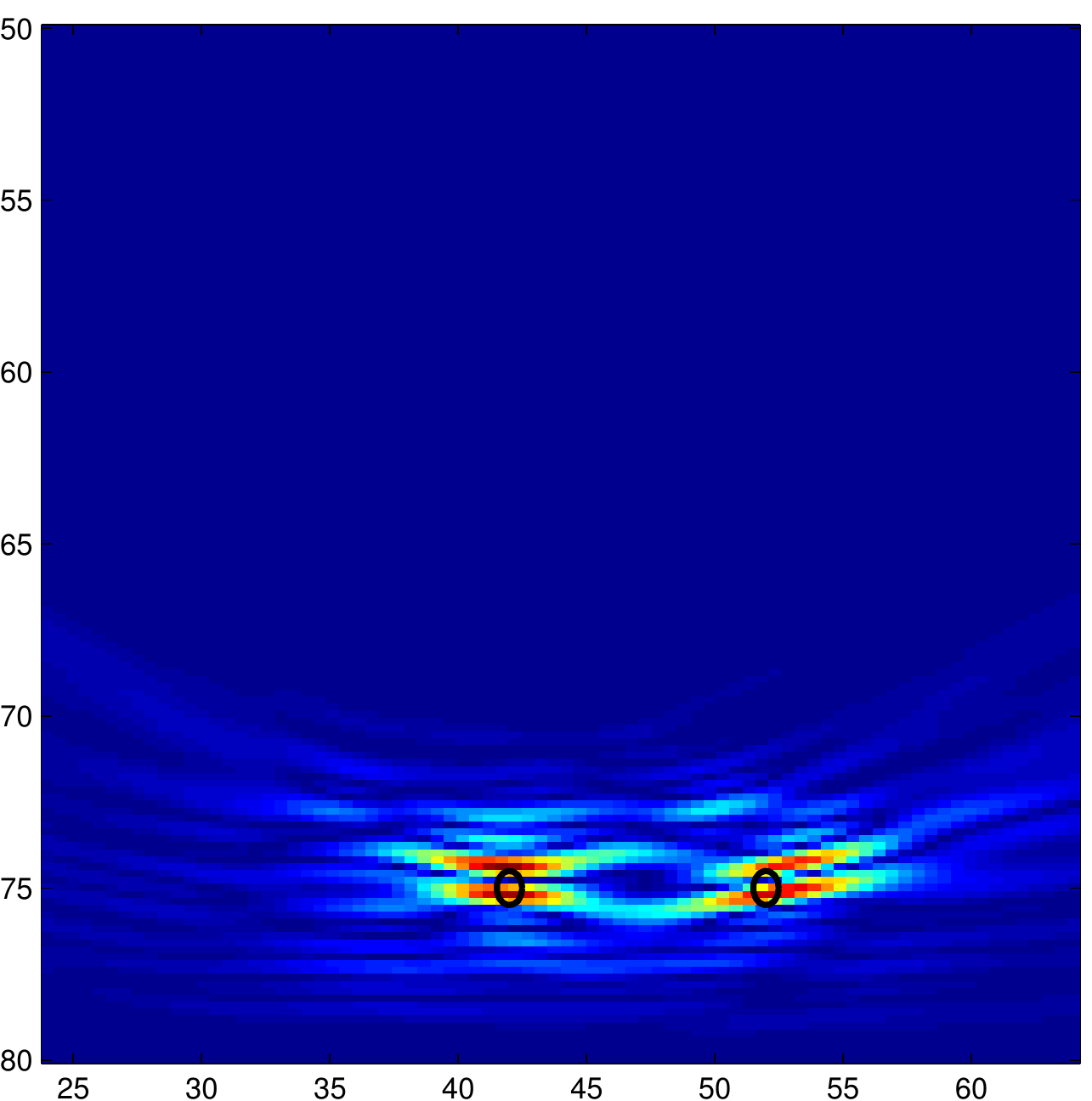}
\includegraphics[width=0.25\textwidth]{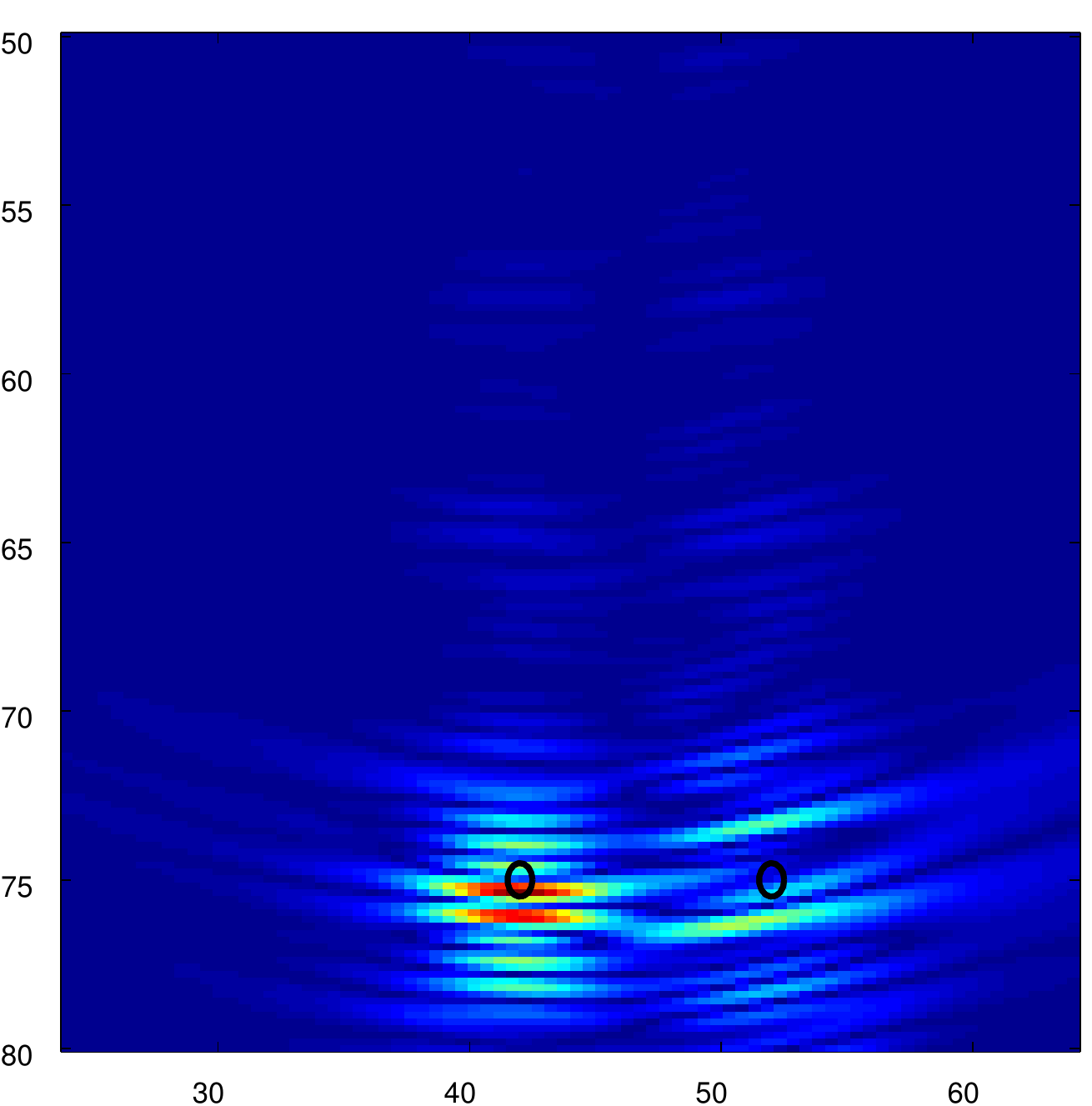}
\includegraphics[width=0.25\textwidth]{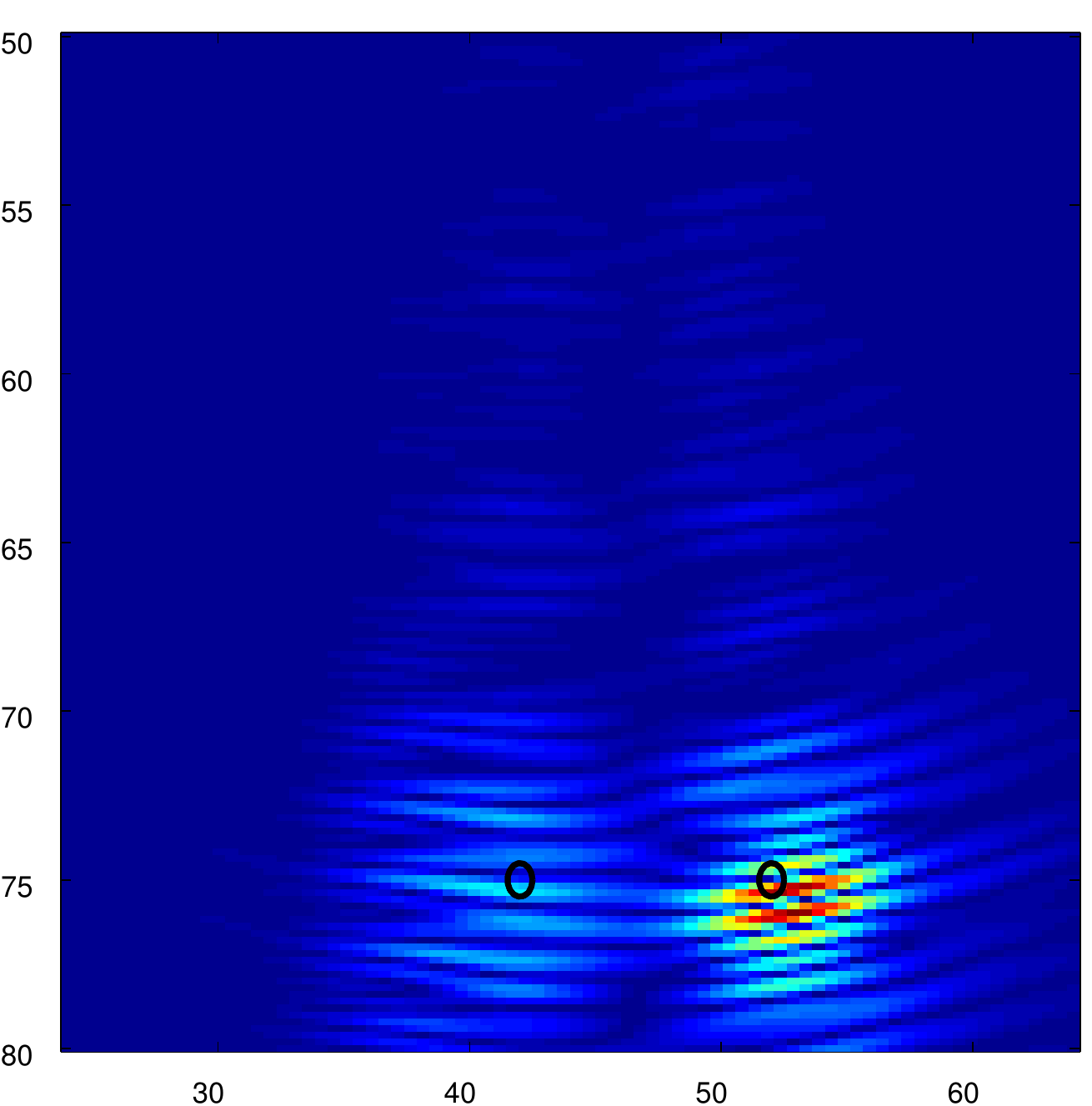}
~\\
\hspace*{0.25\textwidth}
 \includegraphics[width=0.25\textwidth]{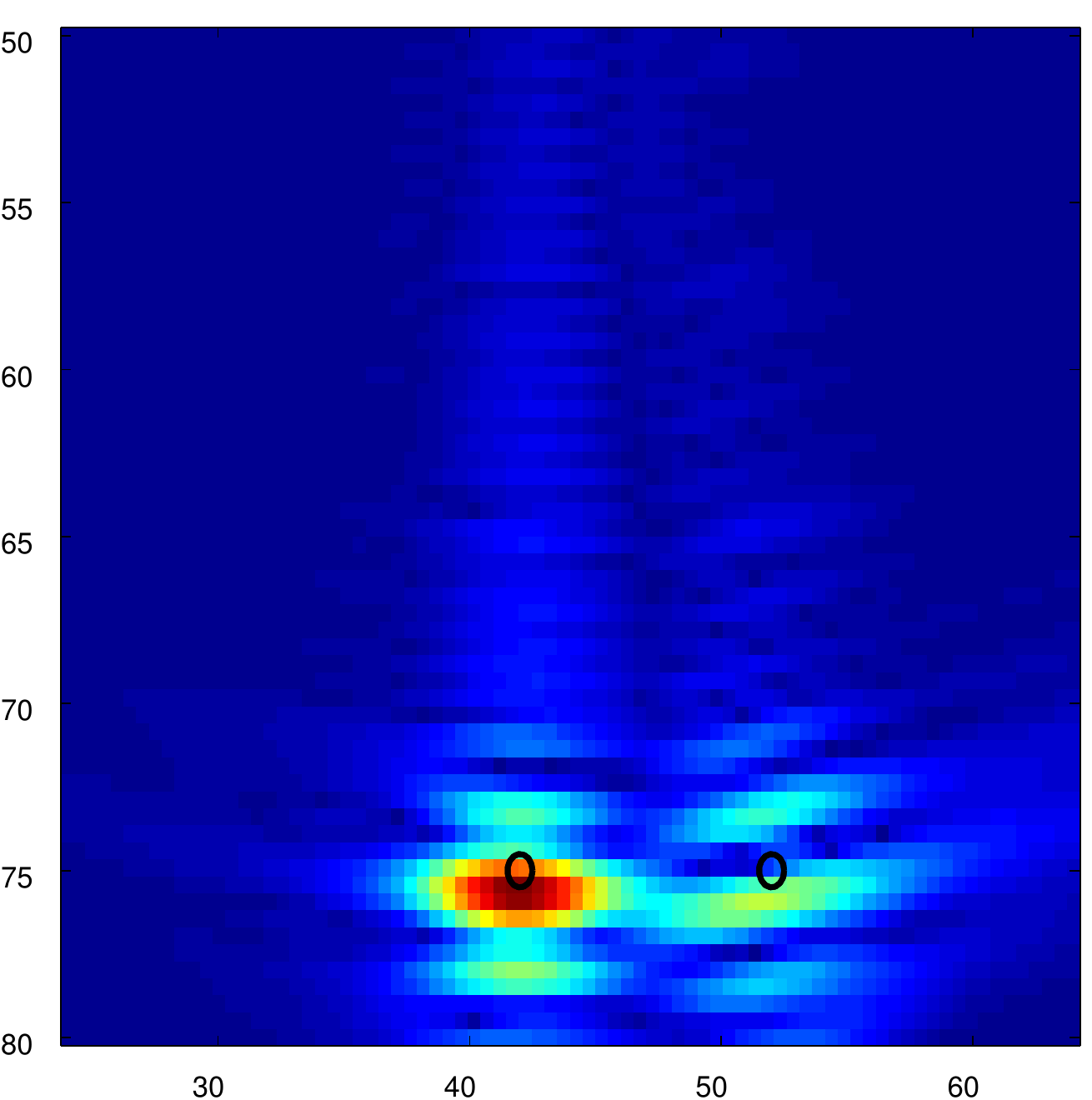}
 \includegraphics[width=0.25\textwidth]{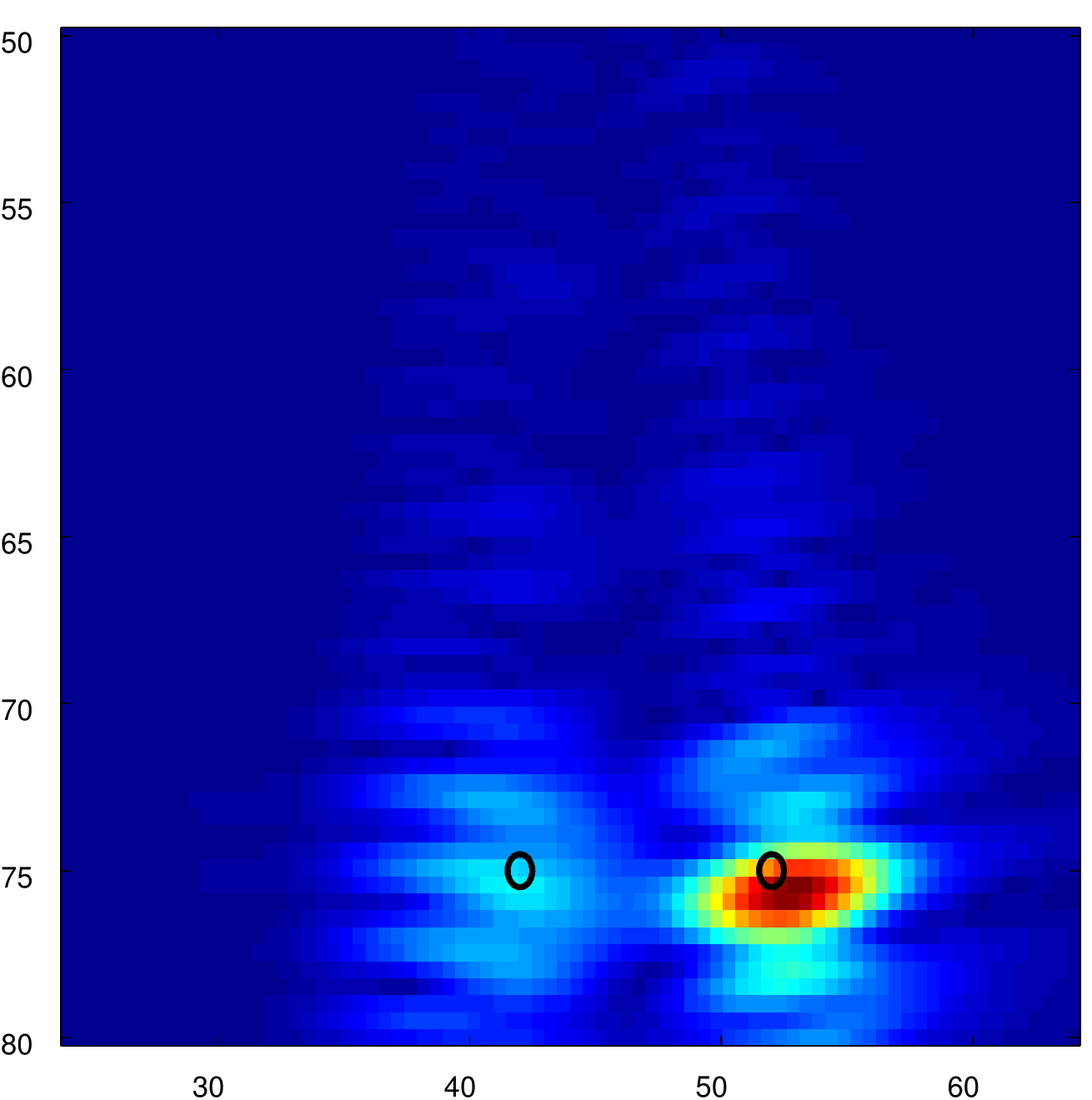}
\end{minipage}~\\
\caption{Imaging results for configuration 2 in the layered clutter. Left column:  KM image formed with $\bP^{^{IN}}\hspace{-0.04in}(t)$.
Top row middle and right column: KM  images formed with the filtered matrices $\bP^{^{OUT}}\hspace{-0.04in}(t)$ for the two selected 
arrival directions. Bottom row: CINT images formed with $\bP^{^{OUT}}\hspace{-0.04in}(t)$ for the two selected 
arrival directions.  The abscissa is 
  cross-range in units of $\la_o$ and the ordinate is range in units of $\la_o$.}
\label{fig:imagL2all}
\end{figure}

\begin{figure}[H]
\begin{minipage}{1\textwidth}
\centering
\includegraphics[width=0.25\textwidth]{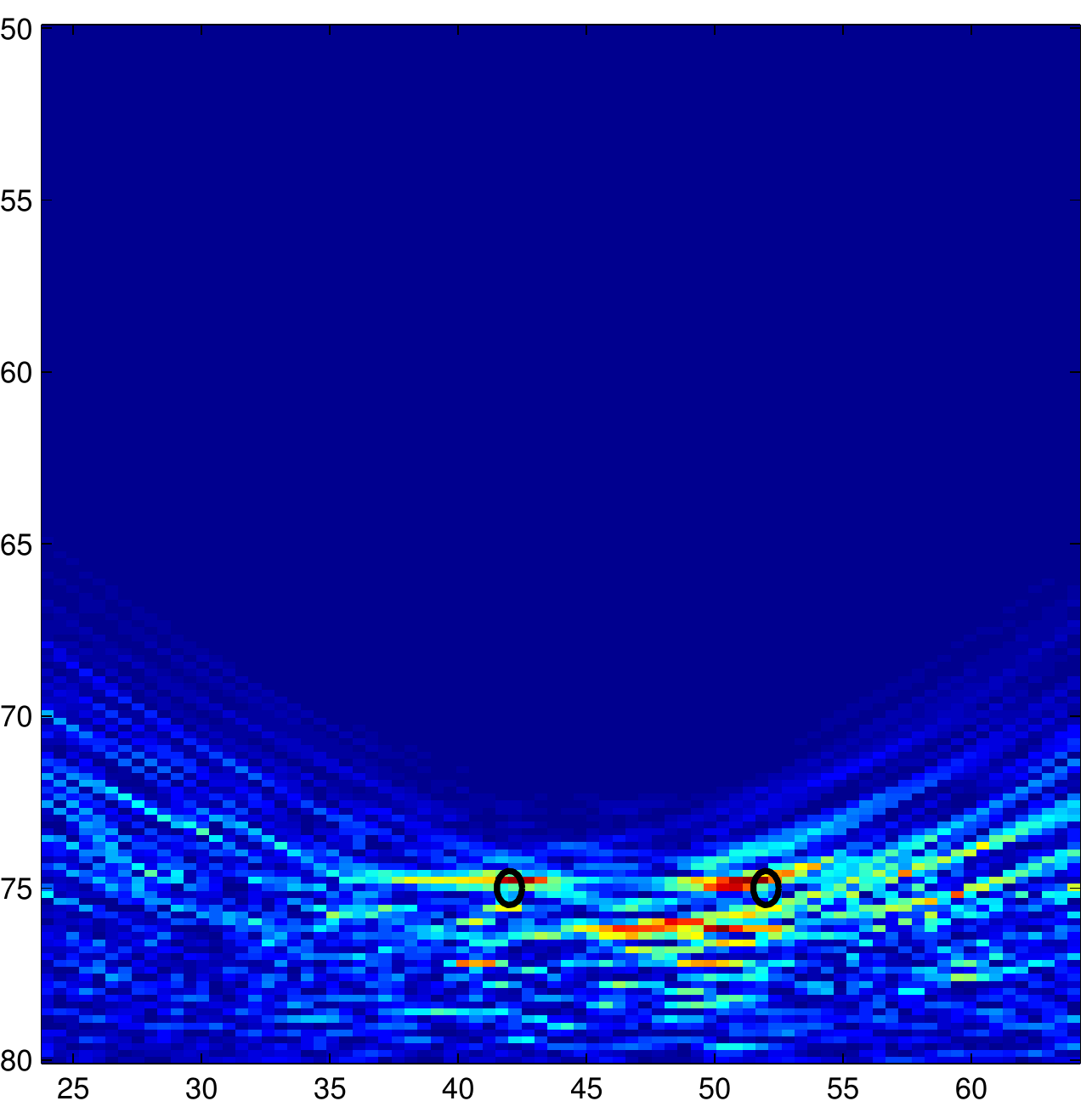}
\includegraphics[width=0.25\textwidth]{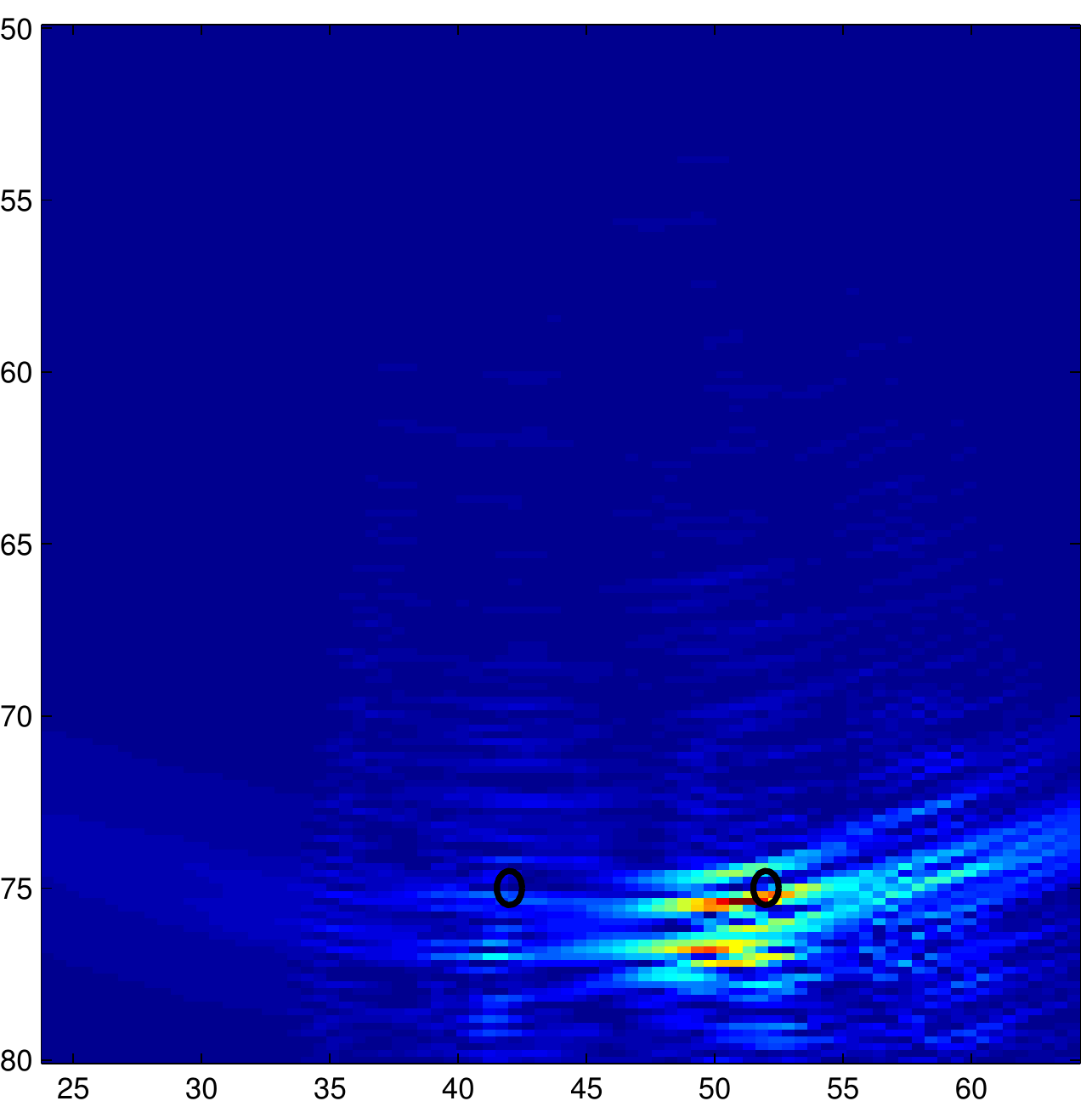}
\includegraphics[width=0.25\textwidth]{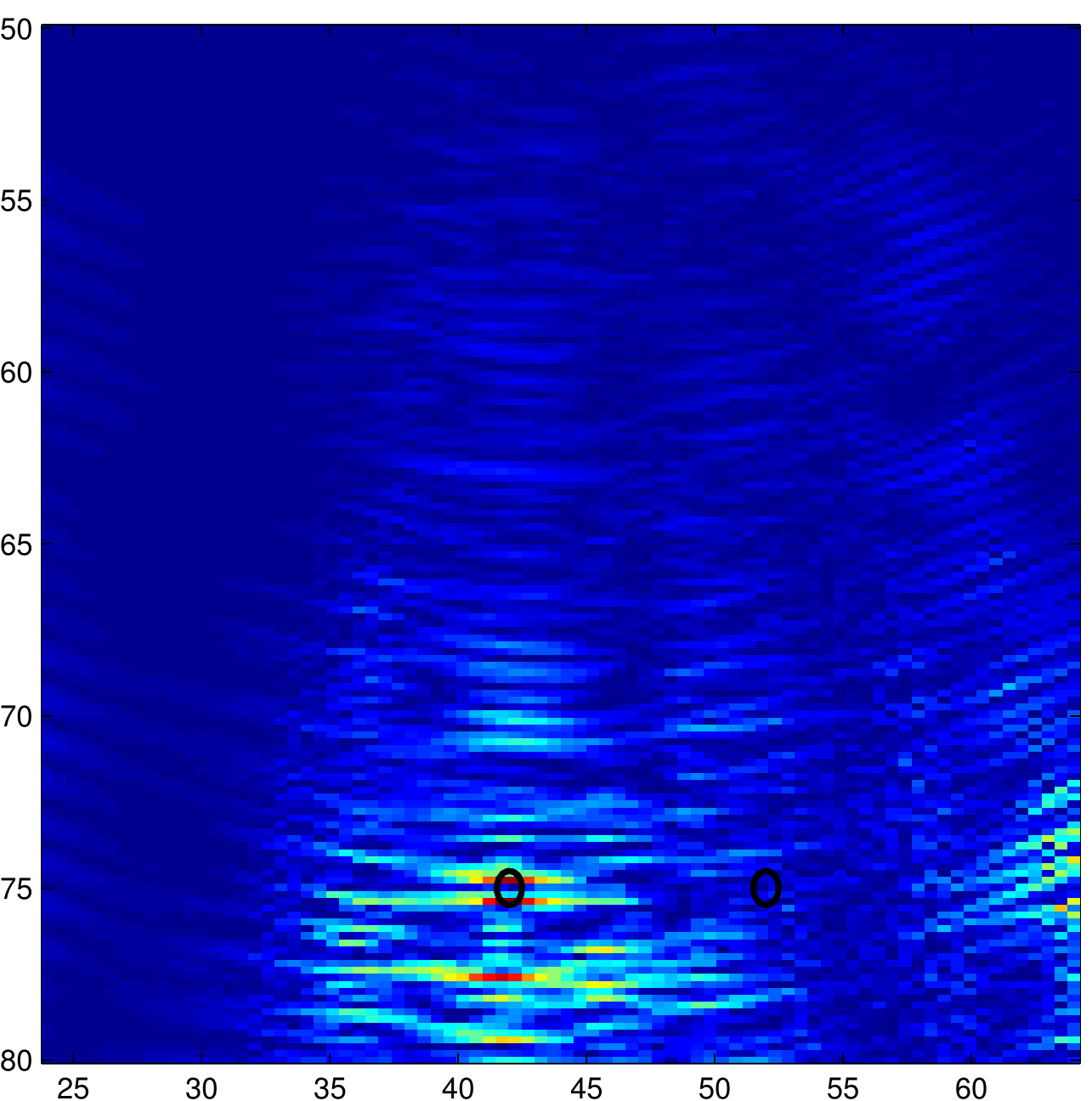}
~\\
\hspace*{0.25\textwidth}
 \includegraphics[width=0.25\textwidth]{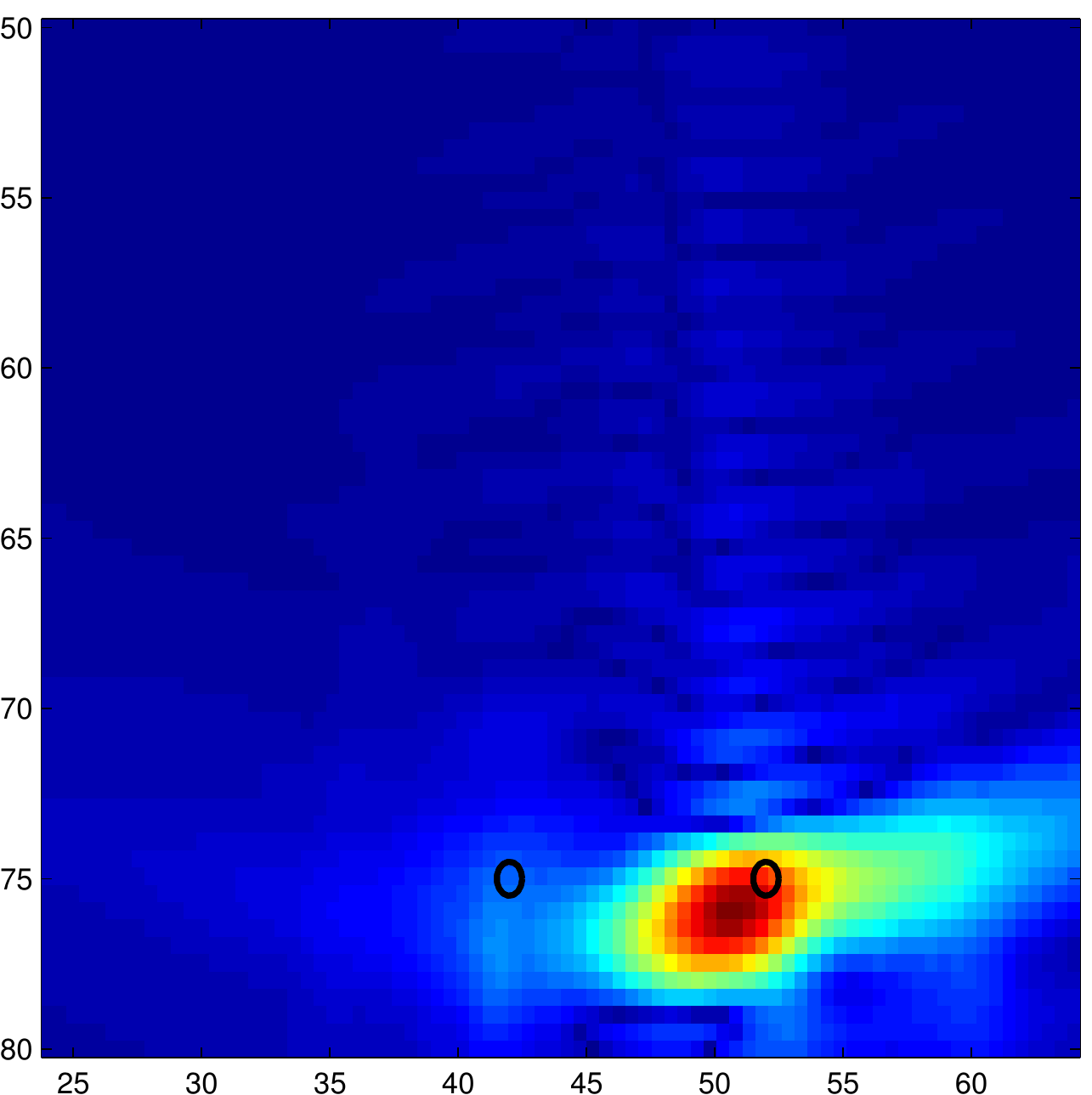}
 \includegraphics[width=0.25\textwidth]{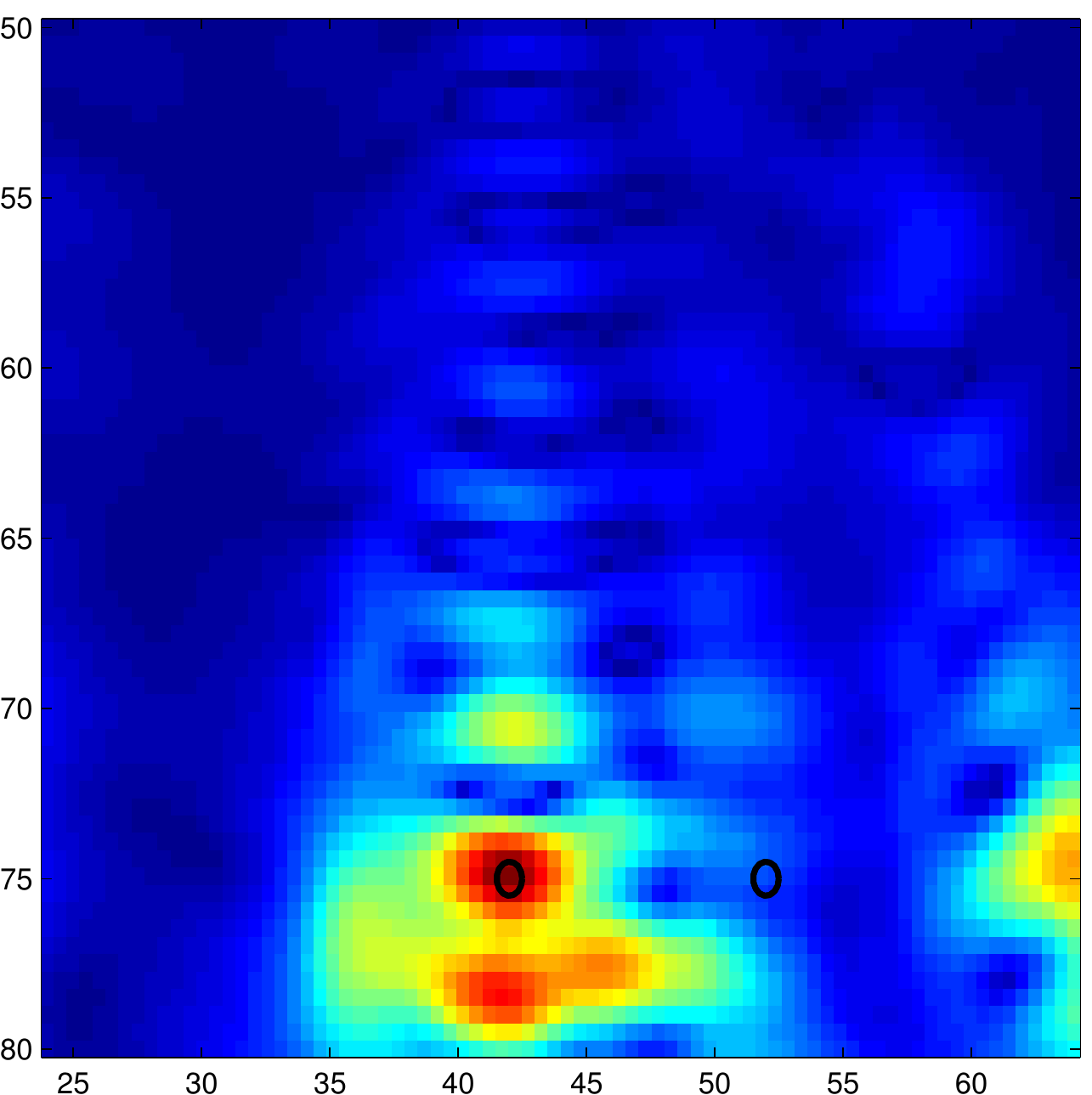}
\end{minipage}~\\
\caption{Imaging results for configuration 2 in the isotropic clutter. Left column:  KM image formed with $\bP^{^{IN}}\hspace{-0.04in}(t)$.
Top row middle and right column: KM  images formed with the filtered matrices $\bP^{^{OUT}}\hspace{-0.04in}(t)$ for the two selected 
arrival directions. Bottom row: CINT images formed with $\bP^{^{OUT}}\hspace{-0.04in}(t)$ for the two selected 
arrival directions.  The abscissa is 
  cross-range in units of $\la_o$ and the ordinate is range in units of $\la_o$.}
\label{fig:imagI2all}
\end{figure}

We display in Figures \ref{fig:imagL2all} and  \ref{fig:imagI2all}  the images of the two reflectors in the 
more difficult configuration 2, in layered and isotropic clutter, respectively. These complement  the images in Figure \ref{fig:imagC2all}. We note that the results are better in 
the layered case, as expected, because scattering in such  clutter does not scramble the direction of the arrivals. 
This is why we can clearly see both reflectors in the KM image formed with the matrix $\bP^{^{IN}}(t)$. 
The direction of arrival filtering does not improve the focusing of the images, it just separates the two reflectors. 
In  the isotropic clutter the imaging is more difficult, and the direction of arrival filtering is essential for focusing the 
images on the reflectors. As was the case in Figure \ref{fig:imagC2all}, CINT gives slightly better images, because
it mitigates the reverberations between the reflectors and the nearby clutter. 
%
%
%
%
%

\section{Summary}
\label{sec:sum}
We introduced and tested with numerical simulations a novel detection and data filtering method 
for coherent array imaging of small reflectors in strongly scattering media, called heavy clutter. The 
array is a collection of $N$ transducers which play the double role of sources and receivers. It uses the sources 
to probe the medium with pulses and records the scattered waves.
The data is organized in the $N\times N$ response matrix $\bP(t)$, which is a function of time. Because the 
medium reverberations (the clutter backscatter) dominate the recordings, it is difficult to distinguish the 
sought-after reflectors in the coherent images 
formed with  $\bP(t)$. These are noisy and difficult to interpret because they change from one clutter to another.

The clutter is not known in imaging applications, which is why we model the uncertainty of the wave speed in 
the medium with a random process. A good imaging method must produce results that are insensitive to 
the realizations of the  random wave speed i.e., be  statistically stable. When the direct (coherent) arrivals of the waves 
scattered at the reflectors are strong enough with respect to the clutter backscatter,  statistically stable 
imaging can be achieved with 
the coherent interferometric method (CINT) \cite{BPT-ADA}. Here we consider much stronger clutter, that cannot be handled by CINT
alone.

The detection and filtering method introduced in this paper is an improvement of that in \cite{BPT-LCT}. It determines
both the arrival time and direction of the weak coherent echoes, and  suppresses all the other arrivals, 
which are clutter backscatter. The arrival time detection involves an adaptive time-frequency analysis 
of the response matrix in sequentially refined time windows, using the singular value decomposition (SVD) of the local 
cosine transform (LCT) of $\bP(t)$. The SVD  of the Fourier transformed matrix $\widehat{\bP}(\om)$ 
has been used to improve imaging in many works, see for example \cite{PF-1994}. However, in our context it is not useful 
by itself, because the clutter backscatter carries most of the energy over the duration of the recordings. Therefore
$\widehat{\bP}(\om)$ is essentially a "noise" matrix, with no distinguishable singular values. Our  method uses the 
SVD in  
combination  with the LCT analysis, to search systematically for the time windows in which the coherent echoes arrive. 
These echoes are distinguishable from the clutter backscatter with the SVD, when the time windows are small 
enough.  

The detection of the direction of arrival of the coherent echoes is carried in the selected time windows, using their 
paraxial approximation. This approximation is justified for array apertures that are small with respect to the distance from the 
array to the reflectors, as is usually the case in practice. To use the paraxial approximation, we localize the data in time by  
backpropagating it to a reference point defined by the center time of the selected time windows, using travel time delays. 
This eliminates the large phase of the coherent echoes and more importantly, it removes their dependence on the source and receiver location offsets in the array. That is to say, it makes the coherent part of the backpropagated data a  Hankel matrix.  
The method exploits this fact by seeking the best approximation of the backpropagated data matrix by a Hankel matrix, and then 
uses plane wave decompositions of the result to detect the direction of arrival of the desired coherent echoes.  
This leads to improved focusing of images, as shown with numerical simulations carried in a realistic setup motivated by the 
application of non-destructive testing.

\section*{Acknowledgements}
The work of L. Borcea was partially supported by a ~\\NAKFI Imaging Science award. Support from NSF grant DMS-1510429 is also gratefully acknowledged.
The work of G. Papanicolaou was partially supported by the AFOSR grant FA9550-14-1-0275.
The work of C. Tsogka was partially supported by the European Research Council Starting Grant, GA 239959, 
the PEFYKA project within the KRIPIS action of the GSRT and the AFOSR grant FA9550-14-1-0275. 
\bibliographystyle{plain}
\bibliography{LCT}
\end{document}